\documentclass{article}
\usepackage{graphicx,color}
\usepackage{bm}
\usepackage{amsmath}
\begin{document}
\begin{center}
{\itshape \large Advances in Physics:}\\[0.3cm]
{\Large \bf Universal Behavior and the}\\[0.1cm]
{\Large \bf Two-component Character of Magnetically}\\[0.1cm] 
{\Large \bf Underdoped Cuprate Superconductors}\\[0.4cm]

{\large Victor Barzykin$^*$ and David Pines$^\dagger$}\\[0.1cm]
{\footnotesize $^*$ National High Magnetic Field Laboratory, 
Florida State University, Tallahassee, FL  32310, USA\\[0.1cm]
$^\dagger$ Department of Physics and Institute for Complex Adaptive Matter,
University of California, Davis, CA 95616, USA}
\end{center}
\begin{abstract}
We present a detailed review of scaling behavior in the magnetically  underdoped cuprate superconductors (hole dopings less than $0.20$) 
and show that it reflects the presence of two coupled components throughout this doping regime: a non-Landau Fermi liquid and a 
spin liquid whose behavior maps onto the theoretical Monte Carlo calculations of the 2D Heisenberg model of localized Cu spins 
for most of its temperature domain. We use this mapping to extract the doping dependence of the strength, $f(x)$  of the spin liquid 
component and the effective interaction, $J_{eff}(x)$ between the remnant localized spins that compose it; we find both decrease 
linearly with x as the doping level increases. We discuss the physical origin of  pseudogap behavior and conclude that it is 
consistent with scenarios in which the both the large energy gaps found in the normal state and their subsequent superconductivity 
are brought about by the coupling between the Fermi liquid quasiparticles  and the spin liquid excitations, and that differences 
in this coupling between the 1-2-3 and 2-1-4 materials can explain the measured differences in 
their superconducting transition temperatures and other properties. 
\end{abstract}

\tableofcontents

\section{Introduction}

Explaining the anomalous normal state properties of the so-called pseudogap 
regime of the underdoped cuprate superconductors is widely 
regarded as an essential step toward understanding the basic physics of these 
materials and unlocking the mechanism of their superconductivity\cite{NPK}. Perhaps the most 
striking aspect of these is the universal, or scaling, behavior, first identified in 
measurements of their temperature-dependent uniform magnetic susceptibility\cite{johnston}, 
and since found in Knight shift, transport, and entropy  measurements. In the present 
article we present a detailed review of scaling behavior in the underdoped cuprates 
that extends previous analyses of its manifestations in both static and low frequency 
dynamic behavior as well as that seen in inelastic neutron scattering (INS) experiments. 
Our review updates our earlier analysis\cite{BPprl} and the results presented in Norman \textit{et al.}\cite{NPK}, 
and complements the recent review of gap behavior presented in H\"{u}fner \textit{et al.}\cite{hufner}. 

We find that from zero hole doping until planar doping levels of  $\sim 0.2$ are reached, 
the scaling behavior seen by probes of magnetic behavior reflects the presence of a spin 
liquid whose behavior maps onto the theoretical Monte Carlo calculations of the 2D Heisenberg 
model of localized $Cu$ spins\cite{MD} for most of its temperature domain. We use this to extract the 
doping dependence of the strength, $f(x)$ of the spin liquid component and the effective interaction, 
$J_{eff}(x)$ between the remnant localized spins that compose it; for $x < 0.18$, $f(x) = 1 - [x/0.20]$ for 
both the 2-1-4 and 1-2-3 materials, while, to first approximation, 
$J_{eff}(x)= J f(x)$,  where $J$ is their interaction at zero doping level. A careful analysis of the 
NMR experiments on both classes of materials makes it possible to identify a quantum critical point 
at a doping level, $x=0.05$ that represents a phase transition from short range to long range 
order in the spin liquid.  It leads to dynamic $z=1$ scaling behavior for a wide range of doping levels 
that extends up to $T^m$, the temperature at which the static susceptibility is maximum, corresponding to 
an antiferromagnetic correlation length of order unity. We find that the extent of this scaling behavior 
is different for the 2-1-4 and 1-2-3 materials: for the former it persists down to temperatures of order the 
superconducting  transition temperature; for the latter it cuts-off at $T^*$, a temperature that it considerably 
greater than $T_c$ over most of the doping range.  

In addition to the spin liquid, whose properties dominate the low frequency magnetic response, bulk susceptibility 
measurements reveal the presence of a second component, a Fermi liquid that makes a temperature independent, 
but doping dependent contribution to this quantity for temperatures greater than $T^*$ and
doping levels of $0.05$ upwards.  We present a simple interpretation of the two fluid description of these 
coupled liquids\cite{BPprl} in terms of the incomplete hybridization of the Cu $d$  and O $p$ bands; 
the spin liquid corresponds to the unhybridized $d-d$ component, while the Fermi liquid has a large Fermi 
surface as a result of the $d-p$ hybridization. We derive the strength of the Fermi liquid component, 
which goes as $[1-f(x)]$ and so is proportional to $x$, and show how the presence of the spin liquid 
is incompatible with the single band Hubbard and Zhang-Rice approximations. 

We conclude that experiment has now provided the answer to the question of the physical mechanism responsible for the 
remarkable pseudogap behavior seen in the underdoped 1-2-3  materials ($x< 0.20$, say). When the present analysis 
is combined with the recent ARPES experiments\cite{campuz} and the STM measurements of the Davis\cite{davis} and 
Yazdani\cite{yazdani} groups, 
a simple physical picture emerges. In the  "normal state", for temperatures above $T^* \simeq T^m/3$,  
one has two quasi-independent components: 
a spin liquid of localized Cu spins described by the 2D Heisenberg model,  whose strength and effective interaction 
become weaker as the doping level increases; and a (non-Landau) Fermi liquid with a large Fermi surface whose strength 
increases with doping and whose  transport properties are determined primarily by its coupling to the 
spin liquid.  At $T^*$ the system makes a transition to a remarkable new quantum state of matter: a state that 
possesses a single  d-wave like gap, with a maximum gap value of order $4 T^*$ , that only becomes superconducting 
at the typically much lower  superconducting transition temperature, $T_c$.  The physical mechanism for the transition 
at $T^*$ (and subsequently at $T_c$) in the 1-2-3 materials is magnetic because the scale of $T^*$ and the gap is set 
by the effective interaction between the localized spins in the spin liquid. Matters are somewhat different for the 2-1-4 
materials and we speculate as to why this is the case.

Our review is organized as follows. In Section 2 we
review the literature on  experimental measurements and corresponding analyses  that indicate universal  scaling behavior.  
In Section 3 we introduce the phenomenological  two-fluid model and use it to  analyze existing  magnetic and thermodynamic  
measurements on the bulk spin susceptibility, the entropy, and the spin fluctuation spectrum revealed in nuclear magnetic 
resonance and  inelastic neutron scattering experiments. In Section 4 we present our conclusions concerning the
interaction between the Fermi liquid quasiparticles and the spin liquid excitations, discuss the similarities and
differences between the 2-1-4 and 1-2-3 materials, and consider the constraints imposed by experiment on microscopic 
theories of their high-temperature superconductivity. We present our conclusions in Section 5.

\section{An overview of experiments suggesting universal behavior}

  The observation of scaling in the cuprates is not new\cite{BP}.
Not long after the discovery of the cuprate superconductors\cite{BM}, 
universal behavior was identified in the magnetic properties of the 2-1-4 materials  
by Johnston\cite{johnston} through an analysis of his  measurements of the bulk spin susceptibility; 
his analysis was later confirmed by Nakano \textit{et al.}\cite{nakano} and Oda \textit{et al.}\cite{oda}. 
The Johnston-Nakano scaling  analysis was 
subsequently extended by Wuyts \textit{et al.}\cite{wuyts} to the Knight shift measurements of 
Alloul \textit{et al.}\cite{alloul} in the 1-2-3 family; more recently it has been shown to be applicable to the 
1-2-4 and several other members of the 1-2-3 family by Curro \textit{et al.}\cite{cfp} and the authors\cite{BPprl}. 
A number of other  experiments also indicate scaling and data collapse. These include electronic heat capacity 
measurements\cite{loram,loram04} for which an analysis of the magnetic entropy found scaling behavior, quantum critical (QC) 
scaling behavior in NMR copper nuclear spin-lattice relaxation rates\cite{imai}, limits on the $T$-linear behavior of 
resistivity\cite{ito,wuyts}, scaling of Hall resistivity\cite{Hwang,carrington,wuyts,GT1}, $\omega/T$ scaling in inelastic neutron 
scattering\cite{keimer,birgeneau,sternlieb,aeppli}, and 
finite-size scaling in the insulating cuprates\cite{cho}. Recently, scaling behavior has been discovered in the 
doping-dependence of  ARPES (angle-resolved photoemission) and STM experiments on the 2-2-1-2 members of 
the 1-2-3 family; similar characteristic temperatures set the scale for the appearance of Fermi arcs\cite{campuz}, 
and  "normal state" gap behavior\cite{campTst}.

In this section we review the above experiments and find that the characteristic scaling temperature 
first identified by Johnston, the temperature, $T^m(x)$, at which the temperature and doping 
dependent bulk magnetic susceptibility reaches its maximum value, provides the common thread
that links these together.

\subsection{Direct measurements of the magnetic susceptibility}

\begin{figure}
\begin{center}
\includegraphics[width=0.75\textwidth]{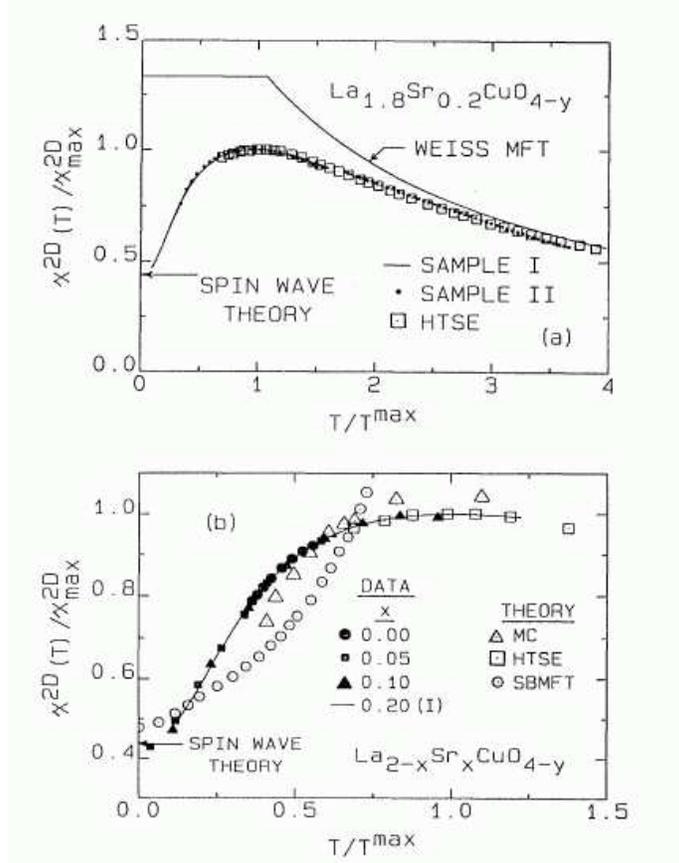}
\caption{Cu$^{2+}$ sublattice susceptibility $\chi^{2D}(T)/\chi^{2D}_{max}$ vs $T/T^{max}$ for five
different samples of La$_{2-z}$Sr$_z$CuO$_{4-y}$ with effective doping levels $x = z-2y = 0.0$, $x = 0.05$,
$x = 0.1$, $x = 0.13$, and $x = 0.2$, compared with theoretical 2D Heisenberg calculations and the Weiss
molecular-field (MFT) prediction. Reproduced with permission from \cite{johnston}.}  
\label{johnstonfig}
\end{center}
\end{figure}

The scaling behavior of the temperature dependent bulk spin susceptibility, $\chi(T)$, was discovered 
empirically  by Johnston\cite{johnston}, who showed that an excellent
collapse (Fig.\ref{johnstonfig}) of his experimental data on five samples of La$_{2-z}$Sr$_z$CuO$_{4-y}$, in which the doping level $x = z - 2 y$
ranged from $x = 0$ to $x = 0.2$, could be obtained if $\chi(x,T)$ had the following scaling form:
\begin{equation}
 \chi(x,T) = \chi_0(x) + [\chi^m(x) - \chi_0(x))] F(T/T^m(x)),
\label{john1}
\end{equation}
where $x$ is the doping level, $\chi_0(x)$ is a doping-dependent, temperature independent term,
$\chi^m(x) = \chi(x,T^m(x))$ is the maximum value of $\chi$ for a given doping level, and
$T^m(x)$ is the doping-dependent temperature at which $\chi(T)$ is maximum.
Johnston concluded that the scaling parameter $T^m$ depends only on the hole
doping level, $x$, in the plane and that the scaling function $F(T/T^m)$ is the same
as that calculated for the 2D Heisenberg model in its spin liquid regime (i.e. at temperatures
above the Ne\'el ordering temperature). In this model $T^m = 0.93 J$, where
$J$ is the nearest neighbor exchange coupling between localized Cu spins.
The temperature independent $\chi_0(x)$ was assumed to include $x$-independent core and Van Vleck 
contributions to $\chi$, and an $x$-dependent Fermi liquid contribution
that grew somewhat slower than linearly with increased doping $x$. 

\begin{figure}
\begin{center}
\includegraphics[width=0.75\textwidth]{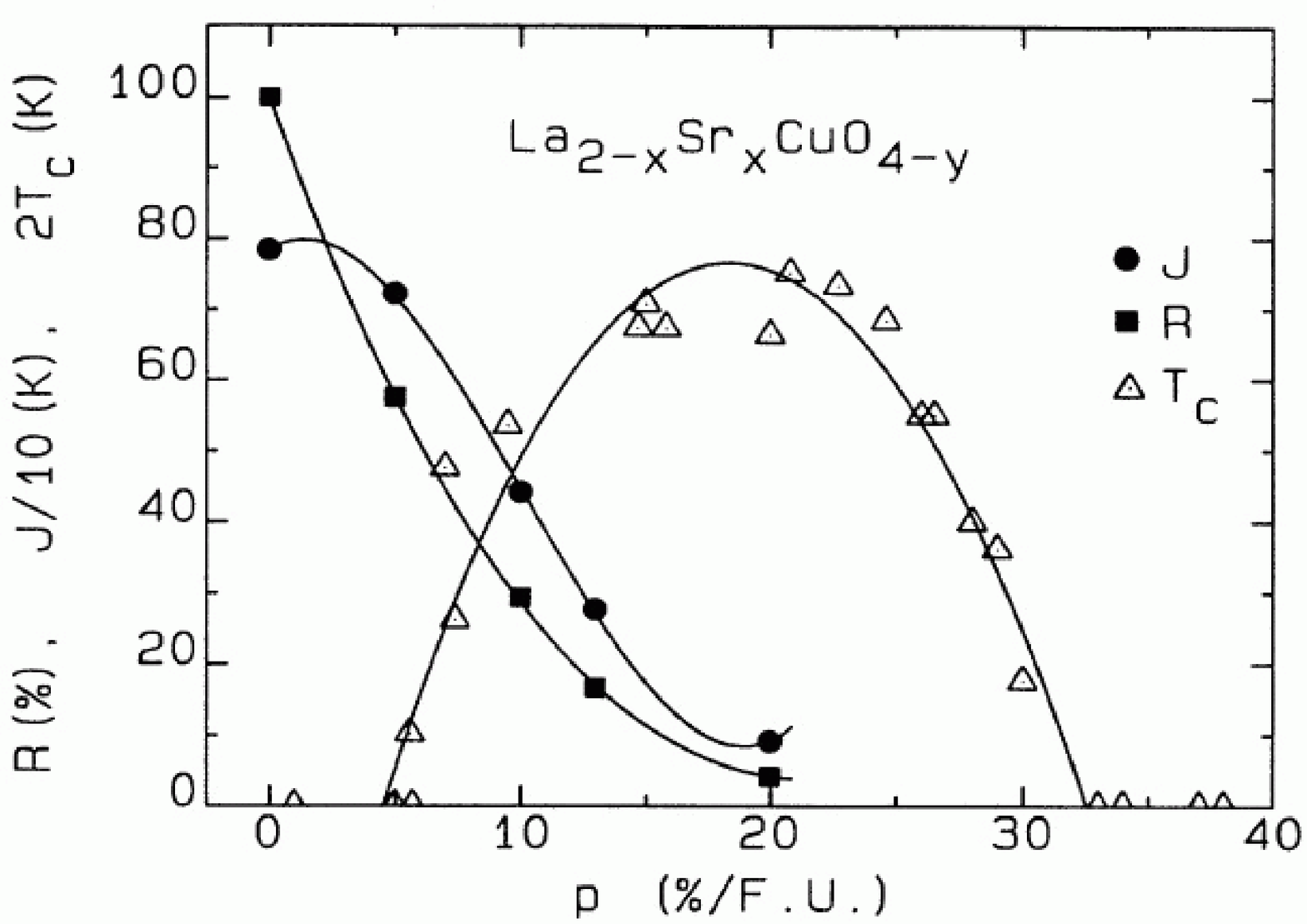}
\caption{Effective intralayer Cu-Cu exchange constant $J$ and the ratio 
$R = \chi^{2D}_{max}/\chi^{2D calc}_{max}$ vs. hole doping $p=x$. $T_c(p)$ is
shown for comparison. The solid curves are guides to the eye. Reproduced with permission from \cite{johnston}.}
\label{johnstonR}
\end{center}
\end{figure}

To explain the scaling behavior of Eq.(\ref{john1}), Johnston\cite{johnston} introduced a doping-dependent 
2D Heisenberg exchange constant, $J(x)$, and the ratio
\begin{equation}
R(x) = \frac{\chi^m(x)}{\chi^m_{calc}(x)}\,,
\end{equation}
where $\chi^m_{calc}(x)$ is the result obtained for the 2D Heisenberg model for a given $J(x)$. He found that
$R(x)$ gradually decreases from $R(x=0) = 1$ to $R(x=0.2) \simeq 0$ (Fig.\ref{johnstonR}).

Oda \textit{et al.}\cite{oda} confirmed Johnston's scaling law for the bulk spin susceptibility
in Sr- and Ba-doped La$_2$CuO$_4$, and extended its applicability to the 2212 family of cuprates, 
Bi$_2$Sr$_{3-y}$Ca$_{y-x}$Cu$_2$O$_8$. The experimental results for both families
displayed excellent data collapse to the theoretical 2D Heisenberg curve. Like Johnston, Oda \textit{et al.} found 
that the weight of the Heisenberg-like contribution decreases with hole doping. To account for this decrease,
Oda \textit{et al.} introduced an effective magnetic moment,
\begin{equation}
\mu_{eff} = g \sqrt{s (s+1)}.
\end{equation}
$g$ is defined in terms of $T^m(x)$ and $\chi^m(x)$  as
\begin{equation}
g^2(x) = 11.6 k_B \frac{T^m(x) \chi^m(x)}{N \mu_B^2}\,.
\end{equation}
They found that both $T^m(x)$ and $g(x)$ decrease linearly with doping,
\begin{equation}
g(x) = 2.2 (1 - 4.5 x) \propto T^m(x).
\label{odda}
\end{equation}
The magnitude of $g(x)$ is that expected from the sum rule for a homogeneous material,
\begin{equation}
\int_0^{\infty} \frac{d \omega}{2 \pi}\, \int \frac{d^3 q}{(2 \pi \hbar)^3} (\chi(\omega, \bm{q}, x) - \chi_0(x)) 
= g^2(x) \frac{s (s+1)}{3}\,,
\end{equation}
but we caution the reader that homogeneity may not be present in the underdoped regime. 

Both the doping dependence of the $x$-dependent Fermi liquid part  and the temperature dependent
spin liquid component were found by Oda \textit{et al.} to be in good agreement with earlier results
of Johnston\cite{johnston}; like Johnston they found that the temperature-dependent scaling part 
of the bulk spin susceptibility disappears at some critical doping value; according to Eq.(\ref{odda}),
that is $x \simeq 0.22$.

Another early study of bulk susceptibility in La$_{2-x}$Sr$_x$CuO$_4$ was performed 
by Yoshizaki \textit{et al.} \cite{yoshizaki}, who give a plot of
$T^m(x) \simeq 1000K (1 - 4.5 x)$, without attempting Johnston scaling. 
Since $1000 K$ does not match the value of the exchange coupling $J$ in the insulator, 
Yoshizaki \textit{et al.} found a jump in $T^m(x)$ at the MI boundary. 

\begin{figure}
\begin{center}
\includegraphics[width=0.75\textwidth]{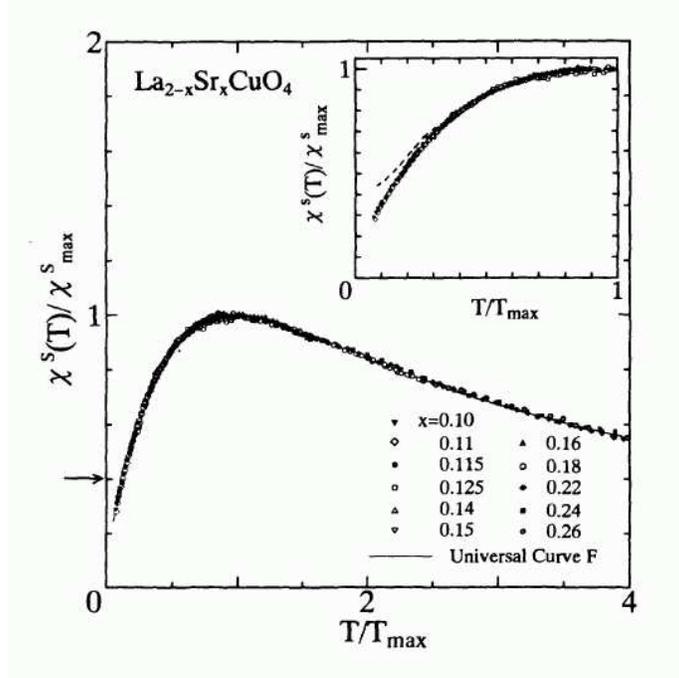}
\caption{$\chi^S(T)/\chi^S_{max}$ versus $T/T_{max}$ for the superconducting samples of La$_{2-x}$Sr$_x$CuO$_4$. The
arrow shows the value of $\chi^S(T=0)/\chi^S_{max}$ predicted by the 2D Heisenberg model calculations. The solid line
is the universal curve. The inset shows that the universal curve decreases more rapidly for $T \ll T_{max}$ than
that obtained by Johnston\cite{johnston} (dashed line). Reproduced with permission from \cite{nakano}}
\label{nakanofig}
\end{center}
\end{figure}

Further confirmation of Johnston scaling of the form Eq.(\ref{john1}) for the bulk spin susceptibility in the 2-1-4 family was 
obtained by Nakano \textit{et al.} \cite{nakano} (Fig. \ref{nakanofig}), who demonstrated an excellent data collapse
for a number of samples of La$_{2-x}$Sr$_x$CuO$_4$, both in the underdoped region and that close to and beyond optimal doping. 
Nakano \textit{et al.} arrived at their scaling law by assuming the presence of additional
temperature-dependent terms: an impurity Curie term $C/T$ and a linear term $B (T-T_a)$ in the
underdoped and overdoped regimes. They did not attempt to fit the 2D Heisenberg model calculations; an
empirical scaling data collapse was constructed instead, with results that were in agreement with 
Oda \textit{et al.}\cite{oda}.

An alternative form of scaling for the bulk spin susceptibility was proposed
by Levin and Quader\cite{LQ1}. Similar to the studies reviewed above, 
they suggested separation of the bulk spin susceptibility in two components, 
the temperature-independent Fermi liquid component and the scaling component. 
However, in their model the scaling component originates from the contribution of a separate itinerant band. 
Levin and Quader\cite{LQ1} took the 2D density of states in the following form:
\begin{equation}
\nu(\epsilon) = \nu_{\xi} + \nu_{\eta} \theta(- \epsilon),
\label{eqLQ}
\end{equation}    
where the index $\xi$ corresponds to a large hole band, which produces the usual Fermi liquid term in the bulk
spin susceptibility. The chemical potential $\mu$ lies very close to the top of the second band $\eta$ and
enters that band at $x=x_0$. 
The contribution of the band $\eta$ to the bulk spin susceptibility becomes temperature-dependent due to thermally 
activated carriers and is a universal function $\chi_{\eta} (\Gamma (x - x_0)/T)$, where $\Gamma \propto \nu_{\xi}^{-1}$
is a Fermi liquid parameter.  The parameter $\Gamma (x - x_0)$ is thus an analog of $T^m(x)$.
The two-band model produced a reasonably successful data fit for the bulk susceptibility 
in TlSr$_2$(Lu$_{1-x}$Ca$_x$)Cu$_2$O$_y$ for both the underdoped and the overdoped regimes.

To summarize, the various independent measurements of the bulk susceptibility in the 2-1-4 and 1-2-3 materials  
can now be seen to be consistent with one another and with the picture first set forth by Johnston: 
that in the underdoped regime, for dopings between $x \sim 0.06$ and $\sim 0.22$, and  $T > T^m/3$, one has two 
independent contributions to the bulk spin susceptibility. One comes from a 2D Heisenberg spin liquid with
a doping-dependent effective interaction, $J_{eff}(x) \sim T^m(x)$; the second 
represents a Fermi liquid contribution  whose strength increases as the doping level is increased. As we shall see, 
the fall-off  of the rescaled spin susceptibility below the Heisenberg spin liquid value at $T \sim T^m/3$ 
appears to be a universal property of the spin liquid in the cuprates.

\subsection{Measurements of the bulk susceptibility using the NMR Knight shift}

\begin{figure}
\begin{center}
\includegraphics[width=0.75\textwidth]{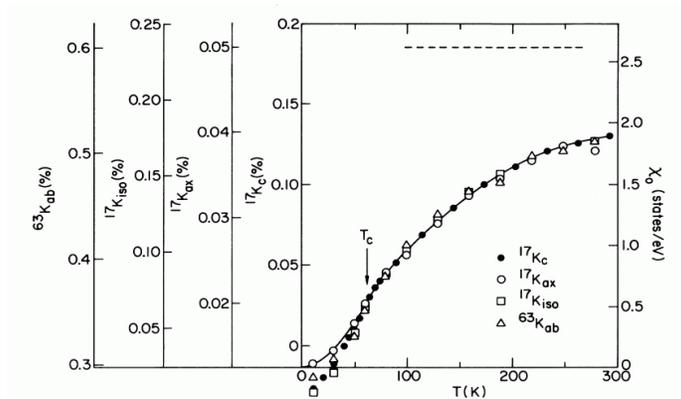}
\caption{The experimental planar Cu and O Knight shifts in YBa$_2$Cu$_3$O$_{6.63}$ plotted vs the temperature $T$. 
The values for the static planar susceptibility $\chi_0(T)/\mu_B^2$ are given on the right-hand scale. 
Different Knight shifts and the static spin susceptibility
have the same temperature dependence. Reproduced with permission from \cite{MPT}.}
\label{MPTfig}
\end{center}
\end{figure}

The Knight shift seen in NMR experiments is a measurement of the bulk spin susceptibility at a particular nuclear 
site\cite{slichter:book}, so that measurements of the  Knight shift for different nuclei serve to supplement direct measurements 
of the bulk susceptibility. Most of the early analysis of the NMR data used a single-component Mila-Rice-Shastry (MRS) 
description\cite{Mila:Rice:Shastry} for which the justification was the observation by Alloul \textit{et al.}\cite{alloul} 
and Takigawa \textit{et al.}\cite{takigawa} that the
$^{63}Cu$, $^{17}O$ and $^{89}Y$ Knight shifts in YBa$_2$Cu$_3$O$_7$ and YBa$_2$Cu$_3$O$_{6.63}$ have the same anomalous temperature 
dependence (Fig.\ref{MPTfig}). 
MRS proposed a hyperfine Hamiltonian that described the coupling of a single magnetic component formed by the system of 
planar Cu$^{2+}$ spins and holes mainly residing on the planar copper sites
to the various nuclei. Most earlier Knight shift experiments\cite{takigawa,MPT} confirmed this  one-component 
Zhang-Rice\cite{ZR} singlet picture, which is basically correct for the parent insulator. However, as first noted by 
Walstedt \textit{et al.}\cite{walst} 
in connection  with spin-lattice relaxation rate measurements, and discussed in detail  below,  the single component 
description turns out to lead to  a number of contradictions in the doped materials, and requires  modification.

\begin{figure}
\begin{center}
\includegraphics[width=0.75\textwidth]{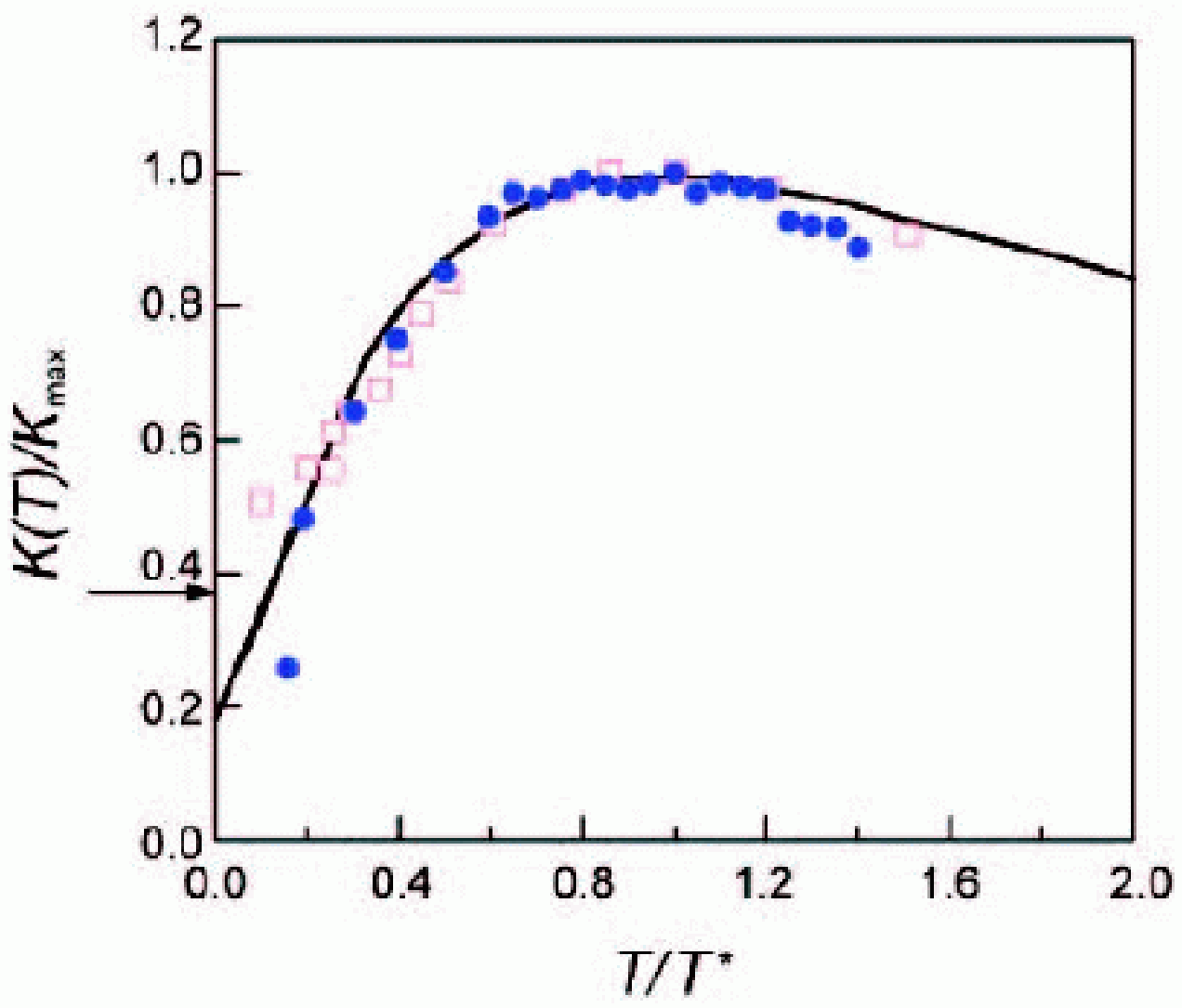}
\caption{The spin shift in YBa$_2$Cu$_4$O$_8$ versus $T/T^m$, showing
the applicability of  Johnston/Nakano scaling and the relevance of the 2D Heisenberg model\cite{cfp}.
The solid line is taken from Ref. \cite{nakano}, the dotted green line is the calculated susceptibility of
the spin-1/2 2D Heisenberg antiferromagnet\cite{MD} and the data points are taken from Ref.\cite{curro97}.
The arrow points to the zero-temperature prediction for the 2D Heisenberg antiferromagnet. Reproduced
with permission from \cite{cfp}.}
\label{currofig}
\end{center}
\end{figure}

	Just as was the case for the bulk spin susceptibility, the temperature-dependent Knight shift data for different
doping levels and different families of cuprates displays Heisenberg model-type scaling. The $^{89}Y$ Knight
shift data from Alloul \textit{et al.}\cite{alloul} for different doping levels
in the 1-2-3 family has been scaled to a single curve with very good data collapse by Wuyts \textit{et al.}\cite{wuyts}, although no
comparison to the 2D Heisenberg model or the Nakano \textit{et al.}\cite{nakano} bulk spin susceptibility scaling curve was provided.
Recently  Curro \textit{et al.}\cite{cfp} and the authors\cite{BPprl} have shown that the two are consistent; as
may be seen in Fig.\ref{currofig}, the NMR
Knight shift follows very well the Johnston-Nakano bulk susceptibility scaling form and the Heisenberg
model for the 1-2-4 member of the 1-2-3 family. Spin liquid scaling behavior determined by $T^m(x)$  
has thus been shown to be  universal in the cuprates.

	What then is the direct experimental evidence from Knight shift measurements for the presence of two distinct  components? 
One material for which a contradiction with the simple one component model was found is YBa$_2$Cu$_4$O$_8$, where a more precise 
measurement\cite{zurich} of the copper Knight shift data in magnetic fields parallel to the c-axis, $^{63}K_{\parallel}$, found that it 
displays a  rather unusual temperature dependence, one that is different from that seen for the bulk spin liquid susceptibility. 
A second example comes from  quite recent measurements by Haase \textit{et al.}\cite{HS} of the planar and apical oxygen Knight shifts  
for  La$_{1.85}$Sr$_{0.15}$CuO$_4$. As we discuss in a later section, these indicate 
that while the temperature-dependent part of the Knight shift on different
nuclei is indeed the same, there is a temperature-independent part of  the Knight shift above the  superconducting temperature, 
$T_c$, that is different for the planar copper and the planar and apical oxygen nuclei. The key signature of this effect, the deviation of various Knight
shifts from the same temperature dependence at temperatures below $T_c$, is clearly visible in Fig.\ref{MPTfig}.
These contradictions suggest that the original MRS hyperfine Hamiltonian has to be modified 
to include these new effects, and that the one-component spin dynamics and the Zhang-Rice singlet picture\cite{ZR} fails at 
moderate doping levels\cite{walst}, a topic to which we return in Section 3.

\subsection{Spin-lattice relaxation rates}

In the MRS Hamiltonian that is described in detail in the following section, wave vector dependent form factors arise from the 
presence of a transferred hyperfine interaction between the Cu spins and the probe nucleus. For probe nuclei other than copper, 
these vanish at the commensurate  wave vector $\bm{Q}=(\pi,\pi)$, so that, for example, an in-plane oxygen nucleus will feel little of a 
spin response that is peaked at the commensurate wave vector. (That this should be the case is obvious if one recalls that such 
oxygen are located midway between copper nuclei, and for an antiferromagnetic array of copper spins, the nearest neighbor 
spins would cancel one another out in their influence on the oxygen site.) The striking difference of the temperature dependence 
and magnitude of the spin-lattice relaxation on the $^{63}Cu$, $^{17}O$, and $^{89}Y$ nuclei led Millis, Monien, 
and Pines (MMP)\cite{MMP} to the conclusion that a 
localized  or nearly localized spin component of the dynamic magnetic response function must be strongly peaked at the 
commensurate wave vector,  as might be expected if one were close to an antiferromagnetic instability. 

Indeed, closer study shows that even a slight deviation from commensurability within the MMP approach based on the MRS one-component 
model will have a significant impact on oxygen relaxation rates that is not seen experimentally\cite{MMP}. Thus, an incommensurate peak 
structure for  $\chi''(\bm{q}, \omega)$, such as has been inferred from inelastic neutron scattering (INS) experiments\cite{cheong,Mason}
on the 2-1-4 materials, 
is inconsistent with this approach. One way to get around this difficulty is to introduce additional transferred hyperfine 
interactions\cite{zha}; a second way, proposed by Slichter\cite{slichter,BPT} , is to note that unlike NMR, 
INS is a global probe of spin 
excitations, so that a suitable  domain structure (regions of commensurate near-antiferromagnetic behavior, separated by domain walls) 
would give rise to the apparent incommensuration inferred from the INS experiments. We adopt this explanation in what follows.

The spin fluctuation response function proposed by MMP was that appropriate to any spin liquid near a commensurate antiferromagnetic 
instability: 
\begin{equation}
\chi_{SL}(\bm{q},\omega) = 
\frac{\chi_{\bm{Q}}}{1 + \xi^2 (\bm{q} - \bm{Q})^2 - i \frac{\omega}{\omega_{SF}}\,}\,,
\label{SLMMP}
\end{equation}
where the peak susceptibility takes the form,
\begin{equation}
\chi_{\bm{Q}} = \alpha \xi^2,
\end{equation} 
with $\xi$ as the magnetic correlation length, and $\alpha(x)$ as a temperature-independent constant.
The copper NMR $^{63}T_1$ and $^{63}T_{2G}$ relaxation rates provide a direct measure of the strength 
and character of the spin liquid response function, measuring as they do the momentum-integrated imaginary and real part of 
the spin response function,  $\chi_{SL}(\bm{q}, \omega)$\cite{SP}.  One finds:
\begin{eqnarray}
\label{T1MMP}
\frac{1}{^{63}T_1}\, & \propto &    \frac{\alpha T}{\omega_{SF}}\,,         \\
\frac{1}{^{63}T_{2G}}\, & \propto &  \alpha \xi.
\label{T2MMP}
\end{eqnarray} 

\begin{figure}
\begin{center}
\includegraphics[width=0.75\textwidth]{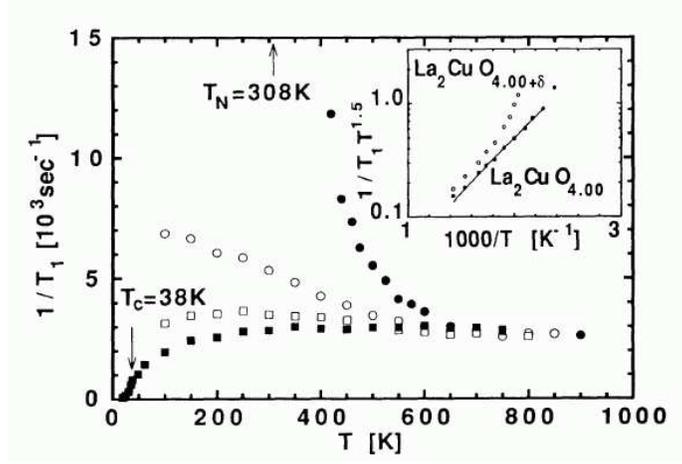}
\caption{Temperature dependence of $1/ ^{63}T_1$ measured by NQR for La$_{2-x}$Sr$_x$CuO$_4$ ($x=0$, $x=0.04$, 
$x=0.075$, $x=0.15$). The inset shows a semilogarithmic plot of $1/ ^{63}T_1 T^{3/2}$ (in units of $sec^{-1}K^{-1.5}$) vs
$1000/T$ for the clean sample of La$_2$CuO$_{4.00}$ and for La$_2$CuO$_{4.00+\delta}$. The solid curve is the best fit
to the theoretical prediction of 2D Heisenberg model. Reproduced with permission from \cite{imai}}
\label{imaifig}
\end{center}
\end{figure}

Imai  \textit{et al.}\cite{imai} in an early scaling analyses (Fig. \ref{imaifig}) of the copper relaxation rates  
for the 2-1-4 family demonstrated experimentally the existence of a universal high temperature limit 
for  $^{63}T_1$  that is temperature- and doping- independent:
\begin{equation}
\frac{1}{^{63} T_1 (T,x)} = const,
\label{hightsc}
\end{equation}
which according to Eq.(\ref{T1MMP}) means that at high temperatures the spin fluctuation energy, 
$\omega_{SF}$ must be proportional to T.
This universal high-temperature behavior of $^{63}T_1$ was
explained in terms of the QC (Quantum Critical)\cite{SP,BP}
behavior seen in the sigma-model\cite{CHN,CSY} that, strictly
speaking, is only applicable to the parent insulating compound.
Indeed the empirical finding of  a temperature- and doping independent $^{63}T_1$ in 
the 2-1-4 family of materials extends to temperatures much higher than those at which
such a theoretical explanation would apply.

\begin{figure}
\begin{center}
\includegraphics[width=0.75\textwidth]{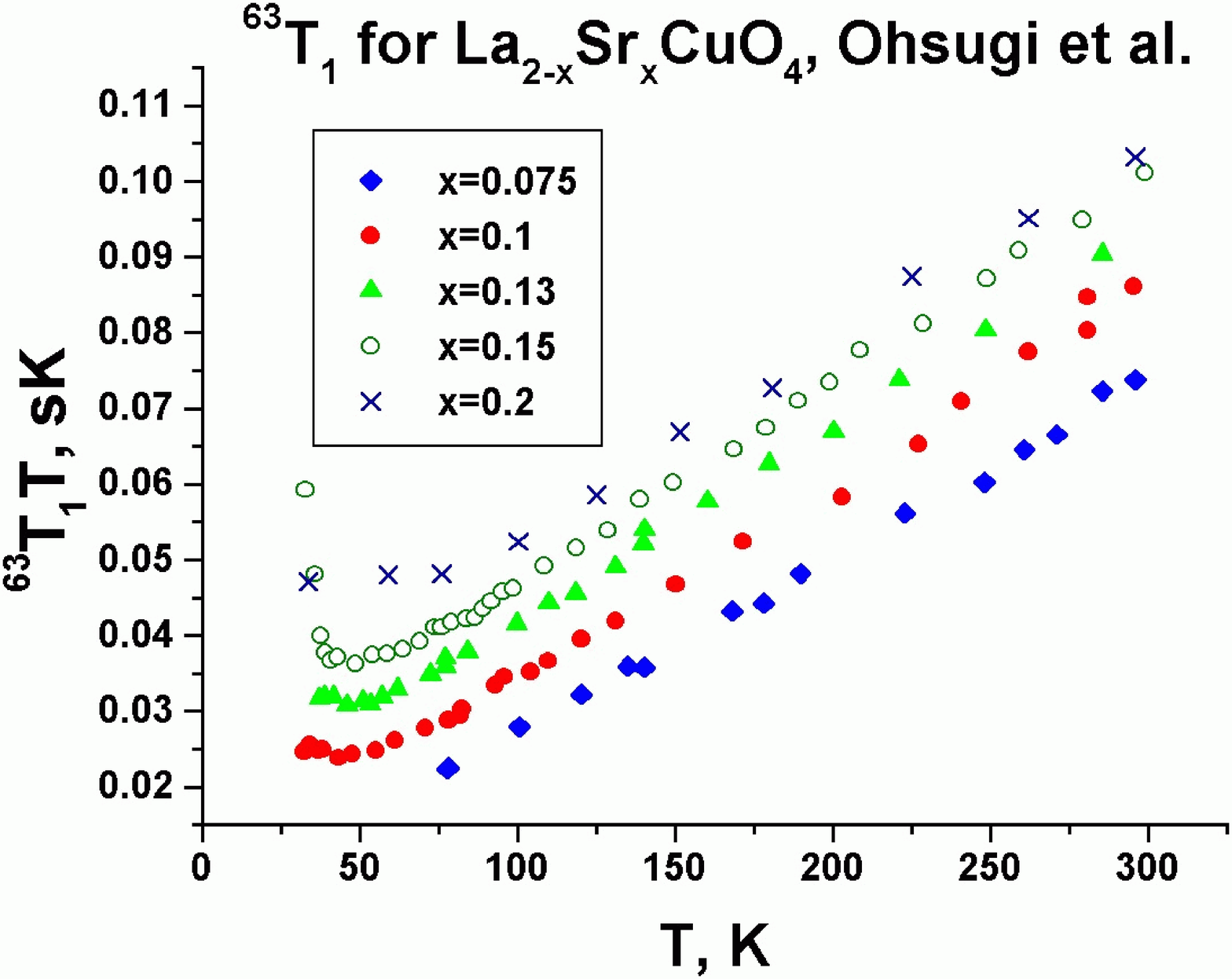}
\caption{At high temperatures the dependence of $^{63}T_1 T$ on temperature in La$_{2-x}$Sr$_x$CuO$_4$\cite{ohsugi} for different
doping levels represents a set of parallel lines.}
\label{kitaokafig}
\end{center}
\end{figure}

Two other high-temperature forms for the relaxation rates based on $z=1$ and $z=2$ dynamical scaling were therefore 
suggested\cite{SP,BP} on the basis of the non-linear sigma model\cite{CHN,CSY} and general scaling arguments, and verified 
experimentally. 
In $z=1$ scaling, $\omega_{SF}$ and $\xi$ are related by
\begin{equation}
\omega_{SF} \propto \Delta = \frac{c}{\xi}\,,
\end{equation}
so the presence of $z=1$ QC scaling in the underdoped materials leads to the simple results:
\begin{equation}
\frac{^{63}T_1 T}{^{63}T_{2G}}\, \propto \omega_{SF} \xi \propto c
\end{equation}
On the other hand, for $z=2$  (mean field) scaling one has
\begin{equation}
\omega_{SF}= \frac{\Gamma}{\xi^2}\,,
\end{equation}
so that if it is present one finds
\begin{equation}
\frac{^{63}T_1 T}{^{63}T_{2G}^2}\,  \propto \omega_{SF} \xi^2 = \Gamma.
\label{z2sc}
\end{equation}
The NMR experimental results of  Curro \textit{et al.}\cite{curro97} show that both forms of scaling are present 
in the  material, YBa$_2$Cu$_4$O$_8$. At high 
temperatures, for $T > T^m \sim 500K$, Eq.(\ref{z2sc}) is valid and one has mean field behavior, 
while  below $T^m$ the spin spectrum displays $z=1$ QC 
behavior down to a temperature $ \sim 150K$, where the $T_1$ measurements suggest that a gap opens up in the spin
liquid spectrum. We return to this finding below.

The relaxation rates for other nuclei, such as oxygen or yttrium, also display anomalous (i.e. non-Korringa) behavior, 
although in much milder form. A modified Korringa-type scaling for oxygen $^{17}T_1$  was suggested by MMP:
\begin{equation}
^{17}T_1 T \chi_0(T) = const.
\label{anom}
\end{equation}
Additional  scaling forms have been proposed for the copper
relaxation rates at lower temperatures. In particular,
the authors\cite{BPprl} have observed recently that  $^{63}T_1 T$ reaches
its universal high-temperature behavior with a different
$x$-dependent offset (Fig.\ref{kitaokafig}), 
\begin{equation}
^{63}T_1 T = B T + A(x),
\label{prlcoll}
\end{equation}
where the constant $B$ is universal, while $A(x)$ changes linearly with doping, 
and its variation suggests the existence of a QC point in the spin liquid at $x_0 \simeq 0.05$, where $A(x_0) = 0$. 
Eq.(\ref{prlcoll}) shows excellent data collapse for the 2-1-4 and the 1-2-3 families.

\begin{figure}
\begin{center}
\includegraphics[width=0.75\textwidth]{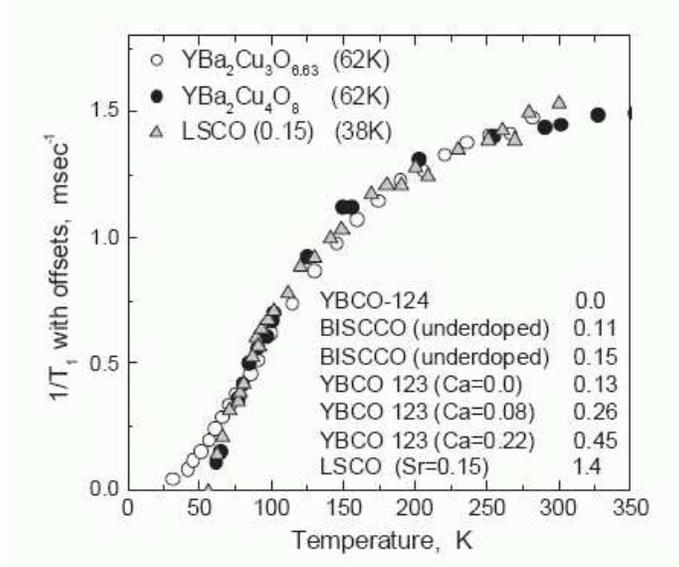}
\caption{Temperature dependence of $1/ ^{63}T_1$ for YBa$_2$Cu$_3$O$_{6.63}$ overlaid with that for YBa$_2$Cu$_4$O$_8$ and
La$_{1.85}$Sr$_{0.15}$CuO$_4$. In the lower right corner: the offset values for different compounds.
Reproduced with permission from \cite{GT}}
\label{GTscaling}
\end{center}
\end{figure}

\begin{figure}
\begin{center}
\includegraphics[width=0.75\textwidth]{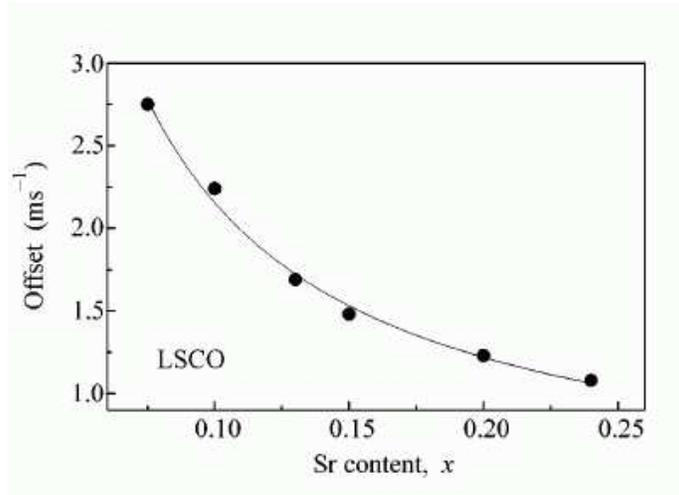}
\caption{The offset $1 /^{63}T_1(x)$ vs Sr content $x$ for LSCO (relative to that for YBa$_2$Cu$_4$O$_8$). The line
is a visual guide. Reproduced with permission from \cite{GT}.} 
\label{GTparam}
\end{center}
\end{figure}

An alternative form of scaling has been suggested to apply at moderate to low temperatures, $T < 300K$, by
Gor'kov and Teitel'baum (GT)\cite{GT} who found excellent data collapse (Fig.\ref{GTscaling}) for the
following scaling form:
\begin{equation}
\frac{1}{^{63}T_1}\, = \frac{1}{^{63}T_1(x)}\, + \frac{1}{^{63}\tilde{T}_1(T)}\,,
\end{equation}
where $\tilde{T}_1(T)$ is a universal function for all high-T$_c$ materials, while
$\frac{1}{^{63}T_1(x)}\,$ varies with $x$. The suggested scaling form
follows from the decomposition of relaxation rate into two different processes.
The x-dependence of $1/^{63}T_1(x)$(Fig.\ref{GTparam})  looks rather unusual\cite{GT},
since the proposed empirical disorder-driven relaxation rate,  $1/^{63}T_1(x)$ decreases with increased doping $x$.
The proposed GT scaling is assumed to be the result of 
intrinsic phase separation, and the pseudogap temperature $T^*(x)$ is proposed as a measure of the onset of such behavior.

\subsection{Finite size effects}

If scaling is present, it is natural to look for the possible presence of finite size effects, as Cho \textit{et al.}\cite{cho} have done 
in the lightly doped La$_{2-x}$Sr$_x$CuO$_4$ material.
They found very good agreement with that expected for  phase separation and  domain formation  on the insulating side of the Mott 
transition. However, there was no direct observation of the domain sizes of the different phases. 
According to finite-size scaling theory\cite{fisher}, the Ne\`el temperature for
a domain of size $L$ takes the form,
\begin{equation}
1 - \frac{T_N(L)}{T_N(L = \infty)}\, \propto L^{-1/\nu},
\end{equation}
where $\nu = 1/2$ in mean field theory.
The experimental data can also be fit well by the expression
\begin{equation}
1 - \frac{T_N(x)}{T_N(x = 0)}\, \propto (x/x_c)^n,
\end{equation}
with $n = 1.90 \pm 0.20 \simeq 2$. Cho \textit{et al.} conclude that 
\begin{equation}
L(x) \propto \frac{1}{x}\,,
\end{equation} 
which means that the width of the domain wall is $x$ independent. They found that
in the antiferromagnetic regime the scaling law takes the form:
\begin{equation}
\chi(x,T) = \chi(F(x)(T - T_N(x))), \ \ F(x) \propto \frac{1}{\left(x + \frac{C}{\xi_0(T_N(x))}\,\right)^2}\,.
\end{equation}
Here $\xi_0(T)$ is the correlation length in the pure Heisenberg model,
\begin{equation}
\xi_0(T) = 0.276 a \exp{(2 \pi \rho_s/k_B T)} \simeq 0.276 a \exp{(J/k_B T)}.
\end{equation}

\subsection{Inelastic neutron scattering}

Inelastic neutron scattering (INS) experiments enable one to explore the extent to which 2D
Heisenberg model captures the momentum dependence and behavior at higher frequencies
of the spin liquid as its properties are altered by doping. There is by now a vast body of 
literature that includes several recent reviews\cite{kastner,stripesreview,birgeneauR}. Our focus here is on the extent that
the doped spin liquid continues to exhibit the quantum critical behavior inferred from the 
ultra low frequency dynamic properties measured in the NMR experiments described in the previous
sections, and the ways in which departures from quantum critical behavior emerge as the frequency
is increased into the multi-time range or one goes into the gapped normal or superconducting
state. As we shall see, one of the most striking features that emerges with doping is the 
observation of peaks whose positions at lower frequencies reflect a doping-dependent incommensuration or
discommensuration that indicates dynamic stripe formation; a second is the appearance of 
resonances and spin gaps in the normal state. Throughout this section we will be concerned,
as we were in our discussion of NMR experiments, with possible universal behavior that is common
to the 1-2-3 and 2-1-4 families.

\subsubsection{$\omega/T$ scaling for the local spin susceptibility}

Perhaps the most direct evidence of quantum critical scaling was provided by the INS measurements
of the local (integrated) dynamic spin susceptibility. The remarkable results obtained in insulating
and metallic low doped samples indicate the presence of the $\omega/T$ scaling expected in the vicinity of 
a $z=1$ quantum critical point. 

\begin{figure}
\begin{center}
\includegraphics[width=0.75\textwidth]{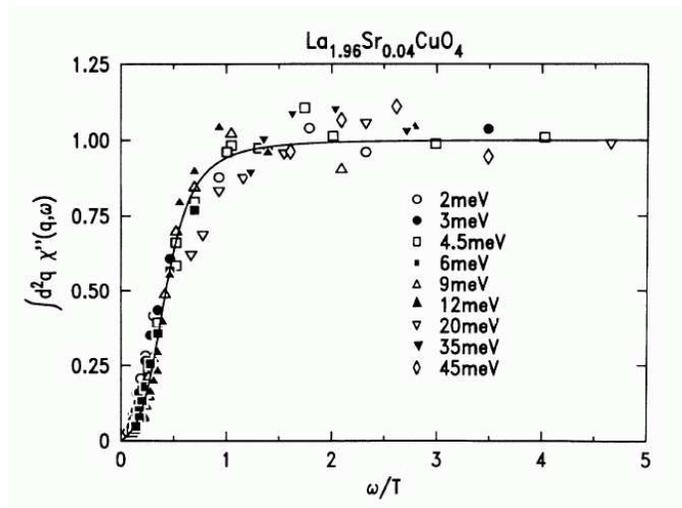}
\caption{Normalized integrated spin susceptibility as a function of the scaling variable $\omega/T$. The solid line is
the function $2/\pi \tan^{-1}[a_1 \omega/T + a_3 (\omega/T)^3]$ with $a_1=0.43$ and $a_3 = 10.5$.
Reproduced with permission from \cite{keimer}.}
\label{Keimerfig}
\end{center}
\end{figure}

Among early neutron scattering experiments that exhibit some form of scaling behavior those of particular interest  
are on lightly doped YBa$_2$Cu$_3$O$_{6+x}$ by Birgeneau \textit{et al.}
\cite{birgeneau}, lightly doped La$_{2-x}$Sr$_x$CuO$_4$ by Keimer \textit{et al.} \cite{keimer}, 
and YBa$_2$Cu$_3$O$_{6.6}$ by Sternlieb\textit{et al.} \cite{sternlieb}. 
Keimer \textit{et al.}\cite{keimer} and Birgeneau \textit{et al.}\cite{birgeneau} 
fit their experimental results for the local (integrated) spin susceptibility to the expression
\begin{equation}
\chi_L''(\omega) \equiv \int \frac{d^2 q}{(2 \pi)^2}\, 
\chi''(\bm{q}, \omega) = C (\omega) tan^{-1}(a_1 (\omega/T) + a_2 (\omega/T)^3 + \cdots )
\label{keimsc}
\end{equation}
The data collapse is good (Fig.\ref{Keimerfig}), but this form does not quite exhibit $\omega/T$ scaling because of 
the $T$-independent amplitude $C(\omega)$.

\begin{figure}
\begin{center}
\includegraphics[width=0.75\textwidth]{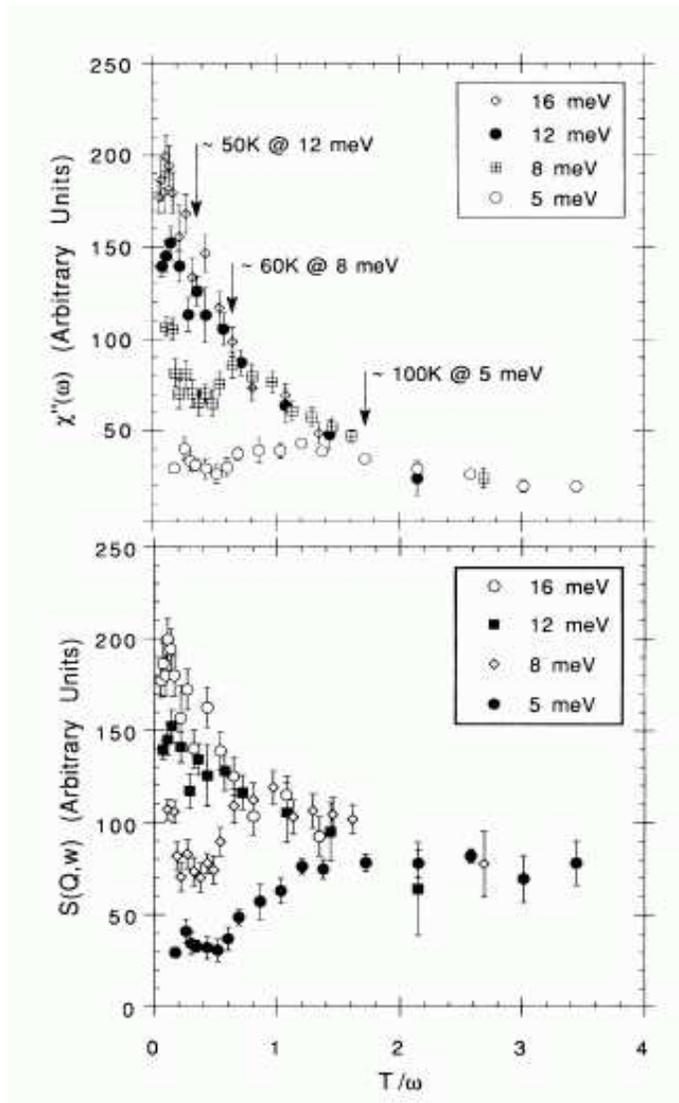}
\caption{(top) The temperature dependence of the local dynamic spin susceptibility $\chi''(\omega)$ for a fixed energy
transfers converges to a universal function of $T/\omega$ at high temperatures. The arrows show the temperature above
which the scaling behavior occurs. (bottom) The corresponding scattering function $S(\omega)$ vs $T/\omega$.
Reproduced with permission from \cite{sternlieb}.}
\label{sternliebfig}
\end{center}
\end{figure}

Sternlieb \textit{et al. } \cite{sternlieb} subsequently found that true $\omega/T$ scaling exists for the local
spin susceptibility in the underdoped material YBa$_2$Cu$_3$O$_{6.6}$. Their experimental results (Fig. \ref{sternliebfig}) 
can be fit to the following simple scaling form:
\begin{equation}
\chi_L''(\omega, T) = A(x) \frac{\omega}{T}\,,
\end{equation}
with a deviation from scaling that occurs at progressively lower temperatures, as the frequency $\omega$ increases.

\begin{figure}
\begin{center}
\includegraphics[width=0.75\textwidth]{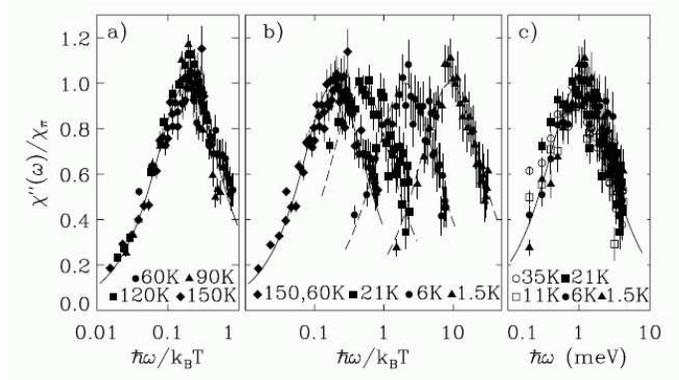}
\caption{ (a) $\omega/T$ scaling is valid for La$_2$Cu$_{0.94}$Li$_{0.06}$O$_4$ in the high temperature QC regime
The solid line is the scaling function, Eq.(\ref{baoscf}). (b) $\omega/T$ scaling becomes invalid in the low temperature
regime. (c) A new scaling for the low temperature regime, with a constant energy scale $\Gamma_0 = 1 meV$. The solid line
is the scaling function Eq.(\ref{baoscqd}). 
Reproduced with permission from \cite{bao}.}
\label{Baosc}
\end{center}
\end{figure}

In more recent experiments, Bao \textit{et al.}\cite{bao} studied experimentally
the destruction of 2D antiferromagnetic order in $Li$-doped La$_2$CuO$_4$.
Since holes are loosely bound by $Li$ impurities, the doped material remains
an insulator even when long-range order is destroyed at $x > 0.03$. However,
unlike the superconducting La$_{2-x}$(Sr,Ba)$_x$CuO$_4$,
there are no additional complications due to the presence of mobile doped holes,
making possible a cleaner measurement of the spin dynamics.
Bao \textit{et al.}\cite{bao} found a commensurate energy spectrum of spin excitations
at $x > 0.03$, where the long-range order is not present, and a characteristic
quantum critical $\omega/T$ scaling for the local spin susceptibility (Fig. \ref{Baosc}). 
The scaling function, however, is different from that obtained by Sternlieb\textit{et. al.}\cite{sternlieb}
In particular, Bao \textit{et al.}\cite{bao} found significant deviations of the scaling function from
linearity, with
\begin{equation}
\chi''_L(\omega, T) = \chi_{\pi} f(\omega/T),
\end{equation}
and
\begin{equation}
f(x) = \frac{0.18 x}{0.18^2 + x^2}.
\label{baoscf}
\end{equation}
Below $T = 50K$ this scaling changes to one with a constant energy scale $\Gamma_0 = 1meV$,
\begin{equation}
\chi''_L(\omega, T) = \chi_{\pi} g(\omega/\Gamma_0),
\end{equation}
where
\begin{equation}
g(x) = \frac{x}{1 + x^2}.
\label{baoscqd}
\end{equation}

Stock \textit{et al.}\cite{stock04} recently confirmed the low-frequency $\omega/T$ scaling for 
$\chi''_L(\omega,T)$ in oxygen ordered ortho-II YBa$_2$Cu$_3$O$_{6.5}$ superconductor found earlier in this 
material in the oxygen-disordered state by Birgeneau \textit{et al.}\cite{birgeneau}. 
Stock \textit{et al.} fit their expression to the form Eq.(\ref{keimsc}), with only a linear term in $\omega/T$ present.
They find that the $\omega/T$ scaling breaks down at higher frequencies, $\omega > 20 meV$, since the amplitude $C(\omega)$
in Eq.(\ref{keimsc}) becomes temperature-dependent.

\subsubsection{Temperature and frequency dependence of the correlation length}

Keimer \textit{et al.}\cite{keimer} find evidence for finite size scaling, since a good fit to their
experimental results can be obtained with the following expression for the correlation length:
\begin{equation}
\xi^{-1}(x,T) = \xi^{-1}(x, T=0) + \xi^{-1}(x=0, T),
\label{necore}
\end{equation}
 with $\xi(x,T=0) \propto 1/x \propto L$, where $L$ is the size of the domain for finite size scaling.

\begin{figure}
\begin{center}
\includegraphics[width=0.75\textwidth]{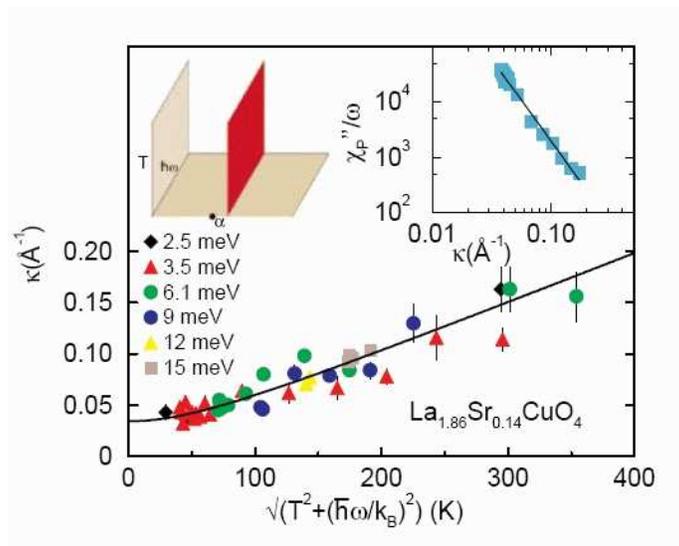}
\caption{Temperature dependence of the inverse correlation length $k(\omega,T)$ at various fixed energy transfers plotted
against $\sqrt{(\hbar \omega/k_B)^2 + T^2}$. The solid line corresponds to a $z=1$ QCP. The graph in the upper right shows how
the peak response depends on $k=k(\omega=0,T)$. The inset in the upper left shows the 3D space defined by $(\omega,T)$ phase
space probed by the $x=0.14$ sample, and the solid circle represents the nearby QCP.
Reproduced with permission from \cite{aeppli}.}
\label{aepplisc}
\end{center}
\end{figure}

Eq.(\ref{necore}) was later generalized to finite frequencies $\omega$ by Aeppli \textit{et al.}\cite{aeppli},
who fit their results for the correlation length in  La$_{1.86}$Sr$_{0.14}$CuO$_4$ (Fig.\ref{aepplisc}) to the following form:
\begin{equation}
\xi^{-1}(T,\omega) = \sqrt{\xi^{-2}(0) + \omega^2 + T^2}.
\end{equation}

\subsubsection{Direct measurements of $\chi''(\bm{q}, \omega)$}

Inelastic neutron scattering experiments provide a direct measurement of the spectrum of spin excitations,
and thus of the spin wave velocity and exchange couplings in the antiferromagnetic insulator\cite{coldea,hay96,reznik0}. 
We shall see that the detailed measurements of the spectrum of the spin excitations in underdoped high-temperature
superconductors that have recently been carried out for both the 
2-1-4\cite{hayden,tranquada2,christensen,vignolle} and the 1-2-3\cite{mook,stock05,reznik} families provide evidence
for its universality. 

\paragraph{Incommensuration/Discommensuration}

Early neutron measurements on both the 2-1-4 and the 1-2-3 families of high-temperature superconductors
revealed an incommensurate spectrum of spin excitations at low frequencies. 
Cheong \textit{et al.}\cite{cheong} and Mason \textit{et al.}\cite{Mason} discovered in their inelastic
neutron scattering measurements on the 2-1-4 family of materials that the position of the low-energy
spin fluctuation peak shifts from $(\pi, \pi)$ in the parent insulating compound 
to $(\pi \pm 2 \pi \delta, \pi)$ and $(\pi, \pi \pm 2 \pi \delta)$ in the underdoped superconductor.
They found that the  incommensurability $\delta$ in the underdoped regime grows approximately linearly with
increased hole concentration $x$, $\delta \simeq x$, at least for $x < 0.12$. For $x > 0.12$, the incommensurability
$\delta(x)$ saturates at a finite value.
Incommensurate peaks in the spin response function $\chi''(\bm{q},\omega)$ at low frequencies were also
discovered in the bilayer system YBa$_2$Cu$_3$O$_{6+x}$\cite{mook98}. As Mook \textit{et al.}\cite{mook98}
have first shown, the incommensurate peaks in the bilayer system are located at the same positions
as in the single-layer La$_{2-x}$Sr$_x$CuO$_4$ system at similar level of hole doping. Thus, they
concluded that the band structure, which is very different in these two families of materials,
plays a minor role in the spectrum of spin excitations at low frequencies. In what follows we will refer
to these peaks as reflecting incommensurate behavior, although, as noted earlier, for these results to
be consistent with NMR, these should rather be regarded as reflecting discommensuration, or dynamic
stripe order. 

\begin{figure}
\begin{center}
\includegraphics[width=0.75\textwidth]{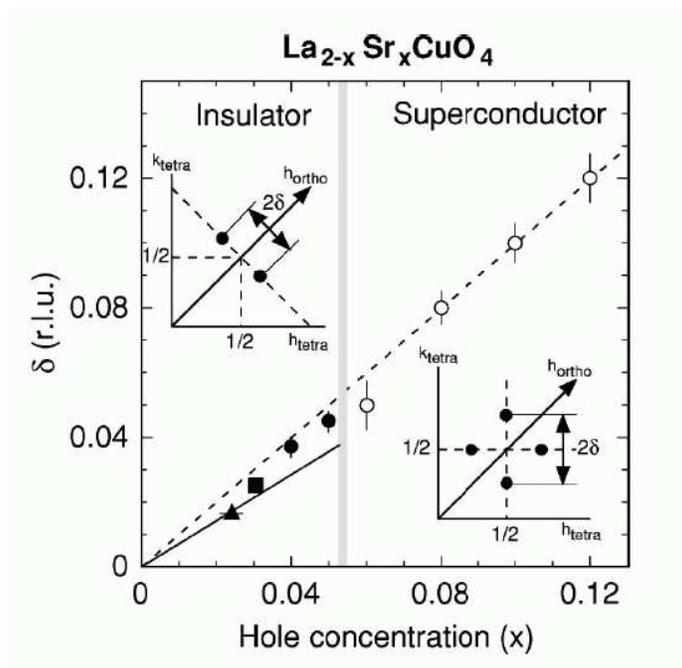}
\caption{Hole concentration $(x)$ dependence of the splitting of the incommensurate peaks $(\delta)$ in
La$_{2-x}$Sr$_x$CuO$_4$ in tetragonal reciprocal lattice units. Open circles indicate the data for the 
inelastic incommensurate peaks reported by Yamada \textit{et al.}\cite{yamada}. Filled circles and 
square are the data for elastic incommensurate peaks reported by Wakimoto \textit{et al.}\cite{wakim}. The
broken and solid lines correspond to $\delta = x$. The insets show the configuration of incommensurate
peaks in the insulating phase (diagonal stripe) and the superconducting phase (collinear stripe) 
Reproduced with permission from \cite{matsuda}.}
\label{matsudafig}
\end{center}
\end{figure}

Wakimoto \textit{et al.}\cite{wakim} and Matsuda \textit{et al.}\cite{matsuda} extended the
measurements of $\chi''(\bm{q},\omega)$ in La$_{2-x}$Sr$_x$CuO$_4$ to the spin glass regime,
$0.02 < x < 0.055$.
They found that while the low-energy spin excitations remain incommensurate with the same $\delta \simeq x$ as in the metallic 
phase, the positions of the incommensurate peaks are rotated by $\pi/4$ relative to those in the metal,
as shown in Fig.\ref{matsudafig}. Unlike in metallic phase, the diagonal stripe structure for the insulating
spin glass phase is seen as Bragg peaks in elastic scattering measurements and thus is \textbf{static} at low 
temperatures\cite{matsuda}. Moreover, the 
incommensurability of the spin fluctuations disappears at higher frequencies or temperatures,
above a certain energy threshold for the incommensurate structure, which we shall see is directly
related to the spin gap.

An interesting linear scaling relationship between the incommensurability
$\delta(x)$ and the superconducting temperature, $T_c(x)$, was suggested by Yamada \textit{et al.}\cite{yamada}. 
which, they find, holds extremely well for
the 2-1-4 family of materials in the underdoped regime, $T_c(x) = \hbar \nu*_{214} \delta(x)$,
with $\hbar \nu*_{214} \simeq 20 meV \AA$.  Balatsky and Bourges\cite{bal:bour} and Dai \textit{et al.}\cite{dai}
found that Yamada scaling is also applicable to the 1-2-3 family of
materials, with $T_c(x) = \hbar \nu*_{123} \delta(x)$ and $\nu*_{123} \simeq 36.6  meV \AA$\cite{dai}. Since
the maximum of $T_c$ is different in different families of cuprates, while the incommensurability is universal
and depends only on hole doping, $\nu*$ is different for different families of materials.

\begin{figure}
\begin{center}
\includegraphics[width=0.75\textwidth]{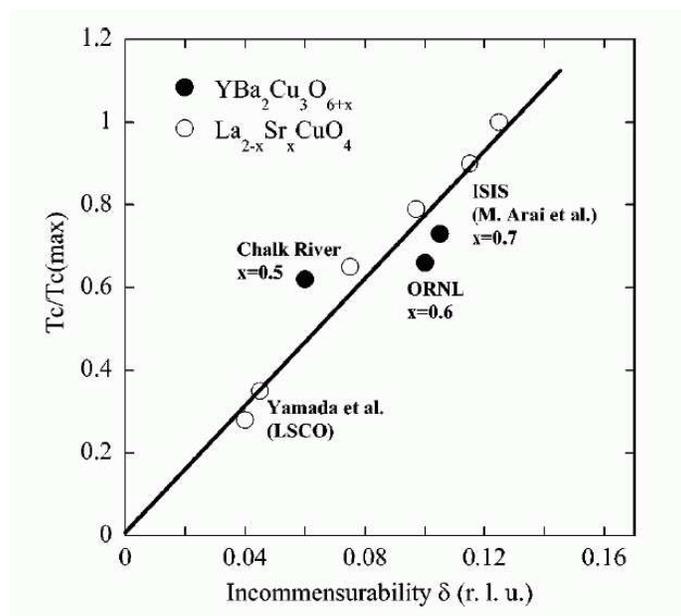}
\caption{Reduced superconducting temperature $T_c/T_c(max)$ as it relates to the incommensurability
$\delta$ in YBa$_2$Cu$_3$O$_{6+x}$ and La$_{2-x}$Sr$_x$CuO$_4$.
Reproduced with permission from \cite{stock04}.}
\label{stockfigA}
\end{center}
\end{figure}

Stock \textit{et al.}\cite{stock04} considered the applicability of Yamada scaling to both 2-1-4 and 1-2-3 families of
high-temperature superconductors.  They plotted the incommensurability $\delta$ as a function of $T_c(x)/T_c^{max}$,
a quantity proportional to the number of holes in the plane in the underdoped regime (see Fig.\ref{stockfigA}).
Their plot thus confirms for the 2-1-4 materials the conclusion of Mook \textit{et al.}\cite{mook98} that the incommensurability
of low-energy spin fluctuations in cuprate superconductors 
depends only on hole concentration in the plane, and not a particular material or band structure.

\paragraph{High-energy spin excitations and the resonance peak}

Early INS studies of high-frequency spin fluctuations were limited due to poor resolution. Many of these
measurements were initially done on the 1-2-3 family of materials, where the resonance peak\cite{fong,dai99} enhances
spin response at higher energies. The resonance peak in the 1-2-3 family of materials corresponds to a sharp
enhancement of the intensity of commensurate spin excitations at high frequencies near the
frequency, $E_{res}(x)$, that depends on hole doping $x$. The observation then raised the question
of whether the peak was a consequence of spin gap in the fermionic excitations or was an intrinsic property of the
spin liquid in the vicinity of $(\pi, \pi)$.  

\begin{figure}
\begin{center}
\includegraphics[width=0.75\textwidth]{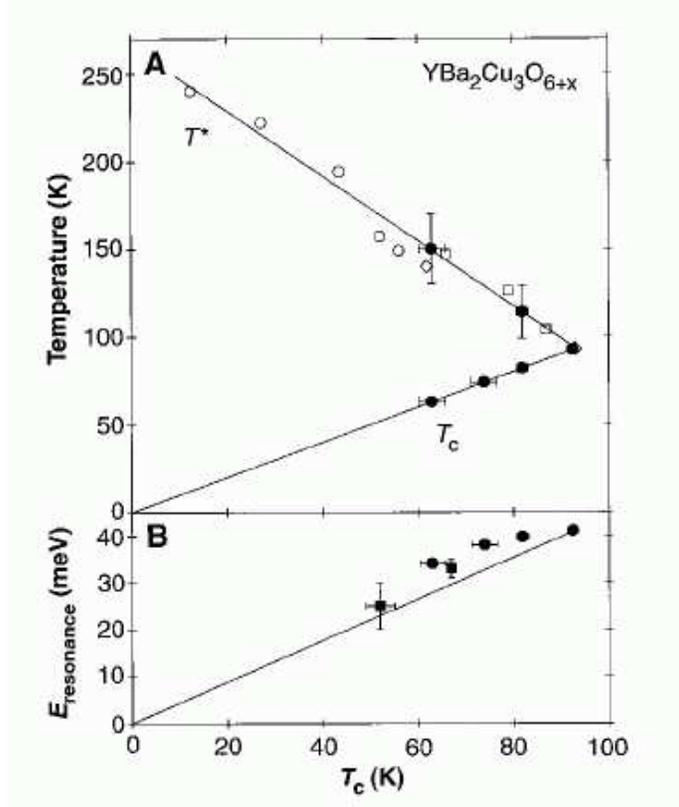}
\caption{$T^*$ and the resonance energy $E_{resonance}$ as a function of $T_c$ in YBa$_2$Cu$_3$O$_{6+x}$. 
The open circles and open squares are temperatures at which $d \rho(T)/dT$ reaches broad maximum.
The open diamonds show the pseudogap temperature $T^*$ determined from NMR measurements\cite{takigawa}. The filled circles correspond
to  $T_c$ and $T^*$, where the resonance first appears in INS measurements. Filled squares are from Fong \textit{et al.}\cite{fong}. Horizontal error bars are superconducting transition widths. The solid lines are
guides to the eye.  Reproduced with permission from \cite{dai99}.}
\label{daireson}
\end{center}
\end{figure}

At first the resonance was only observed
in the superconducting state. Since the resonance peak was not observed in the inelastic neutron scattering
experiments on the 2-1-4 family,
this led to an explanation that this feature most likely arises as a result of the coupling of spin excitations to 
fermions with a d-wave gap or d-wave like gap in the normal state. Indeed, Dai \textit{et al.}\cite{dai99} found that
the resonance peak first appears above $T_c$ at $T^*(x)$, which decreases with increased hole doping,
the resonance energy $E_{res}(x)$ increases with increased doping and tracks the doping dependence of $T_c(x)$, 
as shown in Fig.\ref{daireson}. 

However, more detailed neutron scattering studies revealed that the resonance corresponds to a special
frequency at which high-energy spin excitation spectrum becomes commensurate. As we have seen above, the
spin excitation spectrum depends only on hole doping, and not on a particular material, a conclusion that casts doubt
on the quasiparticle gap explanation of the resonance.
Inelastic neutron scattering studies near the resonance frequency were first done by 
Bourges \textit{et al.}\cite{bourges,bourges2000} in the
YBa$_2$Cu$_3$O$_{6.5}$ material. Bourges \textit{et al.} found that the commensurate resonance peak was broadening in momentum,
both above and below the resonance frequency $E_{res}$. Based on their findings, Bourges\textit{et al.} suggested
that the spin excitations disperse at high frequencies in a way that is similar to that expected for spin waves 
in the parent insulating compound.
Arai \textit{et al.}\cite{arai} later studied the momentum dispersion 
of spin excitations near the resonance peak in YBa$_2$Cu$_3$O$_{6.7}$ in more detail. They found evidence for two modes
that meet near the resonance energy, one that opens downwards and gives rise to incommensurate spin excitations
at low energies, the other, a new mode that opens upwards in energy, giving rise to spin wave-like dispersive
excitations at high energies. Arai \textit{et al.} found that the two modes meet at the frequency $\omega \simeq 41 meV$,
slightly above the characteristic resonance frequency $E_{res} = 36 meV$ for their material.
More recent INS experiments on YBa$_2$Cu$_3$O$_{6.5}$\cite{stock05}, YBa$_2$Cu$_3$O$_{6.6}$\cite{mook} and 
YBa$_2$Cu$_3$O$_{6.95}$\cite{reznik}, however, suggest that the high energy mode and the low energy mode meet at the 
resonance frequency. 

\begin{figure}
\begin{center}
\includegraphics[width=0.5\textwidth]{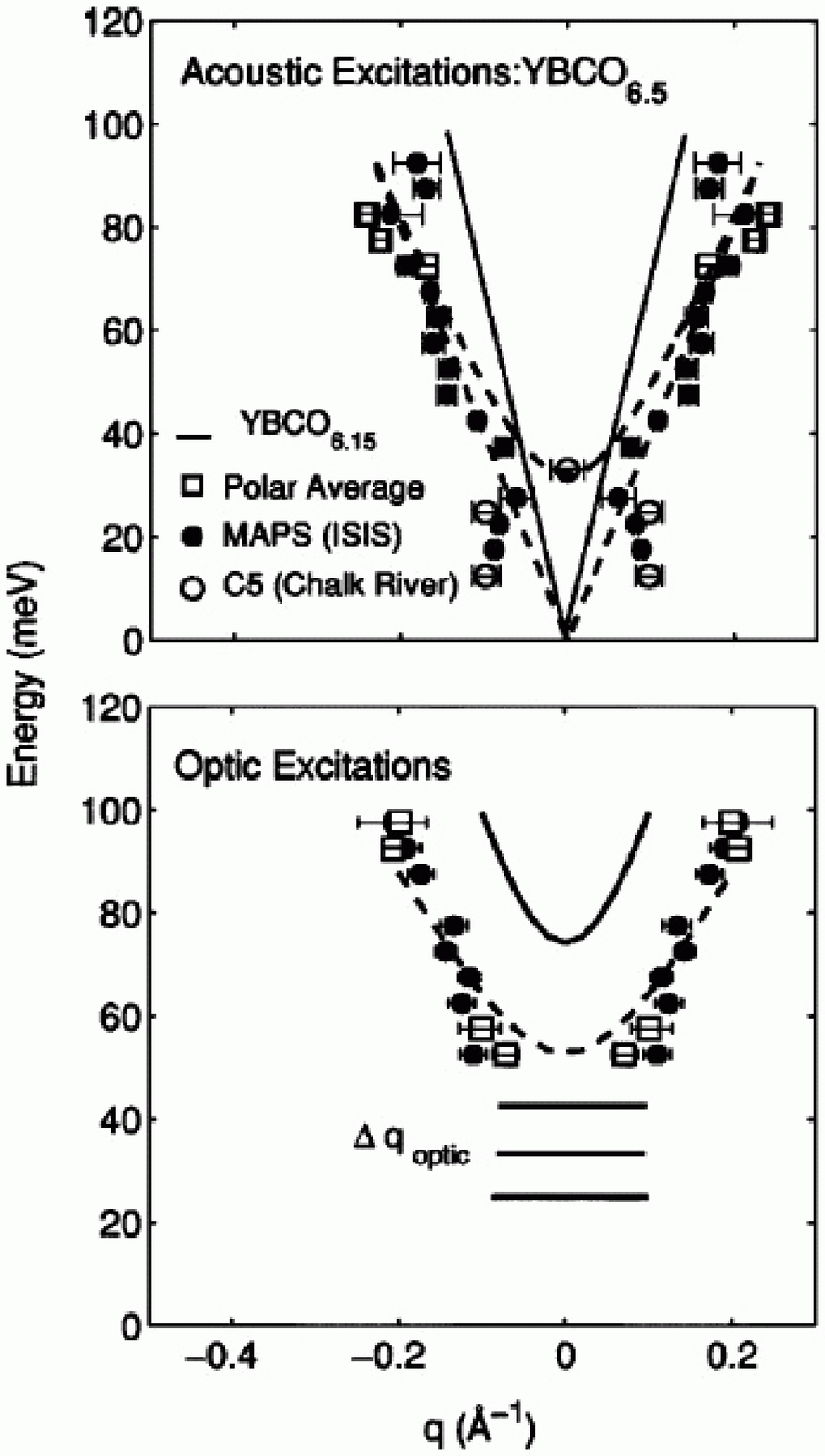}
\caption{The dispersion of acoustic and optic modes in YBa$_2$Cu$_3$O$_{6.5}$ with respect to
commensurate $(\pi, \pi)$ position at $T = 6K$. The filled circles are from two Gaussian fits along the 
$[100]$ and $[010]$ directions above the resonance energy and from $[100]$ direction below the resonance
energy. The open circles represent the positions of the incommensurate peaks found in experiments
conducted at Chalk River. The open squares represent the peak position that results from a polar
average around $(\pi,\pi)$ position. The solid lines schematically represent the dispersion of the
insulating compound as measured by Hayden \textit{et al.}\cite{hay96}. The dashed lines are fits
of the high-energy dispersion to linear spin wave theory for $\omega > 40 meV$. The horizontal
bars show the q-width observed for three optic scans at constant energy. 
Reproduced with permission from \cite{stock05}.}
\label{stockfigC}
\end{center}
\end{figure}

\begin{figure}
\begin{center}
\includegraphics[width=0.5\textwidth]{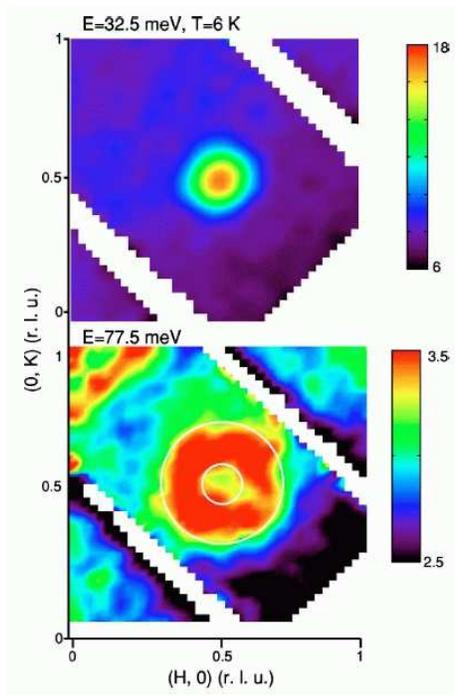}
\caption{Smoothed two-dimensional slices through the correlated response at $32.5 meV$ and $77.5 meV$
energy transfers and integrated $\pm 7.5 meV$ along the energy axis in YBa$_2$Cu$_3$O$_{6.5}$. 
At $77.5 meV$ the ring of scattering shows the intersection of the cone of spin wave dispersion emanating
from the $(1/2,1/2)$ position with a constant energy surface. Within statistics the velocity is isotropic.
The intensity is represented by false color. Reproduced with permission from \cite{stock05}.}
\label{stockfigB}
\end{center}
\end{figure}

A typical energy spectrum of the spin excitations as measured by Stock \textit{et al.}\cite{stock05} in a detwinned ortho-II
sample of YBa$_2$Cu$_3$O$_{6.5}$ is shown in Fig.\ref{stockfigC}. As Mook \textit{et al.}\cite{mook2000} had found earlier for
a detwinned sample of YBa$_2$Cu$_3$O$_{6.6}$, Stock \textit{et al.} found two, rather than four incommensurate peaks below the 
resonance frequency $E_{res} = 33 meV$. This result strongly suggested a one-dimensional character for the spin fluctuations, 
consistent with the
formation of dynamic stripes\cite{tranquada1,tranquada2}. The incommensurability $\delta$ observed at low 
frequencies decreased with increased energy transfer and disappeared at $\omega = E_{res}$. The spin excitations started to 
disperse outward again at high frequencies, $\omega > E_{res}$. Stock \textit{et al.}\cite{stock05} found the dispersion for 
acoustic spin excitations to be isotropic and similar to the spin waves in the insulator, as shown in Fig.\ref{stockfigB}; 
they fit their results to a gapped  spectrum,
\begin{equation}
\epsilon_{ac}(q) = \sqrt{\Delta_{ac}^2 + (c \bm{q})^2}, 
\end{equation}
with $\Delta_{ac} \simeq E_{res} \simeq 33 meV$ and a spin wave velocity $c \simeq 365 meV \AA$,
which is close to the value $c \simeq 400 meV \AA$ obtained from the slope of high energy spin excitations. The spin
wave velocity is dramatically reduced from its value in the insulator, $c \simeq 650 meV \AA$\cite{hay96}.  

Stock \textit{et al.} also measured the width of the ring of high energy spin wave excitations and found that
it increased linearly with frequency at low temperatures. The low-energy spectral
weight was dominated by the resonance peak at $E_{res} = 33 meV$, which had an
asymmetric shape, with a quick drop-off of intensity for $\omega > E_{res}$ and a much slower reduction at $\omega < E_{res}$.
Stock \textit{et al.}\cite{stock04}, however, estimated that the resonance represented only 3 \% of the total spectral weight
of spin excitations; most of its spectral weight would appear only at higher energies. 
Similar to spin waves in the insulating compound, $\chi"_L(\omega)$ approached a constant at high
energies $\omega > E_{res}$, $\chi"_L(\omega > E_{res}) \simeq 5 - 7 \mu_B^2/eV$, somewhat reduced from a similar value
for the insulator, $7 - 8 \mu_B^2/eV$\cite{stock05}. 

A very similar dispersion for spin excitations was found by Reznik\textit{et al.}\cite{reznik}
in a twinned sample of YBa$_2$Cu$_3$O$_{6.95}$. They observed four incommensurate peaks dispersing inward with increased 
frequency at frequencies
below the resonance frequency $E_{res} = 41 meV$ and becoming commensurate at the resonance frequency. The dispersion of spin excitations
was cut off below the resonance frequency at a  characteristic spin gap frequency $E_{gap} \simeq 33 meV$. 
Reznik \textit{et al.} concluded that their data was consistent with the isotropic gapped spin wave-like excitations 
dispersing outward at frequencies above the resonance frequency. As had Stock \textit{et al.}, 
Reznik \textit{et al.} found the $\bm{q}$-integrated
intensity to be approximately constant at high frequencies $\omega > E_{res}$, as expected for spin wave-like excitations.
Because of resolution problems, Reznik \textit{et al.} were not able to calibrate their data in absolute units, 
or estimate the spin wave velocity.

Not all INS studies have found spin wave-like excitations with a reduced effective exchange coupling $J$ at high energies.
For example, Hayden \textit{et al.}\cite{mook} performed measurements  
in a twinned sample of YBa$_2$Cu$_3$O$_{6.6}$ with effective hole doping $x \simeq 0.1$ close to the $x = 1/8$ stripe ordering
instability. As had other groups, Hayden \textit{et al.} observed incommensurate spin excitations that dispersed inward
up to resonance frequency $E_{res} \simeq 34 meV$. However, their measurements at frequencies above $E_{res}$ indicated
a strikingly different picture. Instead of the gapped spin waves found by other groups, Hayden \textit{et al.} resolved 
four peaks at $\omega > E_{res}$ rotated by 45 degrees that dispersed outward in energy. Their measured dispersion for
spin excitations is consistent with that observed earlier in stripe-ordered materials, such as La$_{2-x}$Ni$_x$CuO$_4$\cite{tranquada1}
and La$_{1.875}$Ba$_{0.125}$CuO$_4$\cite{tranquada2}. Hayden \textit{et al.}\cite{mook} found little dispersion of the
rotated incommensurate peaks at high energies, $66 meV < E < 105 meV$. This led them to conclude that the high-energy
spectrum of spin excitations they measured was inconsistent with commensurate gapped spin wave spectrum observed by other groups.

\begin{figure}
\begin{center}
\includegraphics[width=0.5\textwidth]{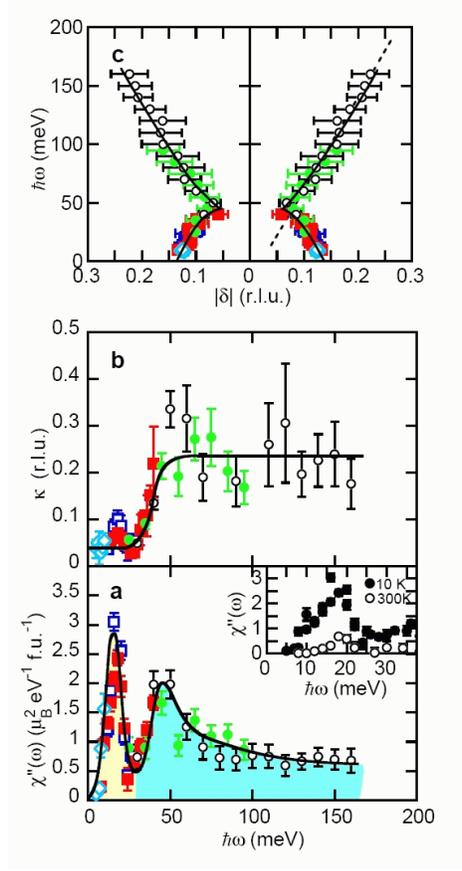}
\caption{Magnetic excitation spectrum and evolution of the form of the magnetic response with energy.
Symbols indicate different incident energies, open diamonds: $E_i = 30 meV$, open squares: $E_i=55 meV$,
filled squares: $E_i = 90 meV$, filled circles: $E_i=160 meV$, open circles: $E_i=240 meV$. 
"peak-dip-hump" structure in the integrated susceptibility $\chi''(\omega)$ suggesting that the 
magnetic response has two components. The inverse correlation length $k(\omega)$ shows significant broadening at high energies.  Reproduced with permission from \cite{vignolle}.}
\label{Vignsp}
\end{center}
\end{figure}

High energy spin excitations have also been studied in several materials belonging to the 2-1-4 family. 
The question of the renormalization of the effective exchange coupling
$J$ and the spin wave velocity $c$ of the insulator by doped carriers
was first investigated by Hayden \textit{et al.}\cite{hayden} for La$_{1.86}$Sr$_{0.14}$CuO$_4$ in
high-energy transfer inelastic neutron scattering experiments. 
They found that the spectrum of spin excitations at high energies fits the Heisenberg model well, 
with an effective exchange coupling $J$ that is only mildly reduced from its value in the insulator: 
for this material, $J_{eff} =130 meV$, $Z_{\chi} = 0.15$, values that are somewhat
reduced from the corresponding values in the insulator, $J=153 meV$, $Z_{\chi} = 0.39$. 
They concluded that quantum fluctuations increase, but that at the high energies they study, $J$ and $c$ do not get strongly
renormalized by doped holes. However, the spectral weight of their measured high-energy spin excitation decreases
very strongly from its value at $x=0$;  the peak of that weight for $x = 0.14$ was found\cite{hayden} to be at $\omega \simeq 22 meV$, 
significantly below the corresponding peak for the insulating compound. The details of the spin excitation spectrum
at low energies, however, could not be resolved.

More recent high resolution inelastic neutron scattering studies of the stripe-ordered compound, 
La$_{1.875}$Ba$_{0.125}$CuO$_4$\cite{tranquada2}, and the optimally doped superconductor, 
La$_{1.84}$Sr$_{0.16}$CuO$_4$\cite{christensen,vignolle},
reveal the details of the spin excitation spectrum at low energies. In their measurements 
on stripe-ordered sample of La$_{1.875}$Ba$_{0.125}$CuO$_4$ Tranquada \textit{et al.}\cite{tranquada2} found two different
branches of spin excitations that meet at a characteristic frequency $\omega_0 \simeq 50-55 meV$, an 
energy spectrum that is similar to that found by other groups, and a sharp feature in the spectrum of spin excitations 
at $\omega = 42 meV$.
Their results are consistent with the behavior seen earlier in other stripe-ordered compounds, such as La$_{2-x}$Ni$_x$CuO$_4$,
where the spectrum of spin excitations was measured earlier\cite{bourges03}. In particular, 
as found by Hayden \textit{et al.}\cite{mook}
and expected on a stripe model\cite{tranquada2},
Tranquada \textit{et al.} found four incommensurate peaks rotated by 45 degrees that disperse outward at high energies $\omega > 50 meV$.
They fit their data to a model of charge stripes separated by 2-leg Heisenberg ladders with $J \simeq 100 meV$, 
a value somewhat reduced from its value $J \simeq 146 meV$\cite{coldea} in the insulator.     
Recent high-resolution neutron scattering studies by Christensen\textit{et al.}\cite{christensen} 
and Vignolle \textit{et al.}\cite{vignolle} of  La$_{1.84}$Sr$_{0.16}$CuO$_4$ provide further important details on the
universality of the spin wave spectrum in high-temperature superconductors. Christensen \textit{et al.}\cite{christensen}
studied the spin excitation spectrum of this material at low frequencies $\omega < 38 meV$ and concluded that the
spectrum of incommensurate spin excitations is dispersive inwards and maps onto the spectrum of acoustic spin excitations observed in
YBa$_2$Cu$_3$O$_{6.85}$, a material with similar planar hole doping. Using this analogy, they concluded that spin excitations
in La$_{1.84}$Sr$_{0.16}$CuO$_4$ should become commensurate at a frequency $E_0 \simeq 41 meV$. However, they did not see
an associated resonance peak. Vignolle \textit{et al.}\cite{vignolle} recently extended these studies to 
higher frequencies; they found the spectrum of spin excitations shown in Fig.\ref{Vignsp}, with a high-frequency
component emerging for $\omega > 40 meV$. Both $J$ and $c$ do get strongly renormalized with doping:
the high-energy excitation spectrum they observe corresponds to commensurate gapped spin waves with
an effective coupling $J_{eff} = 81 meV$, a value that is dramatically reduced from its value $J=146meV$ in the insulator. Below
$\omega = 40 meV$ they find the usual\cite{cheong,Mason} attenuated incommensurate structure. Their analysis of the unusual
double peak structure in the spectral weight suggests the spin excitation spectrum separates into two components - incommensurate 
attenuated spin  excitations for $\omega < 40 meV$, and commensurate gapped spin waves for $\omega > 40 meV$. They find that
the spectral weight at high energies, $\chi''(\omega = 150 meV)$ is roughly 1/3 of the spectral weight observed in the parent 
insulating compound. 

In summary, the energy spectrum of spin excitations as probed by inelastic neutron scattering is approximately universal, 
i.e., for different materials, it reflects only
the hole doping $x$. In particular, the positions of incommensurate peaks at low frequencies are the
same in both 2-1-4 and 1-2-3 families of materials at similar hole doping levels. The spectrum of spin excitations at all frequencies
also turns out to be the same at a similar doping level, at least for the YBa$_2$Cu$_3$O$_{6.85}$ and La$_{1.84}$Sr$_{0.16}$CuO$_4$
pair. The energy spectrum at high energies is consistent overall with what is expected for gapped spin waves, 
and measurements on different materials 
reveal a suppression of the effective exchange constant $J_{eff}(x)$ and the spin wave velocity $c(x)$ with increased doping, 
as well as a suppression of the spectral weight of high-energy spin excitations. However, experiments in materials with the hole
doping close to $1/8$ found a rotated peak structure at high energies, similar to that expected from static stripe ordering.
Overall, the energy spectrum of spin excitations in different families of materials is consistent with the picture of phase
separation and fluctuating stripe order that first appears below $T^*$.

\subsection{Thermodynamics}

The specific heat measurements and electronic entropy analysis of Loram \textit{et al.}\cite{loram} provide another important constraint
on the normal state behavior of underdoped YBa$_2$Cu$_3$O$_{6+y}$. By fitting their data to fermions 
that are assumed to have a ``normal state'' energy gap of the BCS d-wave form,
Loram \textit{et al.}\cite{loram} found that  at $T=110K$, well above the superconducting temperature, $T_c$,
their hypothesized gap, $\Delta(T,x)$,  increases linearly as the doping is decreased,  from being 
$\sim$ zero in a near optimally-doped  sample (y = 1) to 200K in an underdoped sample with  y = 0.7.  
It is clear from their entropy data that their proposed gap size is related to $T^m(x)$ and $J_{eff}(x)$,
that is:
\begin{equation}
\Delta(T=110K) \propto T^m(x) \propto J_{eff}(x).
\end{equation}
In a subsequent paper, Loram \textit{et al.}\cite{loram04} argued that
$S/T$ and the $Y$ Knight shift $\chi_S$  are proportional to each other, with
a Wilson ratio for nearly free electrons,
\begin{equation}
\chi_S(T) \simeq a_W \frac{S}{T}\,. 
\end{equation}

\subsection{Transport measurements}

\begin{figure}
\begin{center}
\includegraphics[width=0.75\textwidth]{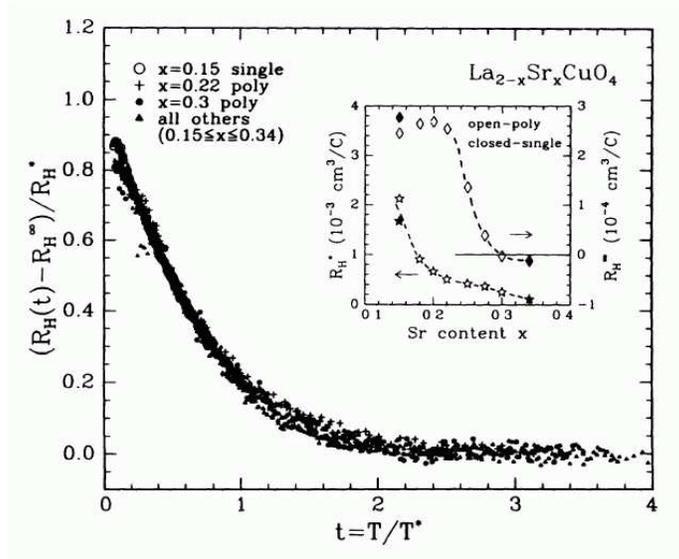}
\caption{The Hall coefficient ($R_H$) for La$_{2-x}$Sr$_x$CuO$_4$ with $0 \ge x \ge 0.34$ plotted rescaled as 
$[R_H(t)-R_H^{\infty}(x)]/R_H^*$ vs $t = T/T^*$. Inset: the parameters $R_H^{\infty}$ and $R_H^*$ vs Sr composition $x$.
The temperature $T^*$ agrees with $T^m$ obtained from the maximum in the bulk spin susceptibility. Reproduced with permission from \cite{Hwang}}
\label{hwangfig}
\end{center}
\end{figure}

The peculiar $T$-linear behavior of the electrical resistivity\cite{resist} that 
has been observed near optimal doping in all families of cuprate superconductors, 
represents one of their main unresolved puzzles.
It has been explained as representing marginal
Fermi liquid behavior arising from the proximity of optimally doped
materials to a quantum critical point\cite{varma1}, with concomitant $\omega/T$ scaling\cite{resist}. 
However, in the underdoped materials, the electrical resistivity shows strong deviations from
linearity in $T$, below a temperature $\sim T^m$. Moreover, scaling with a characteristic 
temperature $T^m$ has been clearly seen in other 
transport measurements, with the best data collapse being that observed
in the Hall measurements of Hwang \textit{et al.}\cite{Hwang}, who use
the following scaling form:
\begin{equation}
R_H = R_H^{\infty}(x) + R_H^m(x) f(T/T^m),
\end{equation}
where $f(T/T^m)$ is a universal function and
$R_H^{\infty}(x)$ and $R_H^m(x)$ are doping-dependent functions(Fig. \ref{hwangfig}).
 
\begin{figure}
\begin{center}
\includegraphics[width=0.75\textwidth]{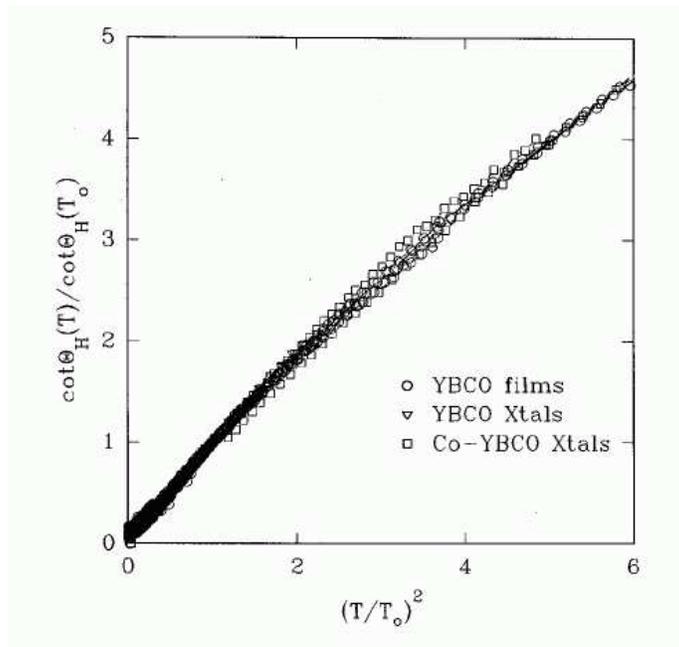}
\caption{Cotangent of the Hall angle $\cot \theta_H$ divided by $\cot \theta_H(T_0)$ vs square of reduced temperature
$(T/T_0)^2$ for 15 sets of thin film data together with reported data for oxygen-deficient and Co-doped YBCO single
crystals. $T_0(x)$ agrees with the temperature obtained by rescaling $Y$ Knight shift. 
Reproduced with permission from \cite{wuyts}.}
\label{WAnfig}
\end{center}
\end{figure}

\begin{figure}
\begin{center}
\includegraphics[width=0.75\textwidth]{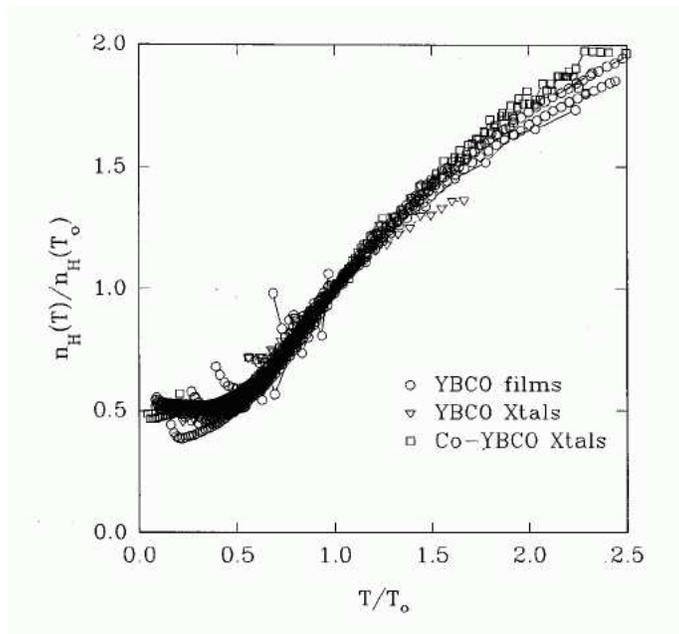}
\caption{Hall number $n_H$ divided by $n_H(T_0)$ vs reduced temperature
$T/T_0$ for 15 sets of thin film data together with reported data for oxygen-deficient and Co-doped YBCO single
crystals. $T_0(x)$ agrees with the temperature obtained by rescaling $Y$ Knight shift.   
Reproduced with permission from \cite{wuyts}.}
\label{WNfig}
\end{center}
\end{figure}

The Hall data for YBa$_2$Cu$_3$O$_{7-y}$ from Ito \textit{et al.}\cite{ito} and Carrington \textit{et al.}\cite{carrington} 
is analyzed in Wuyts \textit{et al.}\cite{wuyts}, who find 
the following scaling forms (Figs. \ref{WAnfig}, \ref{WNfig}) for the Hall angle, $\theta_H(T)$, and the number, $n_H(T)$, 
lead to excellent data collapse:
\begin{equation}
cot(\theta_H(T)) = cot(\theta_H(T^m)) \frac{T^2}{(T^m)^2}\,
\end{equation}
\begin{equation}
n_H(T) = n_H(T^m) \frac{T}{T^m}\,
\end{equation}

\begin{figure}
\begin{center}
\includegraphics[width=0.75\textwidth]{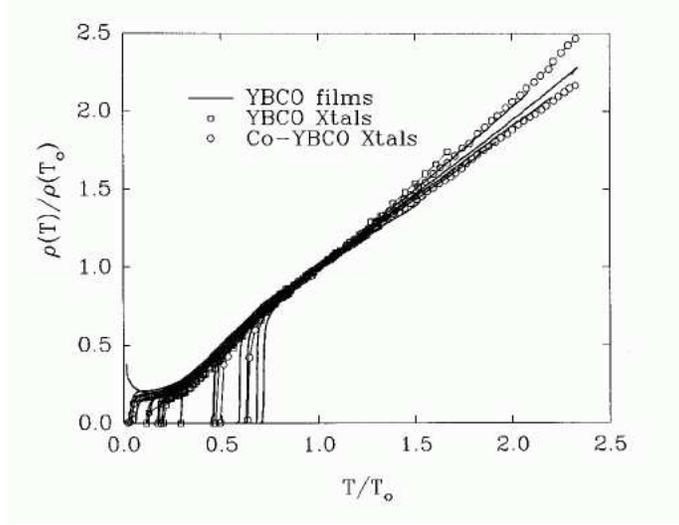}
\caption{Scaled in-plane resistivity $\rho/\rho(T_0)$ vs scaled temperature $T/T_0$  for 15 sets of thin film data together 
with reported data for oxygen-deficient and Co-doped YBCO single
crystals. $T_0(x)$ agrees with the temperature obtained by rescaling $Y$ Knight shift.
Reproduced with permission from \cite{wuyts}.}
\label{WRfig}
\end{center}
\end{figure}

Wuyts \textit{et al.}\cite{wuyts} extended the $T$-linear scaling of resistivity
near optimal doping to the underdoped regime. Their analysis of
Ito \textit{et al.}'s\cite{ito} resistivity data for the YBa$_2$Cu$_3$O$_{7-y}$ family shows
that for temperatures $T$ above $T^m(x)$ the resistivity is universal and linear (Fig. \ref{WRfig})  , and is similar
to the optimally doped samples:
\begin{equation}
\rho(T) = \rho(T^m) \frac{T}{T^m}\,
\label{rhosc}
\end{equation}
The doping dependence of $T^m(x)$ inferred from these scaling forms for the transport coefficients 
is consistent with its direct magnetic measurements.

Levin and Quader analyzed scaling of the Hall and transport\cite{LQ2,LQ3} 
data based on the two-band model Eq.(\ref{eqLQ}) with degenerate ($\xi$) and nondegenerate ($\eta$) carriers.
They found that the $T^2$ dependence of the cotangent of the Hall angle and the deviations from it
can be fit to their model if two lifetimes are introduced for two different bands,
$\tau_{\xi} \propto T^{-1}$ and $\tau_{\eta} \propto T^{-2}$. Their two-band model predicts
three-parameter scaling for the planar resistivity in the form:
\begin{equation}
\rho =  \frac{A(x) T}{\alpha(x) + \Phi(W(x)/T)},
\end{equation}
where $W(x) = \Gamma (x - x_0)$ is the energy scale analogous to $T^*$.

A more recent alternative analysis\cite{GT1} of the high-temperature Hall data\cite{nishikawa,ando,padilla}
for the underdoped La$_{2-x}$Sr$_x$CuO$_4$ family has its basis on the idea of phase separation. It indicates 
that the Hall data can be understood if one assumes an activated form for the
carrier concentration,
\begin{equation}
n_{Hall} = n_0(x) + n_1(x) \exp{(- \Delta(x)/k_B T)},
\label{LPG}
\end{equation}
with a doping-dependent $n_0(x)$ that stays linear in $x$ up to $x \simeq 0.12$, above which strong deviations from
linearity are observed. The behavior of $\Delta(x)$ was found to be consistent with that measured in
photoemission experiments. According to Ref.\cite{GT1}, there is crossover 
behavior when the number of activated carriers becomes approximately equal to the number of doped carriers 
$n_0(x)$; this occurs at a temperature
\begin{equation}
T^*(x) \simeq T_0(x) = - \frac{\Delta(x)}{\ln{x}}\,,
\label{LPG1}
\end{equation} 
that is consistent with characteristic temperatures inferred from other measurements.
At a proposed candidate QCP, $x \simeq 0.2$, the energy gap for activated carriers, $\Delta(x)$, goes
to zero\cite{GT1}.

\subsection{Penetration depth measurements}

\begin{figure}
\begin{center}
\includegraphics[width=0.75\textwidth]{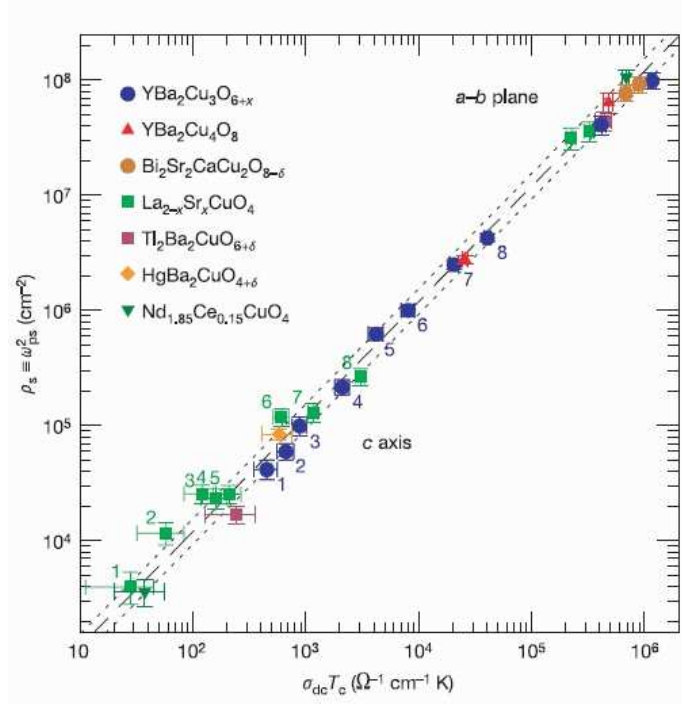}
\caption{Plot of the superfluid density ($\rho_S$) versus the product of d.c. conductivity ($\sigma_{dc}$) and the superconducting
transition temperature ($T_c$) for various cuprate superconductors. $\sigma_{dc}$ is measured just above the transition. Within
error, all of the data fall on the same universal (dashed) line with slope of unity, defined by Eq.(\ref{uem}); the dotted lines
are from $\rho_S = (120 \pm 25) \sigma_{dc} T_c.$ Reproduced with permission from \cite{homes}.}
\label{Uemurafig}
\end{center}
\end{figure}

One of the earliest universal relations that emerged in the high-$T_c$ field is the linear Uemura relation\cite{uemura}
between the superfluid density $\rho_S$, provided by penetration depth measurements, and the superconducting temperature, $T_c$,
valid in the underdoped regime. This relation provides an important constraint on the number of electrons that become
superconducting. Recently\cite{homes} it has been shown (Fig. \ref{Uemurafig}) that a modified Uemura scaling relation,
\begin{equation}
\rho_S = 120 \sigma_{dc} T_c,
\label{uem}
\end{equation}
holds for all cuprate superconductors.

\subsection{Angle-resolved photoemission spectroscopy}

Enormous progress in resolution has been made in angle-resolved photoemission spectroscopy (ARPES) experiments in the past decade 
so that ARPES has now emerged as one of the best experimental probes of the cuprates\cite{per}. 
One of the most important contributions of ARPES  was the detection of the anisotropic normal state gap in
the fermionic spectrum, first by Marshall \textit{et al.}\cite{marshall}, Loeser \textit{et al.}\cite{loeser}, and then
by Ding \textit{et al.}\cite{ding} in Bi$_2$Sr$_2$CaCu$_2$O$_{8+\delta}$. 

By examining their data, Marshall \textit{et al.} concluded that the normal state gap has two energy scales. 
They identified the low energy scale gap ($\sim 20-30 meV$) by the location of the leading-edge midpoint. This 
is a clear gap in the fermionic spectrum that has a $d$-wave-like momentum dependence, with gapless Fermi surface 
arcs near the nodal regions. The high energy scale ($\sim 100-200 meV$) gap appears as a broad 
incoherent feature in the spectrum near the $(\pi,0)$ point, where the low-energy spectral weight is strongly suppressed.
The low energy leading-edge gap first appears below $T^*(x)$, a temperature that has the same doping dependence as 
$T^m(x) \sim J_{eff}(x)$ observed in other measurements. However, $T^*$ inferred from ARPES measurements\cite{per,campTst} 
in  Bi$_2$Sr$_2$CaCu$_2$O$_{8+\delta}$ turns out to be significantly lower than $T^m$,  $T^*(x) \simeq T^m(x)/3 \simeq J_{eff}(x)/3$.  
Both the measured magnitude and the $\bm{k}$-dependence of the normal state leading edge gap are very similar 
to that of the d-wave gap in the superconducting state; the normal state gap smoothly evolves into a superconducting
gap below $T_c$. The doping dependence of the normal state d-wave-like gap amplitude $\Delta(x)$  tracks 
that of $T^*(x)$\cite{per,campTst}.

The ratio of $T^*(x)$ and $T^m(x)$ varies for different families of materials. 
Detailed ARPES investigations of the La$_{2-x}$Sr$_x$CuO$_4$\cite{ino,yoshida} and other families of cuprate
superconductors\cite{per} find the same general form of the spectrum and the same linear doping dependence of the energy gap 
amplitude $\Delta(x) \propto T^m(x)$ as that found in the  Bi$_2$Sr$_2$CaCu$_2$O$_{8+\delta}$ family of materials. 
Following these and other ARPES results,
Damascelli \textit{et al.}\cite{per}  suggest that the value of $T^*(x)$ and $\Delta(x)$ is determined by the maximum value
of $T_c$ for a given material, $T_c^{max}$, rather than exchange coupling $J$:
\begin{equation}
\Delta(x) \propto T^*(x) \propto T_c^{max} 
\end{equation}

The high energy incoherent feature of the ARPES spectra also has a d-wave like dispersion\cite{campTst}. 
It is often claimed to be a remnant of the antiferromagnetic insulator\cite{per}, since it  
exhibits the same dispersion\cite{campTst} along the $(0,0)$-$(\pi,0)$  and $(\pi,\pi)$ - $(\pi,0)$ directions, which are
only equivalent in the reduced antiferromagnetic Brillouin zone, and is similar to the ARPES spectra observed in undoped
antiferromagnetic insulators.

\begin{figure}
\begin{center}
\includegraphics[width=0.75\textwidth]{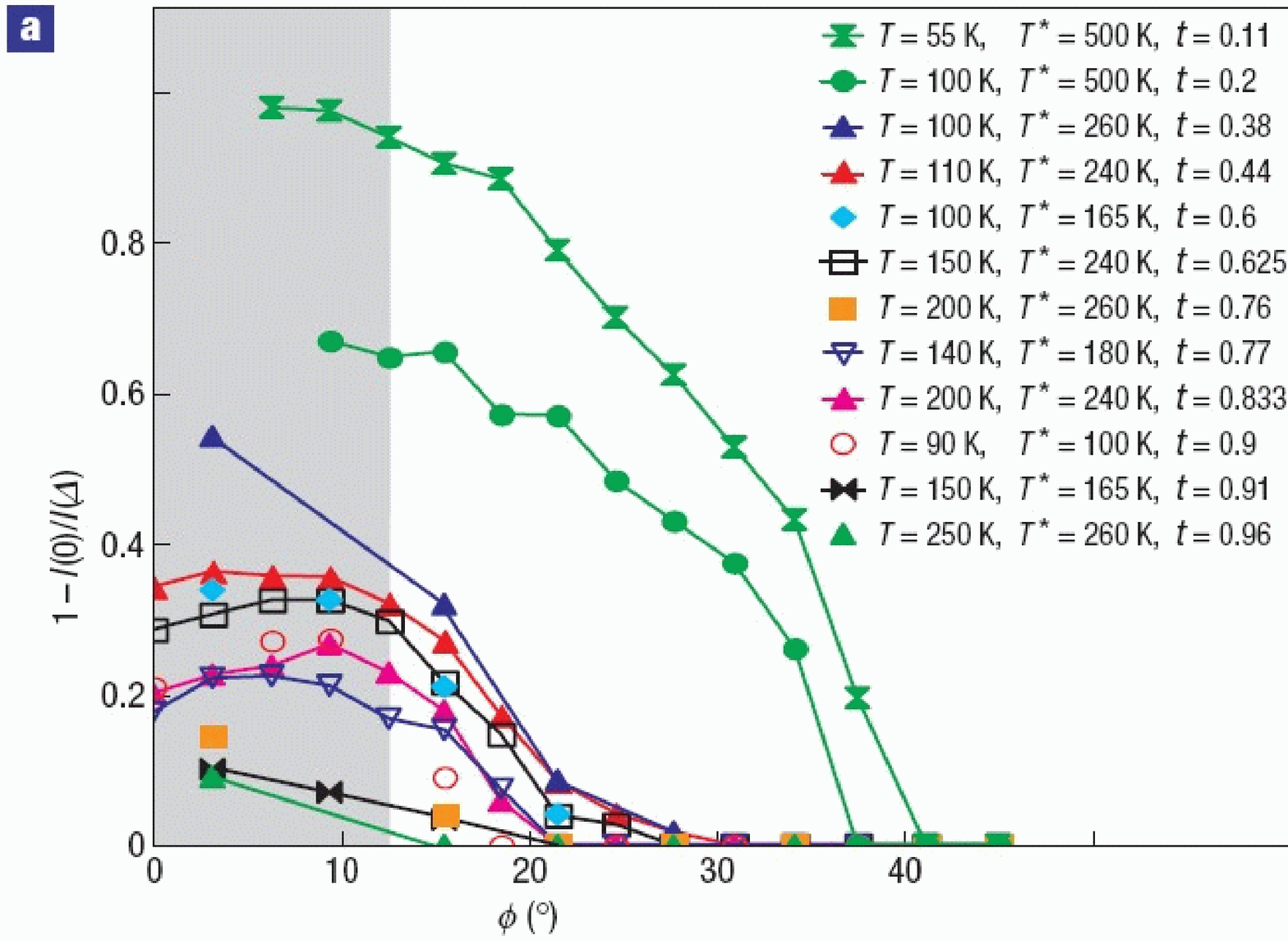}
\includegraphics[width=0.75\textwidth]{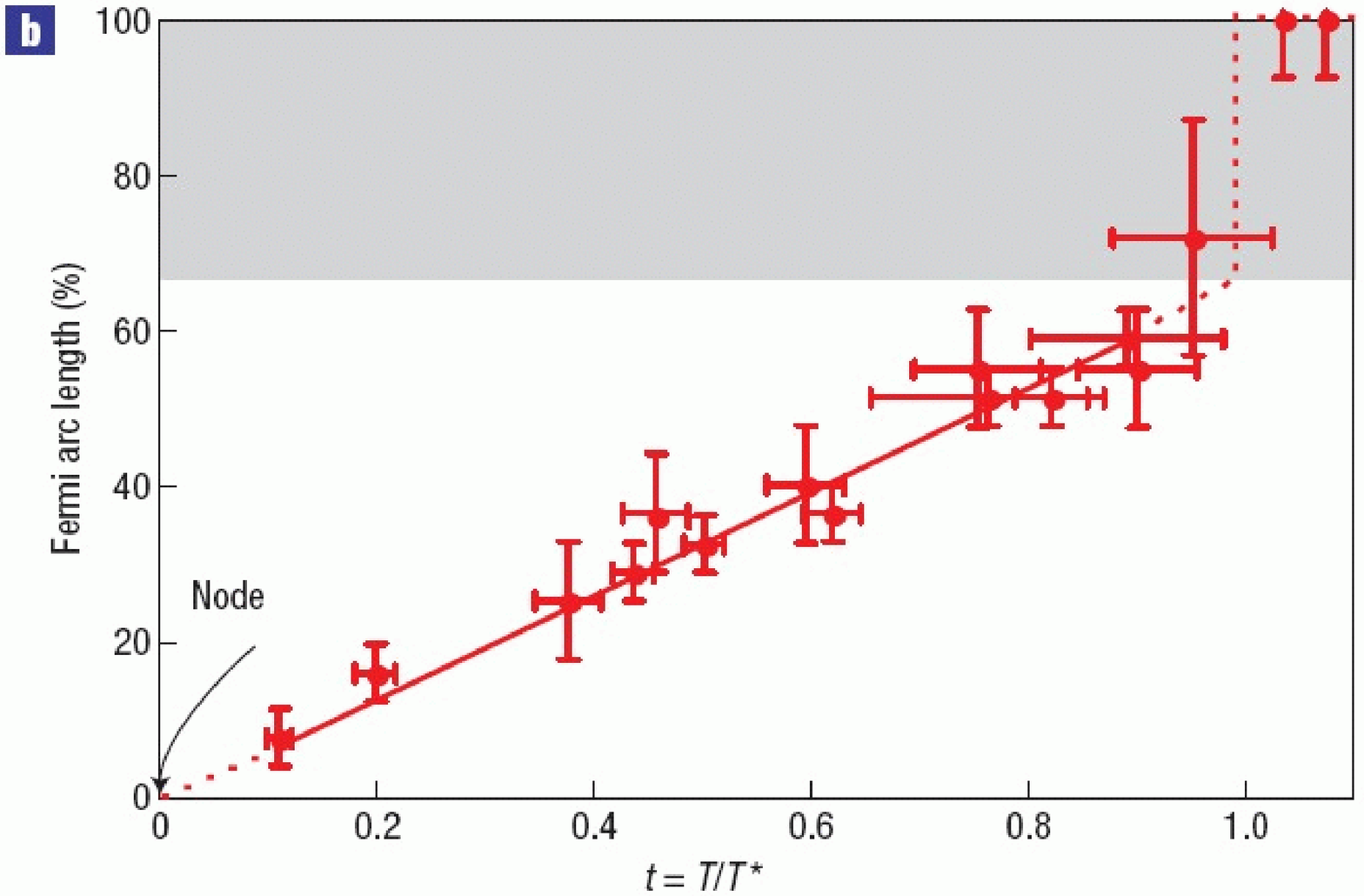}
\caption{(a) Loss of low-energy spectral weight $L(\varphi) = [1 - I(0, \varphi)/I(\Delta,\varphi)]$, where $I(\Delta, \varphi)$ is
the symmetrized intensity at the gap energy at the Fermi point labeled by angle $\varphi$, and $I(0, \varphi)$ is that at the
Fermi energy at the same point. Different symbols correspond to different $t$. The gray area represents the straight
section of the Fermi surface (b) Variation of the arc length with respect to reduced temperature $t = T/T^*$. On the y axis, 0 \%
is the node and 100 \% is the antinode.  Reproduced with permission from \cite{campuz}.}
\label{campscal}
\end{center}
\end{figure}

The fact that the normal state gap closes non-uniformly in momentum space with increased temperature or doping 
has been well established for some time\cite{per}. Recently Kanigel \textit{et al.}\cite{campuz} investigated the
temperature- and doping- dependence of the low energy normal state gap in several samples of Bi2212 and 
discovered a remarkable scaling relation: the anisotropy of the normal state gap and the length of the gapless 
Fermi arcs near the  nodal region for these different samples depend only on a single parameter, a reduced temperature $t = T/T^*(x)$
(Fig. \ref{campscal}).
Kanigel \textit{et al.} also found that the length of gapless Fermi arcs near the nodes above the superconducting temperature $T_c$
is linear in $t$, while below $T_c$ the Fermi arcs develop the usual d-wave superconducting gap with point nodes. 

Is the superconducting gap that develops at $T_c$ in the nodal Fermi arc region the same as the normal state gap, or is it a
different gap? Recently Millis\cite{millis06} argued that this important question has not been resolved for years, 
since different experimental studies yield conflicting results. An experimental observation of two different gaps in
the underdoped regime will confirm that the normal state gap in the underdoped cuprates is driven by a different, competing
nonsuperconducting order, not pairing fluctuations\cite{RVB}.
Since the superconducting gap in the nodal
region is extremely small, answering this question in ARPES requires a very high resolution study. 
So far ARPES studies have proved to be conflicting as well\cite{millis06}. 
For example, a recent leading edge gap ARPES study of Tanaka \textit{et al.}\cite{tanaka} on three different samples of Bi2212 found
that the superconducting gap in the nodal arc region had a doping dependence different from that of $T^*(x)$,
and scaled with $T_c(x)$ instead. On the other hand, Valla \textit{et al.}\cite{valla} reported that the energy gap in 
a non-superconducting sample of La$_{1.875}$Ba$_0.125$CuO$_4$, where the superconducting order is suppressed by
charge ordering, has essentially the same simple d-wave form and magnitude as in the superconducting sample at 
higher doping concentration, thus yielding crucial support to the single gap scenario.

\subsection{STM Experiments}

All STM studies find strong nanoscale inhomogeneity, in which gap magnitudes are observed to vary strongly on the nanometer
length scale,  while 
the recent STM measurements of Alldredge \textit{et al.}\cite{davis}, Gomez \textit{et al.}\cite{yazdani},
and Boyer \textit{et al.}\cite{hudson}
provide detailed evidence on gap formation in the underdoped 2212 and 2201 families of cuprates.

According to Gomez\cite{yazdani}, in the overdoped regime locally there is only one gap 
that has a d-wave symmetry.
The distribution of the energy gaps is, nevertheless, very inhomogeneous, indicating the presence of 
disorder. As the temperature is raised above $T_c$, the total spatial area of ungapped regions increases
until the whole sample becomes ungapped. The local temperature $T_p$ at which the normal state gap
first appears varies strongly in space. Nevertheless,  the ratio of the local gap maximum $\Delta_{max}$ in $\bm{k}$-space 
to the local ordering temperature $T_p$ turns out to be universal for different spatial regions and 
very large\cite{yazdani}, $2 \Delta/k_B T_p = 7.7$, much greater than the d-wave BCS limit of $4.3$, indicating strong coupling.

In the underdoped regime, Gomez \textit{et al.} first see the spatially inhomogeneous formation of the local gaps 
below $T^*$, defined as the temperature  at  which 90 \% of the sample becomes gapped. Below $T_c$, Gomez \textit{et al.}
observe the formation of a smaller kink inside the larger local gap, corresponding to the onset of superconductivity. 
They conclude that in the underdoped regime there are two energy scales, with the lower energy scale corresponding to the 
onset of phase coherence. Boyer \textit{et al.}\cite{hudson} adopted a different approach in their analysis of their STM 
measurements on an overdoped sample, a member of Bi2201 family, (Bi$_{1-y}$Pb$_y$)$_2$Sr$_2$CuO$_{6+x}$ (T$_c = 15 K$).
Since inhomogeneity is unaffected by the onset of superconductivity at $T_c$, Boyer \textit{et al.} look for 
the signature of superconductivity, the emergence of the subgap kink at T$_c$,  by removing the effective background of 
the high-temperature STM spectra, which is largely unaffected by the onset of superconductivity, 
from the low-temperature spectra. As a result of this subtraction, they find a second,  homogeneous superconducting gap, 
$\Delta_{small} = 6.7 \pm 1.6 meV$ that forms at $T_c = 15K$. Their analysis thus strongly suggests that the subgap
kink at $T_c$, also seen by Gomez \textit{et al.}, rather than being an onset of phase coherence, corresponds to the
opening of superconducting gap at $T_c$. 

Alldredge \textit{et al.}\cite{davis} have studied the tunneling density of states in momentum space. 
Their analysis indicates that there is only one gap in the underdoped materials. Alldredge \textit{et al.}
find that a large effective anisotropic scattering rate $\Gamma(\bm{k}) \propto |\Delta(\bm{x}|$ is needed to fit their 
data in the underdoped regime. Their effective scattering rate becomes non-zero at $x \leq 0.22$, and 
increases approximately linearly with decreased doping, similarly to the normal state energy gap. 
Thus, they conclude, they have effectively two kinds of quasiparticles in the underdoped regime:
quasiparticles near the $(\pi,0)$ antinodes in momentum space that are incoherent and almost localized,
and coherent quasiparticles in the nodal region.

\section{A two-fluid analysis of experimental data}

The observations of data collapse reviewed in the previous section provide  two very important constraints on any theory of  
the underdoped cuprate superconductors. They indicate the presence of two distinct fluids whose composition might be 
expected to vary with doping: a spin liquid containing localized Cu spins with a doping-dependent effective interaction 
and a Fermi liquid whose transport properties differ markedly from those of a Landau Fermi liquid. In this section we develop 
a two-fluid  framework for analyzing these experiments and use it to extract their consequences. We present as well the details 
of our earlier analysis\cite{BPprl} of the magnetic measurements of two fluid behavior.

\subsection{Two-Fluid Description}

Our approach is  inspired by the success of the two-fluid phenomenology developed for the 1-1-5 family
of heavy electron materials\cite{NPF,NPFcurro}. For these and other heavy electron materials containing a Kondo lattice 
of localized $f$-electrons coupled to a conduction band, a two-fluid phenomenological model of the hybridization 
of the $f$ electron localized spins with those in the conduction band describes very well the emergence of 
a hybridized  non-Landau  heavy electron Fermi liquid that coexists with unhybridized local f-electrons and 
conduction electrons. The emergence of two components in the bulk spin  susceptibility and the  Knight shift in 
the cuprates  can be understood by writing the total spin of the system as a sum of the
localized $d$-electron and $p$-hole spins,
\begin{equation}
\bm{S}_{tot} = \sum_i \bm{S}^d(\bm{r}_i) +  \sum_j \bm{S}^p(\bm{r}_j),
\label{pdhyb}
\end{equation}
where $\bm{r}_i$ are the positions of copper $d$-electrons, $\bm{r}_j$ the positions
of oxygen p-holes. Quite generally, the coupling between these spins gives rise to
three contributions to spin susceptibility\cite{NPFcurro},
\begin{equation}
\chi = \chi_{dd} + 2 \chi_{dp} + \chi_{pp},
\label{totalchi}
\end{equation}
where $\chi_{dd}$ represents the contribution from the localized Cu spins,
$\chi_{pp}$ represents the part of the oxygen $p$-band that is not hybridized,
while $\chi_{dp}$ corresponds to the magnetic response of the 
hybridized quasiparticles with a large Fermi surface:
\begin{eqnarray}
\chi_{dd} &=& \frac{1}{N}\, \sum_{i,i'} \langle  \bm{S}^d(\bm{r}_i)  \bm{S}^d(\bm{r}_{i'}) \rangle \\
\chi_{dp} &=& \frac{1}{N}\, \sum_{i,j} \langle  \bm{S}^d(\bm{r}_i)  \bm{S}^p(\bm{r}_{j}) \rangle \\
\chi_{pp} &=& \frac{1}{N}\, \sum_{j,j'} \langle  \bm{S}^p(\bm{r}_j)  \bm{S}^p(\bm{r}_{j'}). \rangle 
\end{eqnarray}
The Fermi liquid contribution to spin susceptibility thus arises from $\chi_{pd}$ and $\chi_{pp}$.
Only \textbf{the position} of oxygen holes is important
for the above decomposition; whether or not they form a single hybridized band
is irrelevant. The three parts are present in an
explicit two-band model of Walstedt \textit{et al.}\cite{walst} that produces correct expressions for the relaxation rates in
terms of $\chi_{pp}$, $\chi_{pd}$, and $\chi_{dd}$. It differs from 
the standard Mila-Rice-Shastry/MMP\cite{Mila:Rice:Shastry,MMP} ionic model that assumes
a single spin degree of freedom residing on the copper site, so that
$\chi_{pd} = \chi_{pp} \equiv 0$ and the hybridized $pd$ band description of Millis and Monien\cite{milmon},
in which although $\chi_{pp}$, $\chi_{pd}$, and $\chi_{dd}$ are taken to be non-zero, all three components
are assumed to track the spin susceptibility of a single hybridized $pd$ band.

The scaling results of the previous section tell us that $\chi_{dd}$ must
maintain its local character throughout much of the phase diagram, for
how else could the bulk susceptibility and the low frequency magnetic
response map onto the 2D Heisenberg model? To describe this we
therefore write:
\begin{equation}
\chi_{dd} = f(x) \chi_{SL}(\bm{q}, \omega)
\label{ddpart}
\end{equation}
where $f(x)$ is the fraction of the $dd$ (and the total) response
function that retains its local spin character, while
the remaining portion of the spin-spin response function
describes the response of a (non-Landau) Fermi liquid
with the large Fermi surface that results from hybridization:
\begin{equation}
2 \chi_{dp} + \chi_{pp} = [1-f(x)] \chi_{FL}(\bm{q}, \omega).
\label{pdpart}
\end{equation}
We thus arrive at the two fluid description,
\begin{equation}
\chi(\bm{q}, \omega) = f(x) \chi_{SL}(\bm{q}, \omega) + [1-f(x)] \chi_{FL}(\bm{q}, \omega),
\label{dyn}
\end{equation}
that we will use in our subsequent analysis to extract from experiment 
the doping dependence of both $\chi_{SL}$ and $f(x)$.

The spin liquid contribution completely determines the low frequency dynamic magnetic spin susceptibility measured by NMR 
on copper nuclei. It corresponds to the scaled spin response of the 2D Heisenberg model for localized copper spins, which
in the low frequency limit can be written in a general phenomenological form proposed by 
Millis, Monien, and Pines\cite{MMP}, modified to include the possibility of propagating spin wave excitations \cite{BPST}:
\begin{equation}
\chi_{SL}(\bm{q},\omega) = 
\frac{\alpha \xi^2}{1 + \xi^2 (\bm{q} - \bm{Q})^2 - i \frac{\omega}{\omega_{SF}}\, - \frac{\omega^2}{\Delta_{SL}^2}\,}\,,
\label{SLpar}
\end{equation}
where $\Delta_{SL} = c/\xi$ is the gap in spin excitation spectrum.
We note that in general damping in the Heisenberg model is caused by the spin wave
scattering, so that the relaxational frequency, $\omega_{SF}$, must be, in general, 
frequency-dependent in the underdoped cuprates.
We further note that if one assumes Heisenberg model scaling, then, from dimensional arguments, 
$\alpha(x) = \chi_{\bm{Q}} \xi^{-2}$ is proportional to $1/T^m(x)$.

The Walstedt \textit{et al.} generalization of the Mila-Rice-Shastry hyperfine Hamiltonian for Cu site 
can be written as:
\begin{equation}
^{63}H_{hyp} = \sum_{\rho} \left\{ ^{63}I_{\rho} [A_{\rho} \bm{S}_{\rho}^d  + B \sum_i \bm{S}_{\rho}^d(\bm{r}_i)] +   
^{63}I_{\rho} D \sum_j \bm{S}^p_{\rho}(\bm{r}_j) \right\}
\label{63hyp}
\end{equation}
Here $i$ corresponds to the nearest neighbor Cu sites, $\rho = a, b, c$ while $D$ is a new transferred hyperfine constant
for the spins sitting on oxygen. The sum index $j$ goes over the oxygen sites neighboring the copper site.
Similarly, one can write the hyperfine Hamiltonian for the oxygen nuclei:
\begin{equation}
^{17}H_{hyp} = \sum_{\rho} \left\{ ^{17}I_{\rho} C \sum_i \bm{S}^d_{\rho}(\bm{r}_i) +   ^{17}I_{\rho} E_{\rho} \bm{S}^p_{\rho} 
\right\}
\label{17hyp}
\end{equation}
Here $C$ is the transferred hyperfine coupling constant for localized Cu spins and  
$E_{\rho}$ is the contact hyperfine interaction for the oxygen nucleus; 
the sum index $i$ goes over the NN copper sites. Making use of
\begin{eqnarray}
\mu_B \langle S^p(\bm{r}) \rangle &=& (\chi_{pd} + \chi_{pp}) \bm{B}_{ext} \\
\mu_B \langle S^d(\bm{r}) \rangle &=& (\chi_{pd} + \chi_{dd}) \bm{B}_{ext},
\end{eqnarray} 
where $\bm{B}_{ext}$ is the strength of the external magnetic field and $\mu_B$ is the electron 
magnetic moment, we get, from the above Hamiltonians:
\begin{eqnarray}
\label{K63eq}
^{63}K_{\rho} &=& ^{63}K_{0, \rho} + \frac{A_{\rho} + 4B}{^{63}\gamma_n \mu_B^2 \hbar^2}\, (\chi_{dd} + \chi_{pd}) 
 + \frac{4D}{^{63}\gamma_n \mu_B^2 \hbar^2}\, (\chi_{pd} + \chi_{pp}) \\
^{17}K_{\rho} &=& ^{17}K_{0, \rho} + \frac{2 C}{^{17}\gamma_n \mu_B^2 \hbar^2}\, (\chi_{dd} + \chi_{pd}) + 
\frac{E_{\rho}}{^{17}\gamma_n \mu_B^2 \hbar^2}\, (\chi_{pd} + \chi_{pp}). 
\label{K17eq}
\end{eqnarray} 
Here $\gamma_n$ are the corresponding nuclear gyromagnetic ratios.
On introducing the dynamic spin susceptibilities,
\begin{equation}
\chi_{\mu \nu} (\bm{q}, \omega) = \langle \langle S^{\rho}_{\mu} (\bm{q}) S^{\rho}_{\nu} (- \bm{q}) \rangle \rangle_{\omega},
\end{equation}
where $\mu, \nu = p, d$, we obtain the Walstedt \textit{et al.} expression for relaxation rates:
\begin{equation}
\frac{1}{^{\alpha}T_{1 \gamma} T}\, = \frac{k_B}{2 \mu_B^2 \hbar^2}\,  \lim_{\omega \rightarrow 0} 
\sum_{\bm{q},\mu, \nu = p,d} \, ^{\alpha}F_{\gamma, \mu \nu} (\bm{q}) \frac{Im \chi_{\mu \nu} (\bm{q}, \omega)}{\omega}\,,
\end{equation}
where $\gamma$ is the direction of magnetic field, $\alpha$ is the nuclear site. For $\chi_{dd}$ one gets the
usual MMP form factors:
\begin{equation}
^{63}F_{\parallel, dd} (\bm{q}) = (A_{\perp} + 2 B (\cos(q_x a) + \cos(q_y a))^2,
\end{equation}
\begin{equation}
^{63}F_{\perp, dd} (\bm{q}) = \frac{1}{2}\, ^{63}F_{\parallel, dd} (\bm{q}) + \frac{1}{2}\, (A_{\parallel} + 2B (\cos{q_x a} + \cos{q_y a}))^2,
\end{equation}
\begin{equation}
^{17}F_{\gamma, dd} (\bm{q}) = 2 C^2 (1 + \cos(q_x a)).
\end{equation}
The form factors for the mixed p-d contribution are:
\begin{eqnarray}
& ^{63}F_{\parallel, pd}(\bm{q}) = ^{63}F_{\parallel, dp}(\bm{q}) & \nonumber \\
& = 2 D (\cos{(q_x a/2)} + \cos{(q_y a/2)}) (A_{\perp} + 2B (\cos{q_x a} + \cos{q_y a})), &
\end{eqnarray}
\begin{eqnarray}
& ^{63}F_{\perp, pd}(\bm{q}) = ^{63}F_{\perp, dp}(\bm{q}) & \nonumber \\
& =
D (\cos{(q_x a/2)} + \cos{(q_y a/2)}) (A_{\perp} + A_{\parallel} + 4B (\cos{q_x a} + \cos{q_y a})),&
\end{eqnarray}
\begin{equation}
^{17}F_{\parallel, pd} (\bm{q}) = ^{17}F_{\parallel, dp} (\bm{q}) = 2 C E_{\perp} \cos{(q_x a/2)},
\end{equation}
\begin{equation}
^{17}F_{\perp, pd} (\bm{q}) = ^{17}F_{\perp, dp} (\bm{q}) =  C (E_{\perp} + E_{\parallel}) \cos{(q_x a/2)}.
\end{equation}
Finally, for the p-p part of spin susceptibility:
\begin{equation}
^{63}F_{\gamma, pp} (\bm{q}) = 4 D^2 (\cos{(q_x a/2)} + \cos{(q_y a/2)})^2 
\end{equation}
\begin{equation}
^{17}F_{\parallel, pp} (\bm{q}) = E_{\perp}^2
\end{equation}
\begin{equation}
^{17}F_{\perp, pp} (\bm{q}) = \frac{1}{2}\, E_{\perp}^2 +  \frac{1}{2}\, E_{\parallel}^2 
\end{equation}

Apart from $^{17}F_{pp} (\bm{q})$, the form factors for the p-d and p-p contributions   
vanish at $\bm{Q} = (\pi/a, \pi/a)$:
\begin{equation}
^{17}F_{pd}(\bm{Q}) = ^{63}F_{pd}(\bm{Q}) = ^{63}F_{pp} (\bm{Q}) = 0.
\end{equation}
In what follows we focus on the behavior of planar dynamic and static magnetic response and the thermodynamic behavior, 
as measured by the inelastic neutron scattering, NMR, heat capacity and bulk spin susceptibility.

\subsection{Scaling for bulk spin susceptibility and Knight shift}

We now apply our two-fluid description to an analysis of the bulk spin 
susceptibility and Knight shift measurements in La$_{2-x}$Sr$_x$CuO$_4$
and YBa$_2$Cu$_3$O$_{6+x}$ families of cuprate superconductors. 
For the static susceptibility, the two components in
Eq.(\ref{dyn}) can be written in the following form:
\begin{equation}
\chi(T)  = f(x) \chi_{SL}(T/T^m(x)) + (1 - f(x)) \chi_{FL} + \chi_0,
\label{static}
\end{equation}
where
\begin{equation}
\chi_0 = \chi_{VV} + \chi_{core}.
\end{equation}
Here $\chi_{VV}$ and $\chi_{core}$ are temperature- and doping-independent Van Vleck and core contributions. 
$\chi_0$ is usually taken to be diamagnetic, due to a large core contribution\cite{johnston}.
As noted earlier,
$\chi_{SL}$ follows very well the calculated\cite{MD} bulk 
spin susceptibility for the Heisenberg model with a doping-dependent exchange constant 
$J_{eff}(x) \sim T^m(x)$. Thus, the spin liquid contribution to static magnetic response can be written as
\begin{equation}
f(x) \chi_{SL}(T/T^m(x)) = \chi^m \tilde{\chi}(T/T^m(x)),
\label{fmax}
\end{equation}
where $\chi^m$ is the maximum value of the spin liquid susceptibility and  $\tilde{\chi}(T/T^m(x))$ 
is a universal function of $T/T^m$.  
\begin{figure}
\begin{center}
\includegraphics[width=0.75\textwidth]{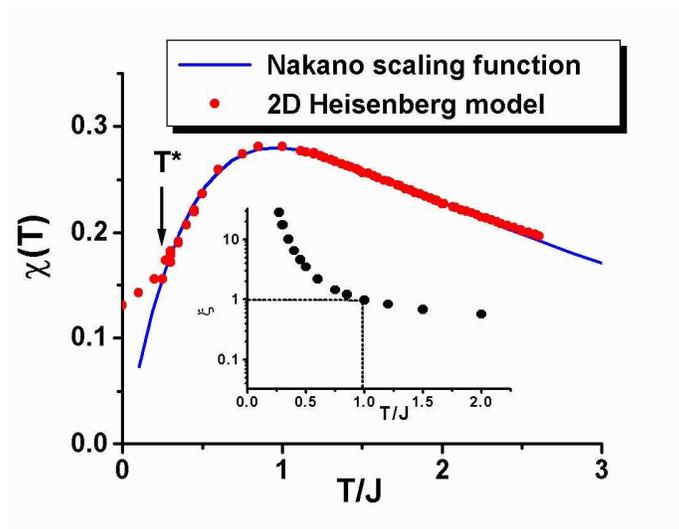}
\caption{Comparison of the scaling curve for bulk spin susceptibility data in underdoped metallic La$_{2-x}$Sr$_x$CuO$_4$ 
from Nakano \textit{et al.}\cite{nakano} to the Heisenberg model numerical calculations of Makivi\'c and Ding\cite{MD}. 
The maximum for spin susceptibility is reached at $T^m \simeq 0.93 J$. Deviations from the Heisenberg model 
results are observed for $T < T^* \simeq T^m/3$.
The inset shows the numerical results for the  correlation  length\cite{MD}, that demonstrate that $\xi \simeq 1$ at 
temperature $T \simeq T^m$.}
\label{hna}
\end{center}
\end{figure}
A comparison of the universal data collapse curve for the bulk spin susceptibility in La$_{2-x}$Sr$_x$CuO$_4$ family found by 
Nakano \textit{et al.}\cite{nakano} to the Heisenberg model calculations\cite{MD} is shown in Fig.\ref{hna}. 
We see that the universal function of Nakano\textit{et al.} deviates significantly from 
the 2D Heisenberg model results at low temperatures, $T < T^* \simeq T^m/3$. 

As noted earlier, the scaling analysis of Nakano \textit{et al.}\cite{nakano}  
differs from that of Johnston\cite{johnston}, because it includes at lower doping levels
a Curie term $C/T$, and a linear term $B(T-T_a)$ in the static spin response. 
While the inclusion of these terms may be justified, it introduces a number of new parameters
that are, in a sense, unnecessary since 
the universal curve in the insulating compound must agree with the 2D Heisenberg model results. 
We therefore omit these terms in our analysis, and we assume, in agreement with recent 
photoemission studies\cite{campuz}, that the Fermi liquid contribution, $\chi_{FL}(T,x)$, is temperature-independent
above $T^*$, but could, for the 1-2-3 materials become temperature-dependent below the temperature, $T^*$, at which a gap starts 
to form in the quasiparticle energy spectrum, leading to the formation of the Fermi arcs.

\begin{figure}
\begin{center}
\includegraphics[width=0.75\textwidth]{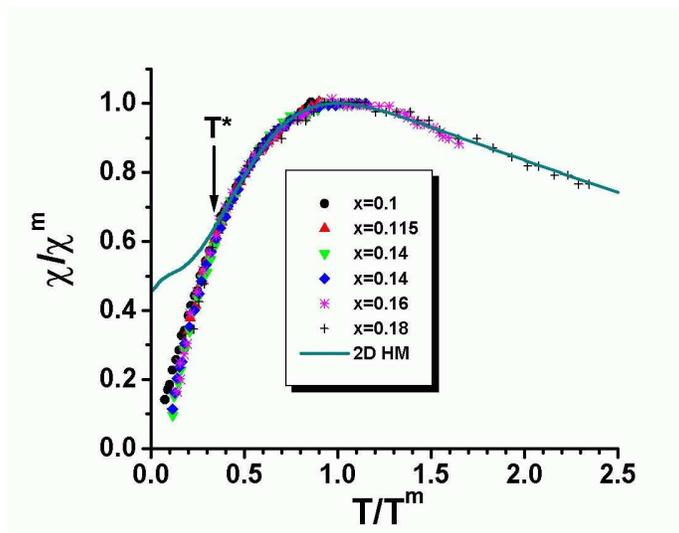}
\caption{Comparison of the bulk spin susceptibility data in underdoped metallic La$_{2-x}$Sr$_x$CuO$_4$ 
from Nakano \textit{et al.}\cite{nakano} to the Heisenberg model numerical calculations of Makivi\'c and Ding\cite{MD}. 
The maximum for spin susceptibility is reached at $T^m \simeq 0.93 J$.
While the data collapse remains reasonably good even at low temperatures, large deviations from 2D Heisenberg model 
prediction are observed for $T < T^m/3$.}
\label{knan1}
\end{center}
\end{figure}

To determine the doping dependence of the parameters in Eq.(\ref{static}),
we analyzed the bulk spin susceptibility data\cite{nakano} for metallic underdoped La$_{2-x}$Sr$_x$CuO$_4$.
With a constant contribution $\chi_0 + \chi_{FL}(x)$ subtracted,
we find, in agreement with the earlier results\cite{johnston,oda,nakano},
the data collapse to the 2D Heisenberg model curve, shown in Fig. \ref{knan1}, 
where we see that the data collapse continues below the temperature  $\sim T^m/3$  
at which one no longer finds agreement with the Heisenberg model.

\begin{figure}
\begin{center}
\includegraphics[width=0.75\textwidth]{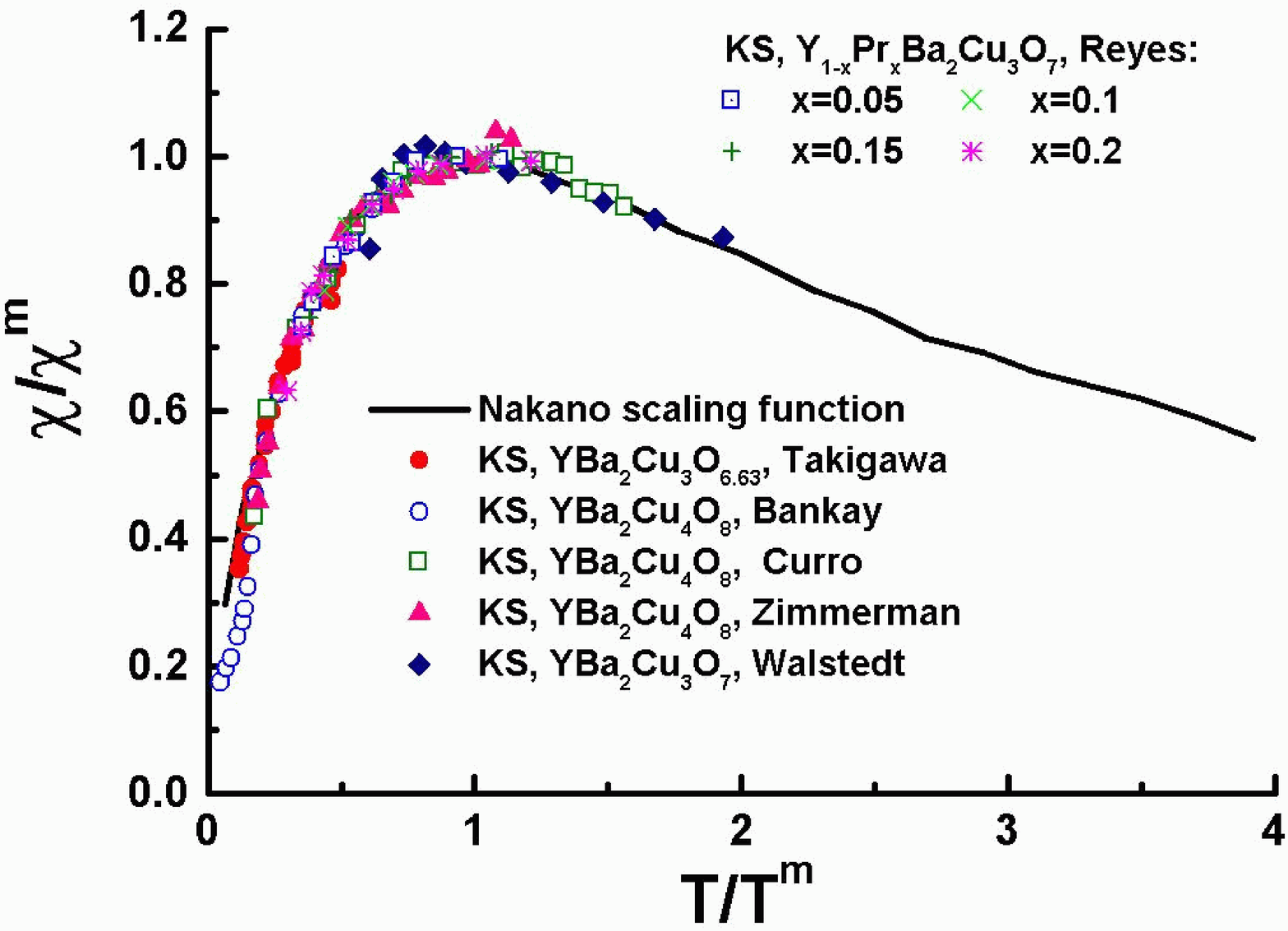}
\caption{Scaling for the $^{63}Cu$ Knight shift\cite{curro97,takigawa,reyes,walstedt,zimmerman,bankay}
 in YBa$_2$Cu$_3$O$_{6+x}$, compared to the scaling function obtained
by Nakano \textit{et al.}\cite{nakano} for the bulk spin susceptibility 
in La$_{2-x}$Sr$_x$CuO$_4$. $T^m$ is the temperature at which the Knight shift has a maximum. }
\label{knan}
\end{center}
\end{figure}

\begin{figure}
\begin{center}
\includegraphics[width=0.75\textwidth]{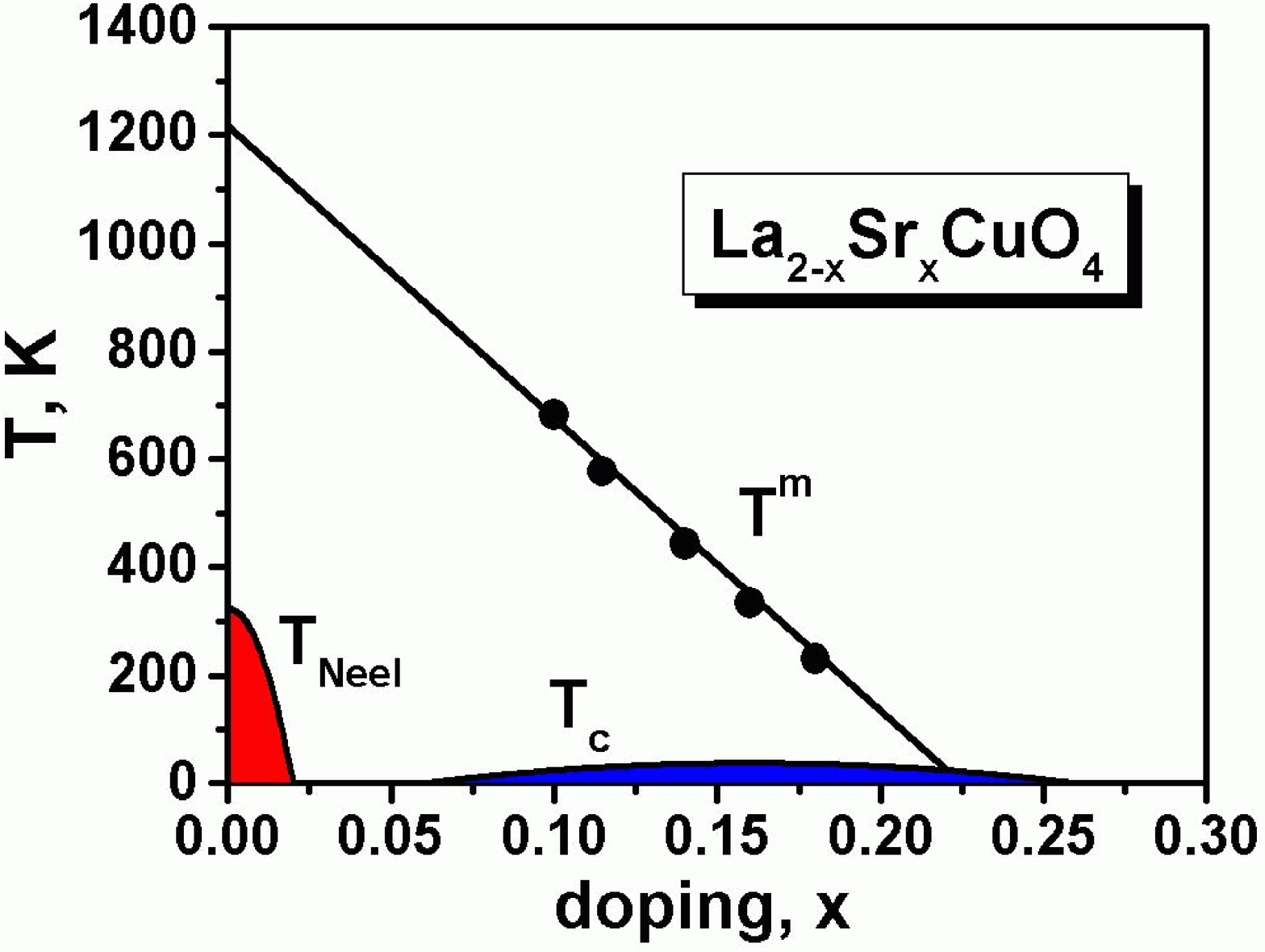}
\includegraphics[width=0.75\textwidth]{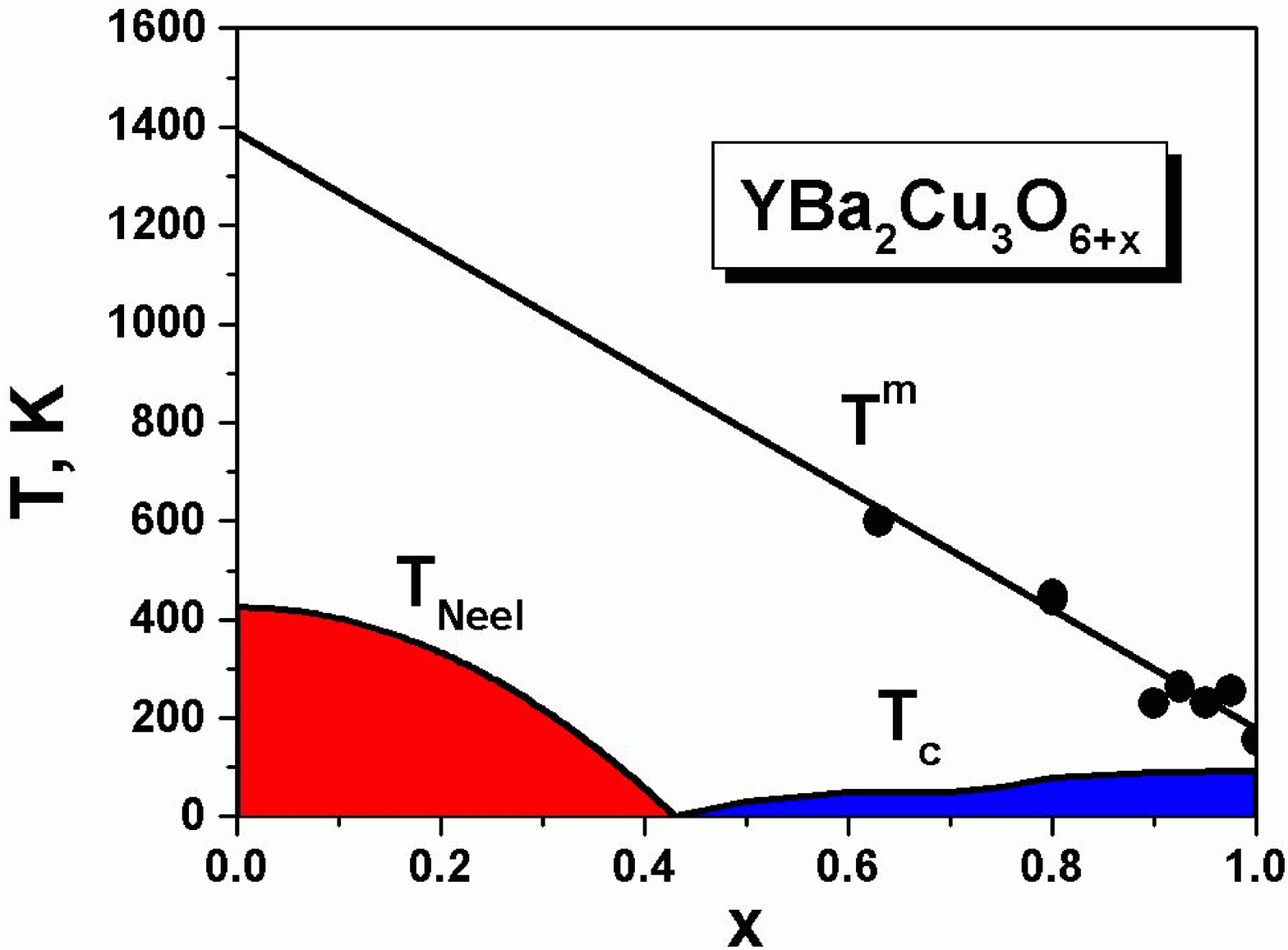}
\caption{(a) The temperature $T^m(x)$ that follows from our analysis of the Nakano \textit{et al.}\cite{nakano} 
bulk spin susceptibility data for La$_{2-x}$Sr$_x$CuO$_4$, and (b) 
the Knight shift data for YBa$_2$Cu$_3$O$_{6+x}$}
\label{nak}
\end{center}
\end{figure}

\begin{figure}
\begin{center}
\includegraphics[width=0.75\textwidth]{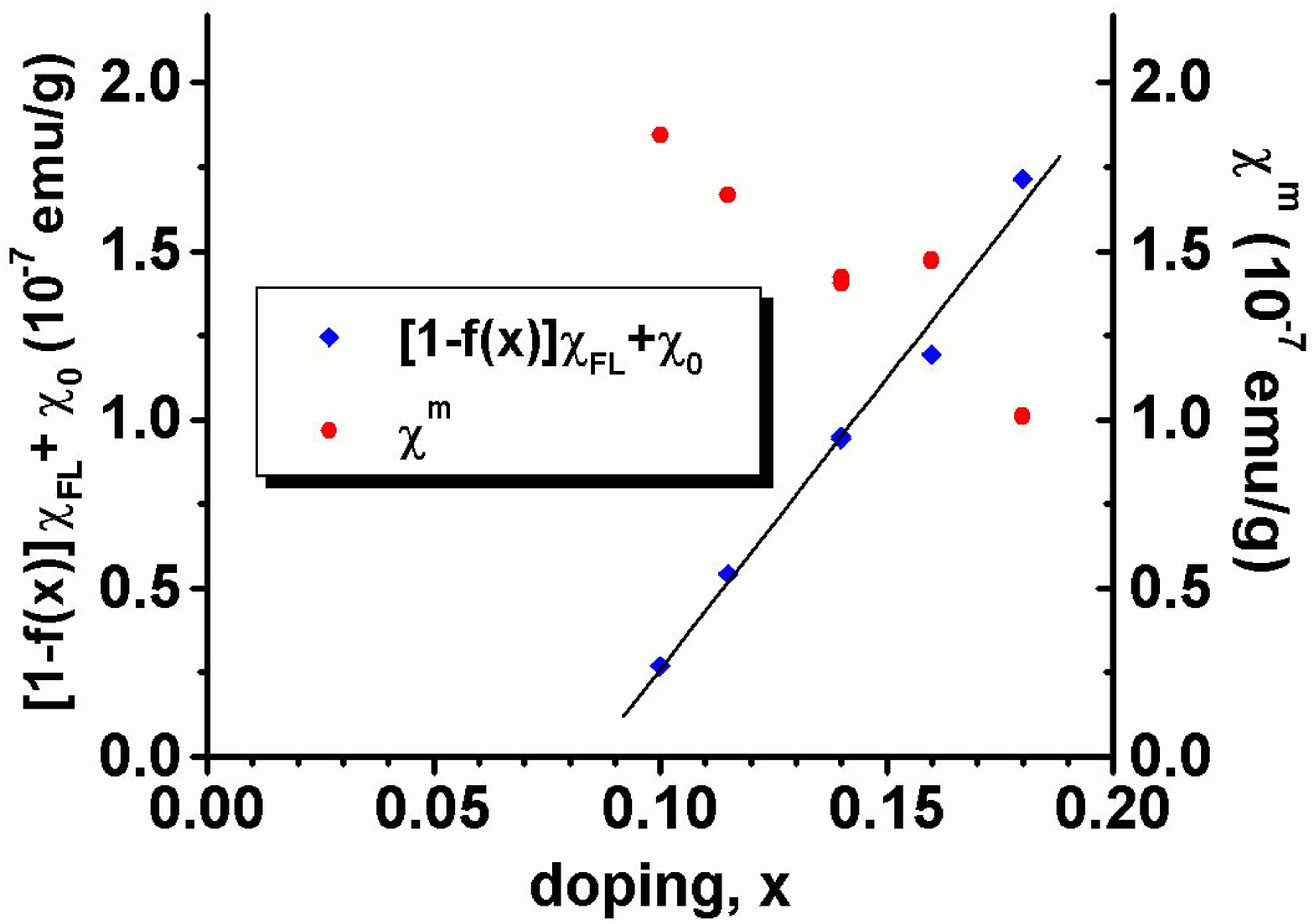}
\caption{$\chi^m(x)$ and $\chi_{FL}(x)$, from our analysis of the bulk susceptibility measurements
in La$_{2-x}$Sr$_x$CuO$_4$. Here $\chi_{FL}(x)$ here also includes a diamagnetic constant, $\chi_0$, hence the
non-zero offset.}
\label{FLcomp}
\end{center}
\end{figure}

A similar scaling analysis for the $^{63}Cu$ Knight shift data 
in 1-2-3 materials is shown in Fig.\ref{knan}. 
The doping dependence of  $T^m(x) \simeq 0.93 J_{eff}(x)$ that follows from the analysis of 
the bulk susceptibility and Knight shift data, is shown in Fig. \ref{nak}
for both La$_{2-x}$Sr$_x$CuO$_4$ and YBa$_2$Cu$_3$O$_{6+x}$ families.  $T^m(x)$ 
falls linearly with increased doping $x$ for both materials. For La$_{2-x}$Sr$_x$CuO$_4$,
and doping levels less than $0.18$,
\begin{equation}
T^m(x) = 1218K (1 - 4.45 x).
\label{Tmresu}
\end{equation}
Our analysis of the bulk susceptibility data in La$_{2-x}$Sr$_x$CuO$_4$ also enables us
to extract the doping dependence for the Fermi liquid and spin liquid components.
The results for $\chi^m(x)$ and $\chi_{FL}(x)$ given by Eqs.(\ref{fmax}),(\ref{static}) 
are shown in Fig.\ref{FLcomp}.

\begin{figure}
\begin{center}
\includegraphics[width=0.75\textwidth]{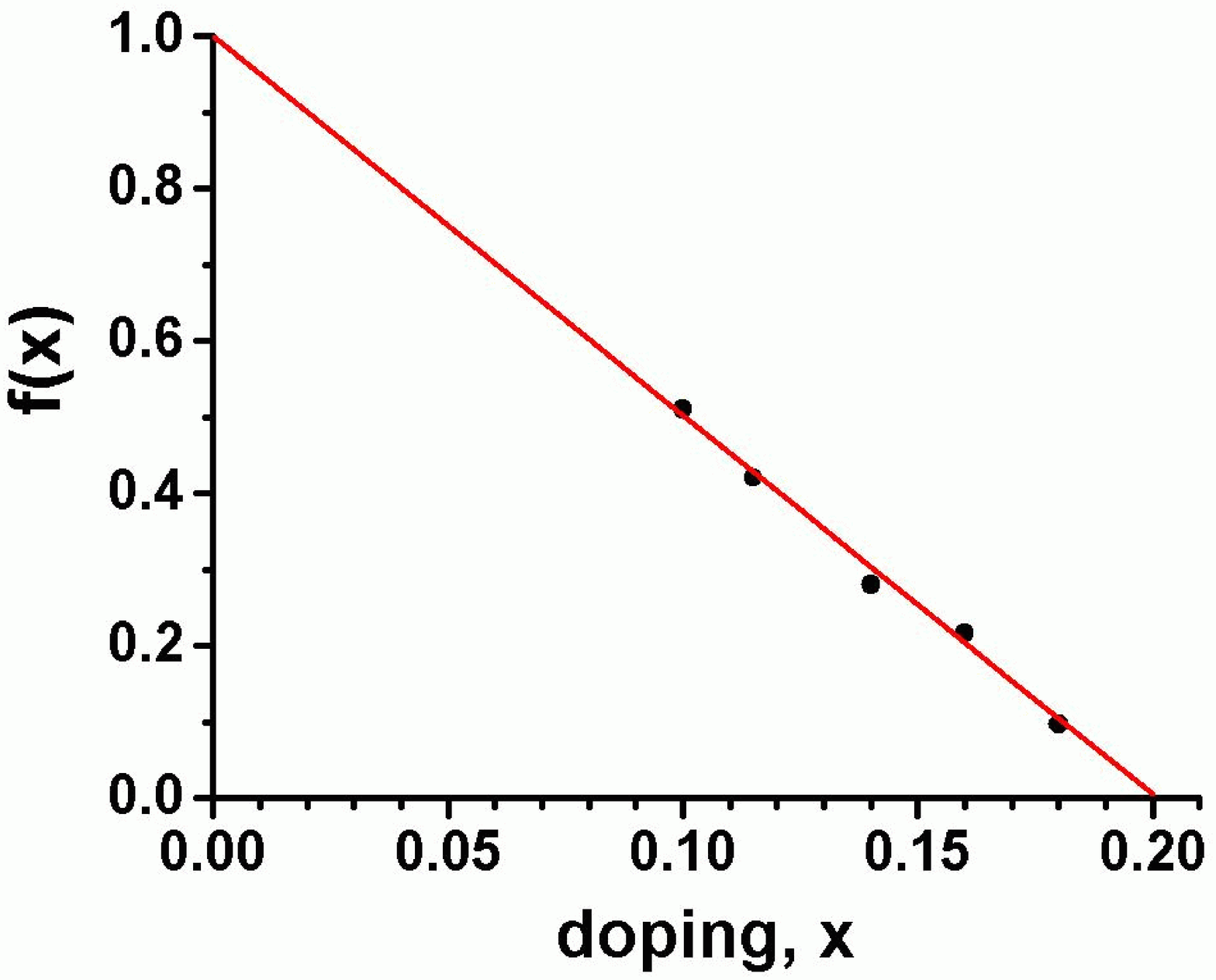}
\caption{The spin liquid fraction $f(x)$, from our analysis of the bulk susceptibility measurements
in La$_{2-x}$Sr$_x$CuO$_4$\cite{nakano}.}
\label{fofx}
\end{center}
\end{figure}

The doping dependence for the fractional occupation, $f(x)$, of the spin liquid state can be determined from 
$T^m(x)$ and $\chi^m(x)$.  The dimension of $\chi$ is $1/energy$. Thus, since
the universal curve for $T>T^*$ coincides with the numerical calculation for the 2D Heisenberg antiferromagnet 
with $T^m \simeq 0.93 J_{eff}$, we can write:
\begin{equation}
\chi_{SL}(T/T^m(x)) \propto \frac{1}{T^m(x)}\, \tilde{\chi}(T/T^m(x)),
\label{scbulk}
\end{equation}
where $\tilde{\chi}(T/T^m(x))$ is a universal function for the Heisenberg antiferromagnet.
Using Eqs(\ref{scbulk}) and (\ref{fmax}), the doping dependence of $f(x)$  is then determined by
the product $\chi^m(x) T^m(x)$:
\begin{equation}
f(x) = \frac{\chi^m(x) T^m(x)}{\chi^m(x=0) T^m(x=0)}\,
\end{equation}
From our data analysis, we find that for the doping levels available for our analysis, $f(x)$ also decreases linearly with $x$, 
\begin{equation}
f(x) = 1 - 4.977 x,
\end{equation}
as shown in Fig. \ref{fofx}, so that 
$T^m(x)$ and $f(x)$ roughly track each other.
Fig. \ref{FLcomp} also shows the doping dependence for the Fermi liquid component, $\chi_{FL}(x) = (1 - f(x)) \chi_{FL} + \chi_0$.
We find that the Fermi liquid component increases linearly with doping in the metallic regime, in agreement with Eq.(\ref{static}).
The offset for this linear dependence is due to the presence of $\chi_0$, which is diamagnetic \cite{johnston}.
Since $f(x)$ also decreases linearly with $x$, we conclude that $\chi_{FL}$ in Eq.(\ref{static}) has only
mild, if any, doping-dependence.

\subsection{Knight shift data}

We now examine the Knight shift data in more detail. As noted in Section 2,
most early Knight shift experiments on different nuclei found the same temperature-dependent contribution
to spin susceptibility\cite{takigawa,MPT,alloulY}, although 
some deviations from this universality (within error bars) can be seen on the universal plot below the superconducting temperature 
$T_c$\cite{takigawa,MPT}. The linear $K-\chi$ and $K-K$ plots above $T_c$, observed in many early Knight shift experiments, 
then provide a measurement of the hyperfine couplings to single copper spin. These measurements had been regarded as a key proof 
of an effective one-component model. 
However, the Cu Knight shift measurements on the 1-2-4 material by Suter \textit{et al.}\cite{zurich} and the recent apical oxygen 
Knight shift measurements in La$_{1.85}$Sr$_{0.15}$CuO$_4$ by Haase \textit{et al.}\cite{HS}, 
tell a different story, and we consider these now.

\begin{figure}
\begin{center}
\includegraphics[width=0.75\textwidth]{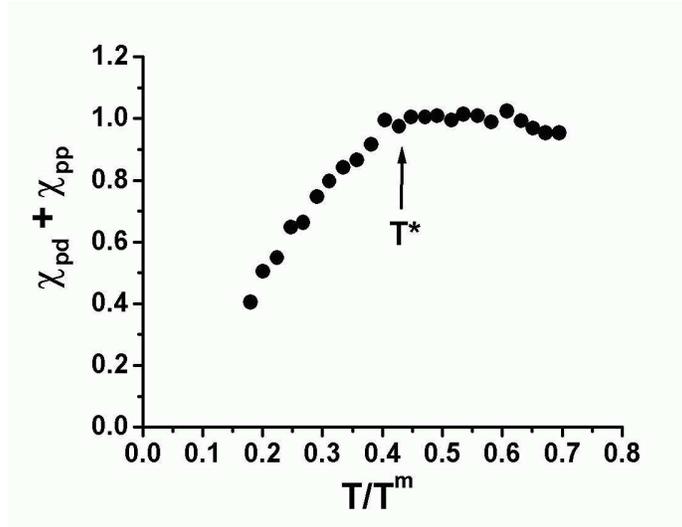}
\caption{Normalized temperature dependence of $\chi_{pd}$ as a function of $T/T^m$, inferred from $^{63}K_c$ measurements\cite{zurich}
in YBa$_2$Cu$_3$O$_8$.}
\label{chippd}
\end{center}
\end{figure}

	The Knight shift measurements of Suter \textit{et al.} were done for an external magnetic field in the c direction, 
where, because of the well-known accidental cancellation\cite{MMP} of the $\chi_{dd}$ form factor, $A_{\parallel} + 4B$, 
in Eq.(\ref{K63eq}), within the single-component approximation, the Knight shift for fields along the c-axis would be of 
purely orbital origin. Within the two-fluid description, we see from Eq.(\ref{K63eq}) that one can, in addition, 
probe Fermi liquid behavior through an isotropic transferred hyperfine interaction $D$ that arises from the  
hybridization of planar oxygen p-orbitals and copper s-orbitals. So to the extent that the Fermi liquid 
susceptibility becomes temperature dependent as a result of a quasiparticle gap opening up at $T^*$,one would expect to see this 
in $^{63}K_{\parallel}$. As may be seen in Fig. \ref{chippd}, where we have extracted the behavior of $\chi_{pp} + \chi_{pd}$ 
as a function of T from their Knight shift measurements, this is just what was found by Suter \textit{et al.}\cite{zurich}. 
The anomalous Fermi liquid behavior begins at a temperature, $T^* \sim 180 K \sim 0.4 T^m$ for YBa$_2$Cu$_3$O$_8$, an onset 
temperature that is consistent with the onset of Fermi arc behavior at a comparable value of $T^*$ seen in the ARPES measurements 
on underdoped 2212 materials.

In their recent re-analysis of the planar $^{63}$Cu and $^{17}$O  Knight shift experiments on La$_{1.85}$Sr$_{0.15}$CuO$_4$ 
that was accompanied by new apical 
oxygen Knight shift experimental results, Haase \textit{et al.}\cite{HS} have found convincing experimental evidence for the existence 
of both a Fermi liquid and a spin liquid component in this material. They show that while all nuclei see the same temperature-dependent 
spin liquid component, the different nuclei see as well a component that is independent of temperature above the superconducting 
transition temperature, $T_c$,  but displays the expected d-wave signature below $T_c$, and hence must be associated with a Fermi 
liquid component, rather than representing an orbital or Van Vleck core term.

	Quite importantly, they combine their results in the superconducting state  (that include correcting for any Meissner 
shielding effects) and the normal state for magnetic fields in and perpendicular to the CuO planes 
with Equations (\ref{K63eq}) and (\ref{K17eq}), supplemented by an equivalent expression for the apical oxygen site, to determine all 
the relevant hyperfine couplings. They show  these can be used to determine the strength and temperature dependence in the 
normal and superconducting states of all three components, $\chi_{dd}$, $\chi_{dp}$, and $\chi_{pp}$. We refer the interested reader to 
their paper for the details of their findings, which place important constraints on any description of the 
hybridization of the Cu $d$ electrons with the oxygen $p$ band. 

While further Knight shift measurements under conditions favorable to observing the Fermi liquid component 
are highly desirable, we believe it is reasonable to conclude that the two-fluid analysis of Knight shift 
in heavy electron materials\cite{NPFcurro} extends naturally to high-$T_c$ superconductors, with 
\begin{equation}
\chi_{dd} = f(x) \chi_{SL},
\end{equation}
and 
\begin{equation}
\chi_{pp}+2 \chi_{pd} = (1 - f(x)) \chi_{FL}.
\label{HSopar}
\end{equation}
Eq. (\ref{HSopar}) can then be combined with the results of Haase \textit{et al.}, and our previously determined value of $f(x)$,   
to determine the doping-independent quantity  $\chi_{FL}$ for the 2-1-4 materials. The results of this analysis can be 
checked by extending the Haase \textit{et al.} analysis to other doping levels.

\subsection{$^{63}Cu$ NMR Relaxation Rates}

In analyzing the NMR relaxation rates, we shall focus on the 
$^{63}Cu$ spin-lattice relaxation rate, $^{63}T_1$, and the spin-echo-decay time, $^{63}T_{2G}$.
As discussed in the beginning of the section, there are different components of the spin
response. However, due to the large peak of the spin-liquid part at $\bm{Q}=(\pi,\pi)$,
and the corresponding non-vanishing form factors for $^{63}Cu$ nuclei, 
the copper relaxation rates primarily probe the low-frequency properties 
of the spin-liquid contribution that arises from $\chi_{dd} = f(x) \chi_{SL}$ part of the spin susceptibility. 
In our modification of MMP\cite{MMP}  
both $\chi_{pd}$ and $\chi_{pp}$ correspond to the contribution of fermions with a large hybridized Fermi surface, or pieces
of such a Fermi surface, in agreement
with the Knight shift measurements of Haase \textit{et al.}\cite{HS}, and therefore have a 
negligible effect on copper relaxation rates.
An alternative possibility, that we mention only briefly, is that the two components 
seen in the bulk measurements are a result of microscopic phase separation and formation of
dynamic stripes. Surprisingly, while the transition points that we obtain from the two-component analysis of
bulk measurements coincide with the well-known region of phase separation in La$_2$CuO$_{4+\delta}$\cite{hammel},
the two-component nature is evident in bulk measurements at temperatures well above the phase separation temperature region.

Where these can both be carried out,  measurements of $^{63}Cu$ spin-lattice relaxation and spin-echo 
decay rates indicate the presence of two different dynamic scaling
regimes that lie above and below $T^m$\cite{BP}. 
Above $T^m$ one is in a mean-field, $z=2$ scaling regime, in which the relaxational frequency, 
$\omega_{SF}$ varies as $1/\xi^2$; below $T^m$ it varies as $1/\xi$ because the spin liquid has entered the $z=1$ dynamic
scaling regime\cite{BP,imai} expected for the quantum critical (QC) regime of a 2D antiferromagnet.
As discussed in Ref.\cite{BP}, one can describe the spin dynamics using the quantum
non-linear sigma model, or spin wave theory\cite{CHN,CSY}. The resulting QC scaling theory\cite{CSY} for the spin
liquid without long-range order gives a linear dependence of the correlation length on temperature,
\begin{equation}
\frac{1}{\xi(T,x)}\, = 1.04 \frac{T}{c}\, + a(x),
\end{equation}
where $c \propto J \sim T^m$ is the spin wave velocity, 
and the offset $a(x) > 0$ goes to zero at $x_c$, a quantum critical point for the spin liquid that marks 
the onset of long-range order.
A similar linear dependence on $T$ can be expected also for
$^{63}T_1T \propto \omega_{SF}$\cite{BP,imai},
\begin{equation}
^{63}T_1T = A(x) + \kappa T^m \frac{T}{T^m}\,,
\label{QCT1T}
\end{equation}
with an offset $A(x)$ that measures the distance from the proposed quantum critical point, while
$\kappa$ is a universal coefficient. 

\begin{figure}
\begin{center}
\includegraphics[width=0.75\textwidth]{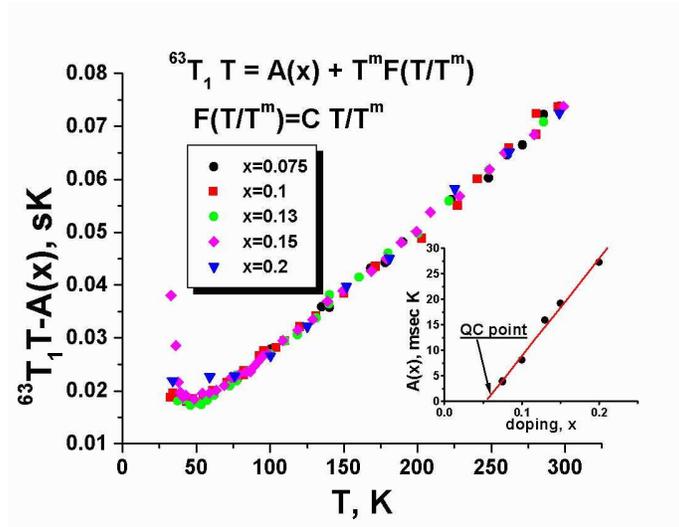}
\caption{Scaling for the NMR relaxation rate $T_1$ in La$_{2-x}$CuO$_4$\cite{ohsugi} 
The offset $A(x)$ shown in the inset depends linearly on $x$, and points to a
QCP in the spin liquid.}
\label{T1TLa}
\end{center}
\end{figure}

\begin{figure}
\begin{center}
\includegraphics[width=0.75\textwidth]{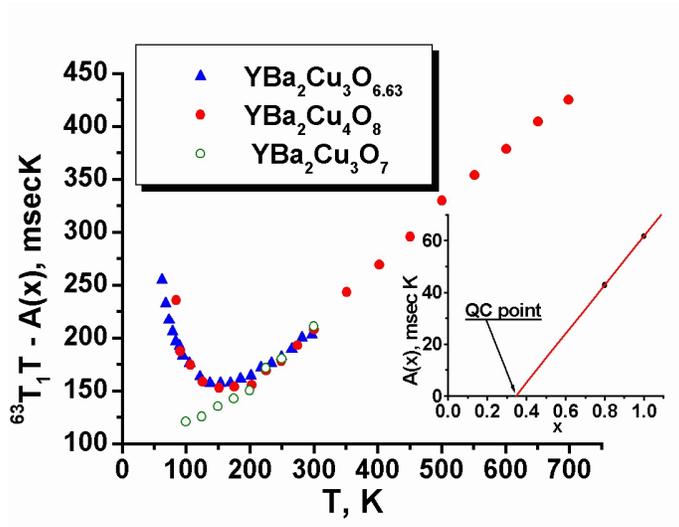}
\caption{Scaling for the NMR relaxation rate $T_1$  in YBa$_2$Cu$_3$O$_{6+x}$.
The offset $A(x)$ shown in the inset depends linearly on $x$, and points to a
QCP in the spin liquid.}
\label{T1TY}
\end{center}
\end{figure}

Quantum critical theory applies in the universal regime, $\xi \gg 1$, thus, well below $T^m$. 
Scaling of the form seen in Eq.(\ref{QCT1T}) is self-evident from the low-temperature $^{63}T_1$ data on 
La$_{2-x}$Sr$_x$CuO$_4$ material of Ohsugi \textit{et al.}\cite{ohsugi}(Fig.\ref{kitaokafig}). When they plot
the product $^{63}T_1 T$ vs temperature in the metallic underdoped regime, they find 
a set of parallel lines for their different doping levels that extend from $T \sim 300K$ down to
a low-temperature upturn in $^{63}T_1T$ that is near $T_c$. We plot in Figs.\ref{T1TLa},\ref{T1TY} the 
scaling behavior for $^{63}T_1$ in the 2-1-4 and the 1-2-3 materials. The $x$ dependence of the off-set, $A(x)$,
of the low-temperature linear behavior in $^{63}T_1T$ points to a
critical point in the spin liquid at $x=0.05$, separating long range order from short range order,
a result that was also suggested earlier in an analysis of the high-temperature $^{63}T_1$ data\cite{imai}. 
The long range magnetic order at $x<0.05$ can be a 2D antiferromagnet or a spin glass\cite{BPprl}.

While the details of the microscopic theory at low doping may vary, the experimental data imply the
presence of a linear spin wave-type excitation spectrum in the spin liquid with a temperature-dependent gap, 
$\Delta_{SL}(x,T)$, that is controlled by its distance from the quantum critical point at $x=0.05$. 
The spin wave excitation spectrum has the following form:
\begin{equation}
\epsilon(\bm{q}) = \sqrt{\Delta_{SL}^2 + c^2 (\bm{q}-\bm{Q})^2}, \ \ \Delta_{SL} = \frac{c}{\xi(T,x)}\,.
\label{spingap}
\end{equation}
Here $c$ is the spin wave velocity. In the $z=1$ QC regime the spin gap $\Delta_{SL}$
tracks $\omega_{SF} = c'/\xi$. For $x > 0.05$ the gap saturates at low temperatures to
$\Delta_0 = c/\xi_0(x)$ at the crossover to the QD gapped spin liquid regime.
We note that the origin of the saturation of the correlation length is
largely irrelevant for data analysis - an energy
gap can appear in the 2D quantum spin liquid\cite{CHN,CSY}, or in 1D stripes of spin liquid\cite{tranquada2} due to
dimensional crossover. In the latter case the correlation length will saturate at the size of the domain or dynamic stripe order, 
$\xi(T=0) \propto L$. Therefore, experiment does not distinguish between these possibilities.

The QC-QD crossover could explain the sharp upturn in $^{63}T_1T$ at low temperatures
observed in NMR experiments on 1-2-3 materials at temperatures well above $T_c$ that is  not found in experiments on 
the 2-1-4 materials.
 Since $c(x) \propto T^m(x)$, and becomes small as one approaches $x=0.22$,
while $1/\xi_0(x)$ increases linearly from $x=0.05$, the zero-temperature spin liquid gap 
$\Delta_0 \equiv \Delta_{SL}(x,0)$ will have a bell-shaped form similar to $T_c(x)$.
The gapping of the spin liquid leads to an exponential decrease of damping with
temperature in the spin liquid.

The detailed properties of the spin liquid can be extracted directly from the NMR $^{63}T_1$ relaxation rates 
measurements using
the ansatz for the correlation length $\xi(T^m) = 1$, or, where these exist, from both $^{63}T_1$ and $^{63}T_{2G}$ 
measurements, as it was shown some time ago by the authors\cite{BP}. In the $z=1$ QC regime,
\begin{eqnarray}
^{63}T_1 T &\propto& \frac{\omega_{SF}}{\alpha(x) f(x)}\,, \\
\frac{1}{^{63}T_{2G}}\, &\propto& \alpha(x) f(x) \xi,
\end{eqnarray}
where all parameters are taken from Eq.(\ref{SLpar}); these equations differ
from those in Ref. \cite{BP} by the presence of an extra factor $f(x)$ in Eq.(\ref{SLpar}).
A combination of these
two measurements\cite{BP} gives a handle on the spin wave velocity $c$:
\begin{equation}
\frac{^{63}T_1 T}{^{63}T_{2G}}\,  \propto \omega_{SF} \xi \equiv c' \simeq 0.55 c 
\end{equation}
The coefficients of proportionality are given by the hyperfine Hamiltonian.
As easily seen from the above equations,
\begin{equation}
^{63}T_1(T^m) T^m \propto \frac{c'}{f(x) \alpha(x)}\,.
\label{tstca}
\end{equation}

The high-temperature scaling for $^{63}T_1$, $^{63}T_1(T) = const$, that was seen by Imai \textit{et al.}\cite{imai} 
in La$_{2-x}$Sr$_x$CuO$_4$, but is not present in the 1-2-3 materials, at first sight seems to imply $x$-independent 
exchange integral, $J$, or $c'$,
which contradicts the bulk susceptibility data for this material. 
However, a closer look shows there is no contradiction.
As we have seen in Section 3.1, $f(x)$ and $T^m(x)$ both depend linearly on
$x$.  Thus, on general grounds, one can write:
\begin{equation}
c'(x) \propto c(x) \propto T^m(x) \propto \alpha^{-1}(x) \propto f(x). 
\label{empir}
\end{equation}
The Imai \textit{et al.}\cite{imai} scaling then follows from Eq.(\ref{tstca}), since one
can approximately write:
\begin{equation}
\alpha(x) f(x) \simeq \alpha(0) \simeq 11.3 states/eV
\label{alpaf}
\end{equation}

\begin{figure}
\begin{center}
\includegraphics[width=0.75\textwidth]{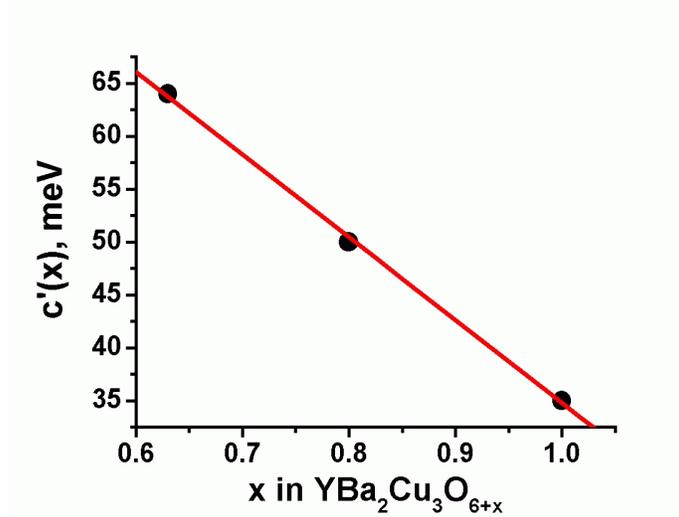}
\caption{$c'(x)$ from NMR experiments in YBa$_2$Cu$_3$O$_{6+x}$}
\label{cx}
\end{center}
\end{figure}

\begin{figure}
\begin{center}
\includegraphics[width=0.75\textwidth]{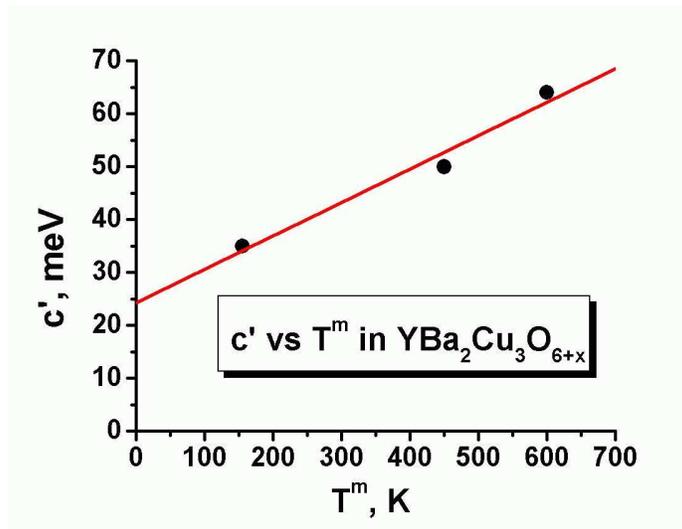}
\caption{$c'(x)$ vs $T^m(x)$ from NMR experiments in YBa$_2$Cu$_3$O$_{6+x}$}
\label{ctst}
\end{center}
\end{figure}

In general, the empirical scaling of Eq.(\ref{empir}) is somewhat accidental, 
since the low-energy properties of the spin liquid can be renormalized
by interaction with fermionic quasiparticle excitations. 
This renormalization appears to be small for the 2-1-4 materials but appears to be important 
for the 1-2-3 materials, where the 
Ohsugi \textit{et al.}\cite{ohsugi} results and the Imai \textit{et al.}\cite{imai}
high-temperature scaling of $^{63}T_1$ are not observed. 
$f(x)$, $T^m(x)$, and $c'(x)$ 
can in principle be independently determined from NMR experiments.
Since $^{63}T_{2G}$ has only been measured for the YBa$_2$Cu$_3$O$_{6+x}$ family,
we plot the doping dependence of $c'$ for these materials in Fig. \ref{cx}, assuming \textit{commensurate}
local spin response.
It is evident that $c'(x)$ decreases approximately linearly with doping.
However, a comparison of $c'(x)$ with $T^m(x)$
shows that the two quantities do not track each other, violating the empirical scaling of Eq.(\ref{empir}). 
While the doping dependence for both is linear,
it is evident from Fig.\ref{ctst} that they are not proportional to one another. 
The violation of the Imai \textit{et al.}\cite{imai} scaling relation by this
family of materials is therefore to be expected.

\begin{figure}
\begin{center}
\includegraphics[width=0.75\textwidth]{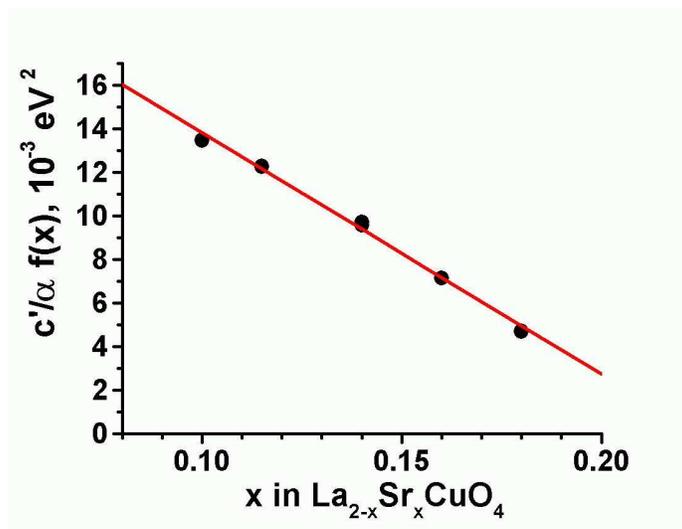}
\caption{$c'(x)/f(x) \alpha(x)$ from NMR experiments in La$_{2-x}$Sr$_x$CuO$_4$}
\label{calpha}
\end{center}
\end{figure}

\begin{figure}
\begin{center}
\includegraphics[width=0.75\textwidth]{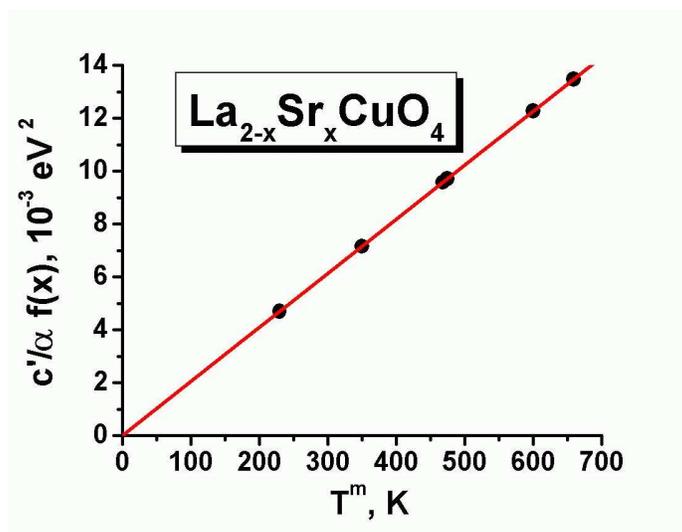}
\caption{$c'(x)/f(x) \alpha(x)$ vs $T^m(x)$ from NMR experiments in La$_{2-x}$Sr$_x$CuO$_4$}
\label{caltst}
\end{center}
\end{figure}

The doping dependence of the parameter $c'(x)/f(x) \alpha(x)$, that follows from NMR $^{63}T_1$ measurements only, 
is shown in Figs \ref{calpha},\ref{caltst}. 
We see that the empirical scaling relation Eq.(\ref{alpaf}) holds reasonably well for
the La$_{2-x}$Sr$_x$CuO$_4$ family, so that the doping dependence of this parameter also yields the doping dependence
of $c(x)$.

\subsection{Thermodynamics}

We now turn to an analysis of thermodynamic measurements of the electronic entropy\cite{loram}. 
These are quite difficult to carry out, 
since the phonon contribution accounts for 98-99 \% of the measured value of the entropy,
and must be subtracted to a high degree of accuracy to get the electronic contribution.
The subtraction procedure is a complicated process, involving subtracting data for a reference material, such
as YBa$_2$Cu$_3$O$_6$\cite{loram}, and making additional corrections for
the modification of the phonon DOS with doping. The reference material is usually
an insulator, either the parent compound, or a compound in which the Fermi
liquid is suppressed with impurities such as Zn. The difficulty with such a subtraction procedure
for analyzing the data is that the spin liquid contribution exists even when the
Fermi liquid is suppressed. Thus, a consistent data analysis requires that the reference material be known.
For this reason we restrict our attention to the data of Ref.\cite{loram}, where the
reference material is  YBa$_2$Cu$_3$O$_6$. The spin liquid contribution of
the parent compound is subtracted from the electronic entropy in this analysis. Nevertheless, at moderate
temperatures $T < 300K$ it remains comparatively small. 

Our analysis of the data of Ref.\cite{loram}
shows that the entropy displays scaling and two component behavior in the
form:
\begin{equation}
S = f(x) S_{SL} (T/T^m) + (1 - f(x)) \gamma T,
\label{entrfact}
\end{equation}
with a scaling function $S_{SL}(T/T^m)$ that corresponds to the theoretically calculated
entropy for the Heisenberg model\cite{MD}.
\begin{figure}
\begin{center}
\includegraphics[width=0.75\textwidth]{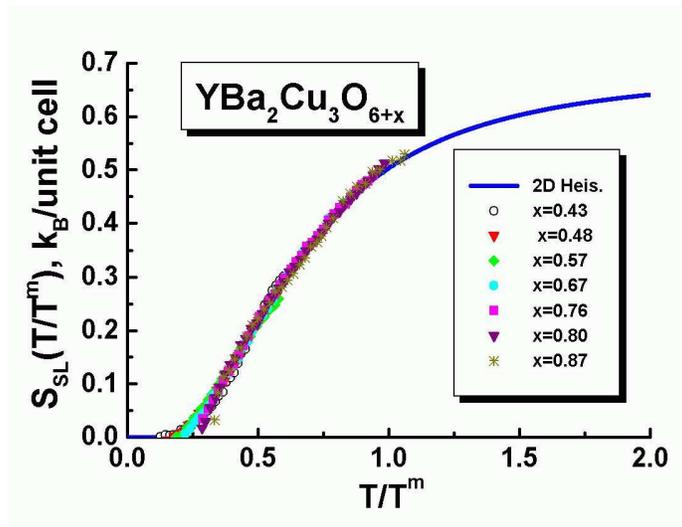}
\caption{Scaling for the spin liquid part of the entropy $S_{SL}(T/T^m)$ in YBa$_2$Cu$_3$O$_{6+x}$. The entropy
data is from Ref.\cite{loram} in the intermediate doping regime, where the spin liquid part of the
entropy is not subtracted. $T^m(x)$ and $f(x)$ obtained from this analysis are consistent with our analysis of
the NMR Knight shift and the bulk spin susceptibility.
The solid line is the integrated numerical result for the heat capacity of the 
2D Heisenberg model taken from Ref. \cite{MD}.}
\label{entr}
\end{center}
\end{figure}
We have integrated the numerical results for
the heat capacity of the Heisenberg model from Ref.\cite{MD} to get the expected entropy. 
We show the results of this analysis and the data 
collapse on the Heisenberg scaling curve in Fig.\ref{entr}.
We find that at small doping  scaling becomes progressively worse, as might be expected because of
the subtraction procedure followed.

To understand our result, we note that on general grounds, one can write the entropy in the form:
\begin{equation}
S = f(x) S_{SL} + (1-f(x))S_{FL} + S_{config},
\end{equation}
where $S_{config}$ is the configurational entropy of the fluctuating stripe or domain order discussed
in the next subsection. The configurational part, however, is not seen in the data factorization Eq.(\ref{entrfact})
We believe, the reason for this is that the configurational contribution of the fluctuating stripes could
also be linear in temperature\cite{pokr}. Indeed, the domain walls
can be treated as solitons, with the Hamiltonian
\begin{equation}
H_s = \frac{1}{2}\, \epsilon_0 \int dy \left(\frac{du}{dy}\, \right)^2,
\end{equation}
where $y$ is the direction along the stripe, while $u$ is the displacement of the wall perpendicular
to its equilibrium direction; $\epsilon_0$ is the linear tension of the domain wall (soliton).
This model maps on a 1D fermion model\cite{pokr,emkivf}, and the resulting configurational entropy has
a linear form,
\begin{equation}
 S_{config} = \frac{\pi^2}{3}\, \frac{T}{\epsilon_0 l^2}\,,
\end{equation}
where $l$ is the mean distance between solitons. Such a linear contribution to the entropy is indistinguishable
from the Fermi liquid contribution.

\subsection{Inelastic neutron scattering}

Our analysis also casts light on the issue of intrinsic spatial inhomogeneity\cite{stripesreview} 
in the pseudogap state\cite{tranquada1,tranquada2}. 
The issue is how to reconcile the commensurate spin susceptibility peaks required by the spin liquid and the fit to the NMR 
data\cite{walst}, with the incommensurate peaks seen in inelastic neutron scattering experiments\cite{aeppli}. 
The oxygen relaxation rate is known to be very small compared to copper, with a very different modified Korringa-type temperature
dependence. An attempt to explain it within the one-component model with incommensurate spin liquid peaks\cite{walst,BPT}  in
La$_{1.85}$Sr$_{0.15}$CuO$_4$ leads to an oxygen relaxation rate that is a factor of three too large compared with actual
experimental data, and has the wrong temperature dependence. 

Similar inconsistencies occur with relaxation rates on other
nuclei, such as $O$ or $Y$ in YBa$_2$Cu$_3$O$_{6+x}$. The contradiction cannot be
resolved if one assumes two different components sitting on oxygen and copper.
Indeed, since the Knight
shift on different nuclei has an identical temperature dependence\cite{MPT,HS}, it 
requires the existence of a transferred hyperfine coupling from the copper\cite{Mila:Rice:Shastry}.
Such a transferred hyperfine interaction assumed by MMP will, however,
induce a "leakage" spin liquid contribution from the spin liquid in the two-component picture
for any incommensurate spin liquid response. 

The resolution was suggested some time ago by Slichter\cite{slichter}, who pointed out that if, 
as a result of Coulomb interaction between the holes, regions that were hole rich formed domain walls between regions
in which one had nearly antiferromagnetic commensurate behavior, the two kinds of experiments were compatible.
This issue has been independently confirmed and clarified by 
Tranquada\textit{et al.}\cite{tranquada1,tranquada2}, who 
found that the spin wave dispersion seen in cuprate superconductors by inelastic neutron scattering 
is consistent with commensurate spin response of a spin ladder, while the incommensurate structure appears
as a result of the fluctuating stripe order.

Such a scenario implies finite-size scaling, and 
saturation of the antiferromagnetic correlation length at low temperatures at a value of the order of the size of the domain. 
Estimating the size of the domain from the discommensuration seen in neutron scattering experiments\cite{aeppli} 
in La$_{1.86}$Sr$_{0.14}$CuO$_4$, $\delta q = \pi \delta$, $\delta = 0.245$, we can write:
\begin{equation}
\delta q L = 2 \pi = \pi \delta L
\end{equation}
Thus,
\begin{equation}
L/a = \frac{2}{\delta}\, = 8.16,
\end{equation}
where $a$ is the lattice spacing. Aeppli \textit{et al.}\cite{aeppli} find the saturated correlation length
in La$_{2-x}$Sr$_x$CuO$_4$ at a doping level of $x=0.14$ is
\begin{equation}
\xi_0 = 29.4 \AA = 7.7 a,
\end{equation} 
a value that is very close to our estimate of $L$. We conclude that both magnetic probes imply the presence of
a fluctuating striped or other domain order.
Apart from the issue of incommensurate peaks seen in INS experiments and its possible resolution in 
terms of a fluctuating domain order, the
spin wave spectrum and the correlation length observed in inelastic neutron scattering experiments is
completely consistent with our proposed two-fluid scenario, as we show below.

We return now to the magnetic scaling seen in the inelastic neutron scattering experiments which, like the
$^{63}Cu$ NMR relaxation rates, probe spin liquid behavior. We begin with the experiments of Sternlieb \textit{et al.}\cite{sternlieb}
that imply the presence of a true $\omega/T$ scaling for the local spin susceptibility in the underdoped 
material YBa$_2$Cu$_3$O$_{6.6}$. The data collapse observed experimentally
can be fit to the following simple scaling form:
\begin{equation}
\chi_L''(\omega, T) = A(x) \frac{\omega}{T}\,,
\end{equation}
with deviations from scaling that occur at progressively lower temperatures, as the frequency $\omega$ increases.

\begin{figure}
\begin{center}
\includegraphics[width=0.75\textwidth]{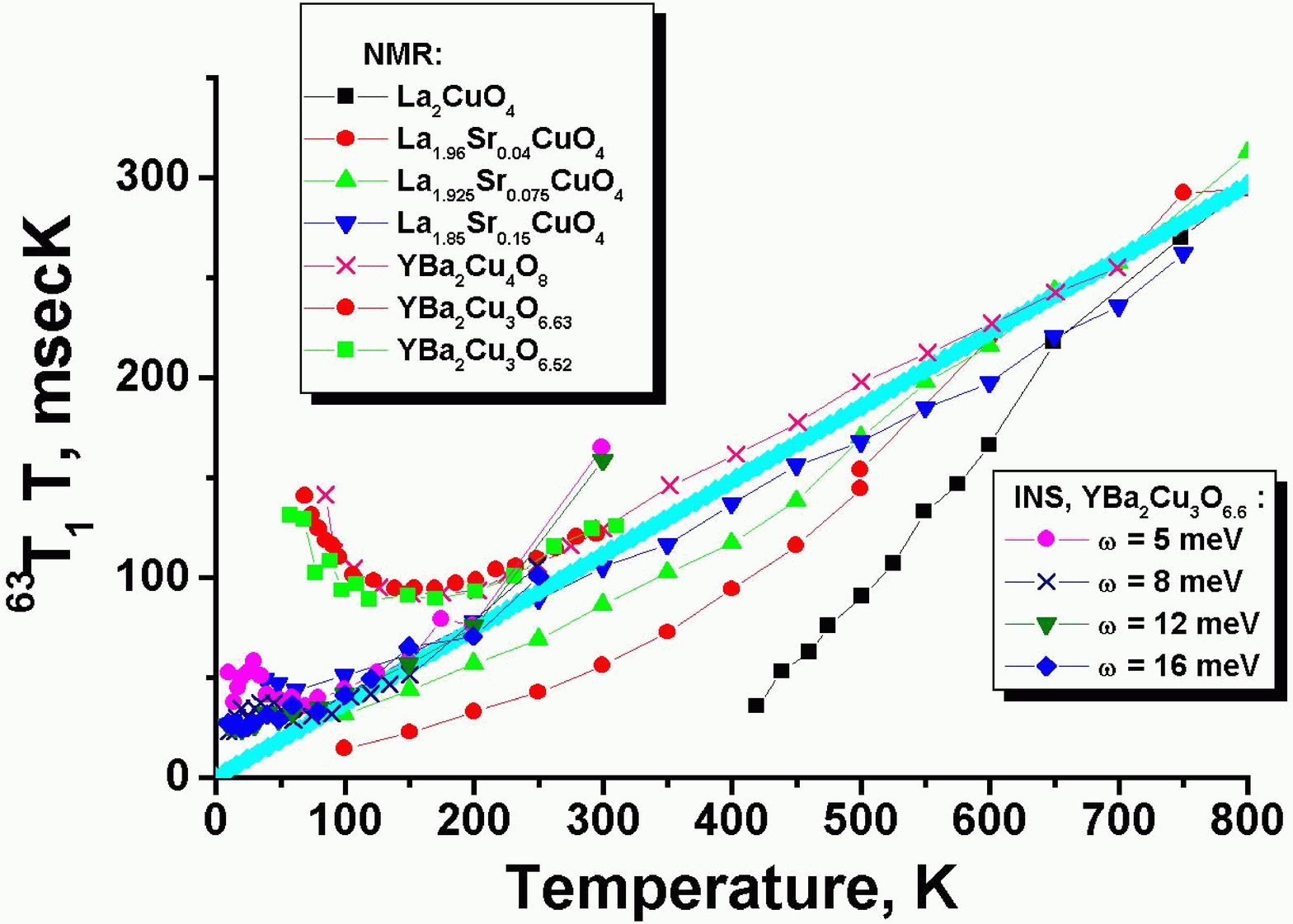}
\caption{Local spin response seen in inelastic neutron scattering in YBa$_2$Cu$_3$O$_{6.6}$, 
$\omega/\chi''_L(\omega, T)$ , scaled to NMR $^{63}T_1T$ in La$_{2-x}$Sr$_x$CuO$_4$ and YBa$_2$Cu$_3$O$_{6+x}$.}
\label{insnmr}
\end{center}
\end{figure}

This scaling confirms the high-temperature scaling seen in NMR\cite{imai}. To clarify this point,
we plot both experimental results in the same units in Fig. \ref{insnmr}. The quantity measured in
neutron scattering experiments analogous to $^{63}T_1T \propto \omega_{SF}$ is a function of the local spin response, 
$\omega /\chi''_L(\omega, T)$, typically given in counts/minute. We therefore fix the absolute value of
that response to be consistent with high-temperature NMR. 
Both experiments reveal linear in temperature $\omega_{SF}$, a signature of magnetic quantum critical behavior.
At finite frequencies this regime extends to lower temperatures, as one may expect on general grounds.

The presence of $z=1$ quantum critical behavior is verified further by the INS measurements of 
the temperature dependence of the correlation length
seen in La$_{1.86}$Sr$_{0.14}$CuO$_4$\cite{aeppli}. Aeppli \textit{et al.}\cite{aeppli}
show that 
\begin{equation}
\frac{1}{\xi^2}\, = \frac{1}{\xi_0^2}\, + \frac{a_0^{-2}}{\Gamma^2}\, [(k_B T)^2 + (\hbar \omega)^2],
\end{equation}
where $\xi_0 = 29.4 \AA$, $\Gamma \simeq 47 meV$.  For the QC scaling in the $\sigma$-model\cite{CHN,CSY},
\begin{equation}
\frac{c}{\xi}\, = 1.04 k_B T.
\end{equation}
On the other hand, the Aeppli \textit{et al.}\cite{aeppli} result gives, if $\xi$ is measured in lattice constants:
\begin{equation}
\frac{1}{\xi}\, \simeq \frac{k_B T}{\Gamma}\,,
\end{equation}
Thus, for $x = 0.14$ we find
\begin{equation}
c = 1.04 \Gamma = 49 meV,
\end{equation}
a value that is much less than that found for the insulator value $c_{ins} = 224 meV$, and  
is qualitatively consistent with the behavior of the effective $J$ inferred from 
the bulk spin susceptibility measurements.

Another comparison of theory with experiment on the spin liquid phase diagram comes from
the neutron scattering measurements of the magnetic correlation length.
For the Heisenberg spin liquid, the correlation length $\xi=1$ at $T \simeq T^m$.
On using the results for $\xi$ obtained by Aeppli \textit{et al.}, we find:
\begin{equation}
T(\xi=1) = T^m(x=0.14) = 540K  \pm 100K.
\end{equation}
In good agreement with the result, $T^m =460K$,  at  $x=0.14$ doping obtained using Eq.(\ref{Tmresu}).

\begin{figure}
\begin{center}
\includegraphics[width=0.75\textwidth]{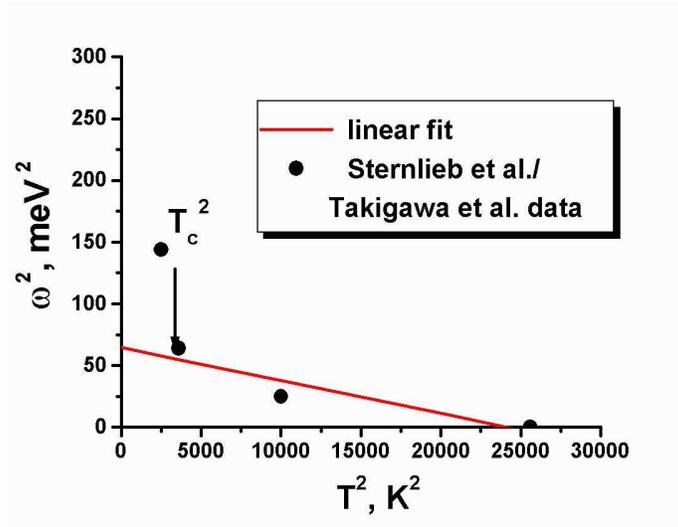}
\caption{Temperature-frequency analysis of scaling failure, data from Sternlieb \textit{et al.} and Takigawa (6.63 NMR)}
\label{st1}
\end{center}
\end{figure}

A further important point that follows from the results of
Aeppli \textit{et al.}\cite{aeppli} concerns the breakdown of scaling, which should occur at a universal scale
$T_a$, such that
\begin{equation}
k_B T_a = \sqrt{(\hbar \omega)^2 + (k_B T_{br}(\omega))^2},
\end{equation}
where  $T_{br}$ is the temperature of the breakdown of $\omega/T$ scaling seen in the neutron
scattering experiments at a fixed frequency $\omega$. NMR experiments just measure $T_a$ directly. We analyze
the data of Sternlieb \textit{et al.}\cite{sternlieb} using this framework to
see if it is consistent with the results shown in Fig. \ref{st1}.
According to Aeppli \textit{et al.}, the breakdown of scaling frequency/temperature can be fit to the  form:
\begin{equation}
\omega^2_{br} = E_{br}^2 - A T_{br}^2
\label{break}
\end{equation}
From this analysis we find, for YBaCuO$_{6.6}$:
\begin{equation}
E_{br}= 8.03 meV, \ \ A= - 0.00256 (meV^2/K^2),
\end{equation}
or
\begin{eqnarray}
\omega_{br}(T=0) &=& 8.03 meV = 93 K, \\ 
T_{br}(\omega = 0) = T^m/3 &=& 159 K, \\ 
\frac{\omega_{br}}{T_{br}}\, &=& 0.58.
\end{eqnarray}
Overall, this experimental picture is self-consistent, although the experimental finding of the simple form for the
breakdown Eq. (\ref{break}) is rather surprising.

\begin{figure}
\begin{center}
\includegraphics[width=0.75\textwidth]{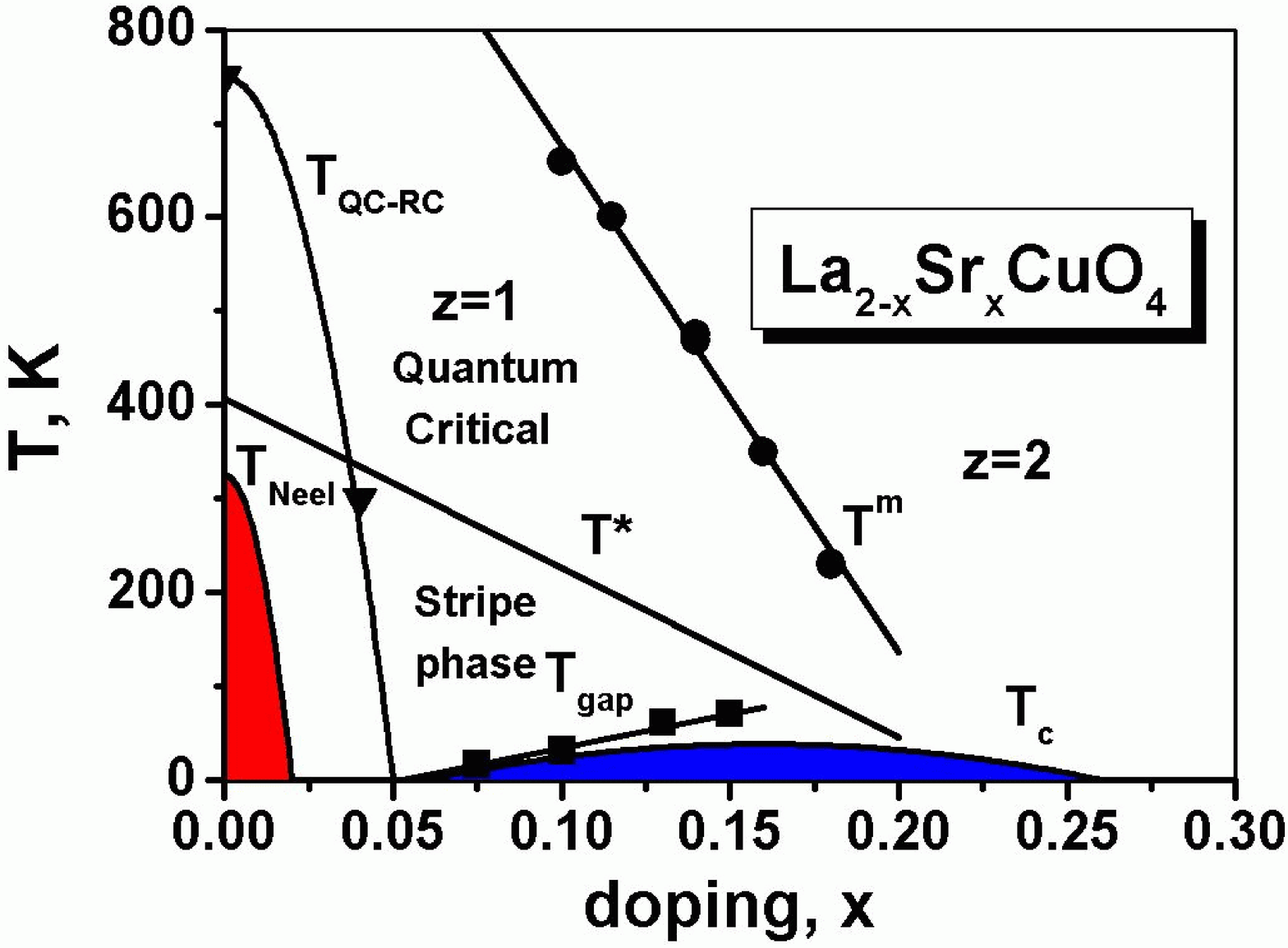}
\includegraphics[width=0.75\textwidth]{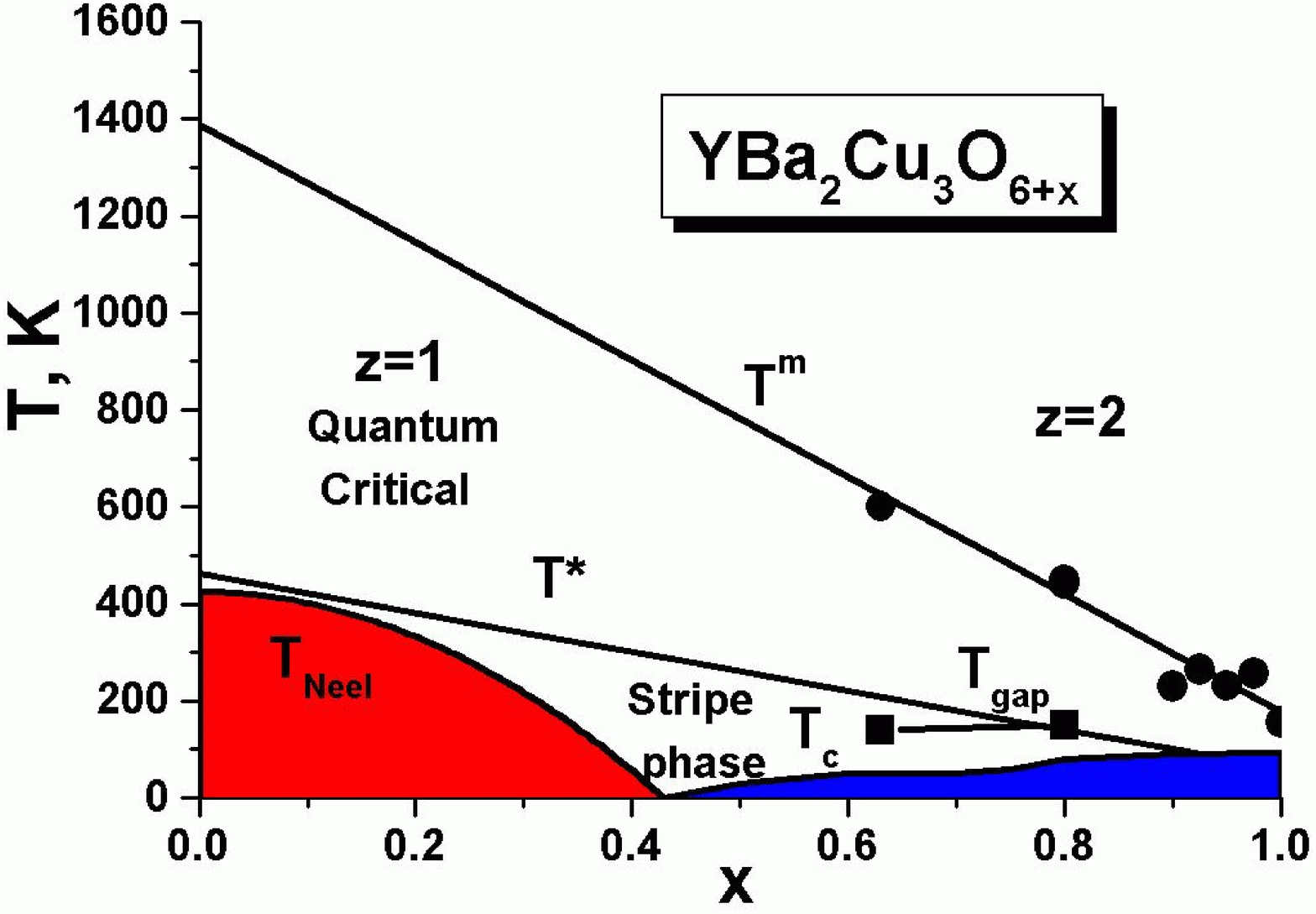}
\caption{(a) The spin liquid phase diagram for La$_{2-x}$Sr$_x$CuO$_4$. 
$T^m(x)$ corresponds to a crossover  at $\xi(T,x)=1$ to the strongly correlated  phase in the spin liquid. 
$T^*(x) \simeq T^m(x)/3$ corresponds to the appearance of fluctuating stripe order
 $T_{gap}$ marks the spin liquid crossover from the quantum critical 
(QC) to quantum disordered (QD), or gapped spin liquid, regime in the stripe phase. $T_{QC-RC}$ is the spin liquid crossover from 
the QC to renormalized classical (RC) regime.
        (b) The spin liquid phase diagram for YBa$_2$Cu$_3$O$_{6+x}$ displays similar crossovers.}
\label{phased}
\end{center}
\end{figure}

We conclude this section with a brief discussion of phase diagram for the spin liquid that is
consistent with the experimental constraints discussed above.
Our proposed phase diagram\cite{BPprl} is shown in Fig.\ref{phased}.
 
The phase transition or a crossover at a doping level $\sim 0.22$ is that 
from the conducting Fermi liquid on the right-hand side to pseudogap matter on the left-hand side, in which 
a portion of the quasiparticle Fermi surface has lost its low-frequency spectral weight, becoming localized in an insulating
spin liquid that coexists with the remaining Fermi liquid component. The two interacting components are present at all temperatures
in the underdoped regime, $x < 0.22$. At high temperatures $T > T^* \simeq T^m/3$ the interaction leads to a modification
of $J$, the effective exchange interaction in the spin liquid, while the Fermi liquid component remains intact. 
As the temperature is lowered, the growth of correlation length in the spin liquid results in stronger renormalization
of the Fermi liquid component, and the formation of the Fermi arcs 
at temperatures $T < T^* \simeq T^m/3$.  The antiferromagnetic 
correlation length also becomes finite at low temperatures, which results in a gap for the spin excitation
spectrum at $\bm{Q} = (\pi, \pi)$. 
The quantum critical point at $x \sim 0.05$ corresponds to a quantum transition between two different
states of the spin liquid. 

We now return to the question of how close the measured spin liquid maps onto the Heisenberg spin liquid. While
there is a one to one correspondence for the static spin susceptibility above $T \simeq T^*$, the difference
becomes large at lower temperatures, where the correlation length $\xi$ and the spin liquid gap $\Delta_{SL}$
saturate at a finite value.   Presumably
it is this difference that is responsible for the mild departure in concentration dependences deduced for $f(x)$
and $T^m(x)$.

\section{Discussion}

Given the presence of both a spin liquid and a Fermi liquid in the normal state of underdoped superconductors, it is natural 
to consider their interaction as being responsible for their anomalous normal state properties and the transition to a gapped 
quantum state at $T^*$ and to superconductivity at $T_c$. We begin by considering the extent to which the observed scaling 
behavior reflects that interaction, and what might be some of its  other consequences.

\begin{figure}
\begin{center}
\includegraphics[width=0.75\textwidth]{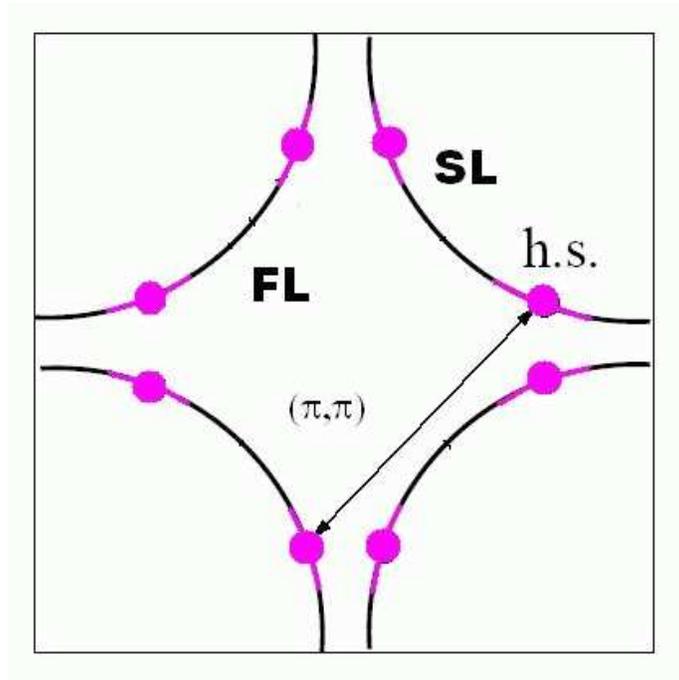}
\caption{Hot regions near spots on the Fermi surface separated by wavevector  $\bm{Q} = (\pi, \pi)$ and cold regions away
from the hot spots}
\label{hotcold}
\end{center}
\end{figure}

From an examination of the experimental results presented in the previous sections we draw the following conclusions:

(i) The scaling with $T^m$ of various transport properties of the non-Landau Fermi liquid tells us that the transport properties 
in the normal state of the cuprate superconductors are dominated by the magnetic coupling between the quasiparticles and the 
spin liquid excitations

(ii) In the 1-2-3 materials, the scaling of  $T^*$ and $\Delta_g$, the d-wave energy gap found below $T^*$ for quasiparticles 
located in the "hot" regions of the Fermi surface, and below $T_c$ for the remaining "cold" or nodal quasiparticles, 
tells us that the physical mechanism for their gaps and subsequent superconductivity must be magnetic.

(iii) As may be seen in Fig.\ref{hotcold}, the hot regions seen in the ARPES experiments are defined by their proximity to points on 
the Fermi surface that are separated by commensurate wavevectors, $\bm{Q} = (\pi, \pi)$; their physical location tells us 
that the magnetic interactions responsible for the gap must be peaked at $\bm{Q}$. This is to be expected if the "glue" 
for the  quasiparticle interaction is provided by the spin liquid excitations of the Heisenberg model that exhibit the MMP 
form with a strong peak at $\bm{Q}$, since their coupling to the Fermi liquid yields an induced quasiparticle interaction 
that is likewise peaked at $\bm{Q}$.  Moreover, in agreement with the early models of spin-fluctuation-induced 
superconductivity\cite{MBP,Mor,MP} the resulting  energy gaps will be d-wave.

 (iv) The penetration-depth measurements of the doping-dependence of the  of the superconducting condensate tell us that where 
present in the underdoped materials, the spin liquid does not participate in their superconductivity.
 
 (v) The fact that the scaling relations below $T^m$ for a pure Heisenberg model are much better obeyed in the 2-1-4 materials 
than in the 1-2-3 materials tells us that the influence of the hot quasiparticles on the  spin liquid is considerably weaker 
in the 2-1-4 materials.

 (vi) Since the spin liquid excitations in the two classes of materials are quite similar, these differences plausibly arise from 
band-structure effects\cite{Ole,Ole1} that reduce the importance of hot spots on their Fermi surface. 

 (vii) These last conclusions concerning the relative strengths of the coupling of hot quasiparticles to the spin liquid and their 
spin-fluctuation-induced quasiparticle interaction in the 1-2-3 and 2-1-4 materials provide a simple physical explanation of some 
of the major differences between their physical behavior of the two classes of materials. It is why the 2-1-4 materials show no 
indications of either a spin fluctuation gap or a quasiparticle gap opening up at $T^*$, and may be related to their  significantly 
lower superconducting transition temperatures. 

 (viii) There are many indications that "stripy" behavior is more prevalent in the 2-1-4 materials; this may be due to the fact that 
the fluctuating  domains become well established only when fairly strong AF correlations are present. For the 1-2-3 materials gaps 
open up at $T \sim T^m/3$, so the maximum AF correlation length one expects to find is $\sim 3$ lattice spacings; for the 
2-1-4 materials, on the other hand, one finds AF correlation lengths $\sim 8$  lattice spacings near optimal doping and still 
longer correlation lengths are to be expected for the underdoped materials.

 (ix) Experiment indicates that at low temperatures $T \sim T^* \simeq T^m/3$, charge fluctuations, domains, and tendencies toward 
phase separation all play an important role, although the two-component nature is present in the underdoped regime at all 
temperatures, and is therefore intrinsic.

 (x) Our two-fluid model provides a natural way to develop a phenomenological description of  stripe formation at low frequencies, 
since all that is required is to let $f(x)$ vary periodically in space and time. 

We conclude this section with a brief  discussion of some necessary elements of a microscopic description of the spin and Fermi liquids. 
We note first that for the spin liquid, the AF critical point at $x \simeq  0.05$ is responsible for the observed $z = 1$ QC behavior. 
The spin liquid component at temperatures $T > T^*$  is well described to first approximation by the standard bosonic  
$\phi^4$ Hertz-Millis theory of quantum critical behavior\cite{herts,millis}. For   $\xi > 1$ one is outside the critical region, 
in the vicinity of weak-coupling Gaussian fixed point, where fluctuation corrections are not important. 
The standard damping due to its  interaction with fermions gives rise to 
the  $z = 2$ mean field Ornstein-Zernike/MMP form. In the critical region  $\phi^4$ theory becomes equivalent to the
QNL$\sigma$ model\cite{CHN,CSY} (the two theories differ by an irrelevant
operator). However, at low correlation lengths the irrelevant operator of the $\phi^4$ theory plays a crucial role, 
inducing a crossover to the mean field Gaussian regime\cite{BP} with   $\xi \propto T^{-1/2}$ for $N = \infty$ 
version\cite{zinn-justin}. Such $z = 1$ to $z = 2$ mean field crossover cannot be obtained in the 
quantum non-linear $\sigma$-model with fermionic damping\cite{SCS}.

There appear to be two kinds of departures from the simple Heisenberg description. The first concerns the static bulk susceptibility 
that falls below the currently calculated values for the Heisenberg model for $T <  T^m/3$. Because this fall-off is essentially the 
same in both the 2-1-4 and 1-2-3 materials, it is likely not brought about by spin-fermion coupling, since that coupling is quite 
different for the two materials. It could represent an inadequacy in the calculations of the expected behavior of the Heisenberg model 
at these low temperatures. The second concerns the opening up of a gap in the low frequency spin liquid excitation 
spectrum at $(\pi, \pi)$ that is seen as a minimum in $T_1 T$ at $150K \simeq 0.31 T^m$ in the NMR experiments on the 1-2-4 material 
and at $140K \simeq 0.23 T^m$ in the  YBa$_2$Cu$_3$O$_{6.63}$ material. It is plausible to conclude that it reflects a reduction 
in quasiparticle  damping brought about by the onset of the hot quasi-particle gapped behavior at $T^*$. However,it would then seem 
that the  transition to this new quantum state is not universal, i.e. does not always occur at a fixed fraction of $T^m$. 
Indeed these NMR-determined onset temperatures, $T^*$, scale, to first approximation with $T_c$; thus for the 1-2-4 material one 
has $T^* \simeq 2.1 T_c$, while for YBa$_2$Cu$_3$0$_{6.63}$ one finds $T^* \simeq 2.3 T_c$. Based on these estimates, 
and those obtained from the ARPES and STM experiments on 2212 members of the  1-2-3 family , our candidate phase diagram for 
the 1-2-3 materials is given in Fig. \ref{phased}b. It differs from the corresponding diagram for the 214 materials in the presence of a 
"high temperature" transition at $T^*$ to the gapped quantum state.  

In considering  microscopic  models for the evolution of  Fermi liquid behavior with doping, we begin by recalling that high-T$_c$ 
superconductors are doped charge-transfer insulators\cite{zaanen1,zaanen2}. 
Thus, an adequate starting point for a description of the CuO$_2$ planes follows from their chemistry, 
and is the three-band Hubbard model composed of 3 d$_{x^2-y^2}$ orbitals on the Cu site and 2 p$_{x,y}$ orbitals 
on the O sites\cite{emery,GS,varma}.  The two key parameters of the three-band
Hubbard model are are the charge-transfer energy $E_{CT} = \epsilon_p - \epsilon_d$, and the copper on-site
Coulomb repulsion, $U_d$.
In the limit of strong p-d hybridization and very strong on-site Cu Coulomb
interaction, the model gives rise to 5 different bands: an antibonding band, further split into upper Hubbard band 
(Cu $d^{10}$ configuration), and lower Hubbard band (Cu $d^{8}$ configuration) by the Coulomb interaction,
a bonding band, further split into Zhang-Rice (ZR)\cite{ZR} singlet and triplet bands, and the non-bonding oxygen band. 

 Following Anderson\cite{RVB}, most theoretical studies have considered a simpler one band Hubbard model, which has fewer parameters
and assumes that the low-energy physics of the electronic states in the vicinity of the charge-transfer gap is described
well by taking a relevant low-energy subset of this complex band structure, the ZR singlet band and the upper Hubbard band, 
separated by $U \simeq E_{CT} \sim 1.7 eV$ of the order of the charge-transfer gap. 
Numerical studies of the three-band Hubbard model\cite{hybersten} support such
reduction. The two low-energy bands of the effective single-band Hubbard model are strongly correlated hybridized
copper-oxygen bands, with the upper Hubbard band having mostly oxygen character\cite{saw1,saw2,Ole,Ole1}.

However, on the basis of numerical LDA calculations, it has been 
suggested that a single-band tight binding model may not provide a sufficiently good qualitative description
of the many bands involved\cite{Ole,Ole1}. Moreover, we have seen that $\chi_{dd}$ contains the spin liquid
component localized on the copper sites, while $\chi_{pd}$ and $\chi_{pp}$ pick up contributions from the holes
on the oxygen sites. The difference in relevant form factors could potentially explain many anomalies observed
for NMR relaxation rates\cite{walst} and inelastic neutron scattering experiments, as we have
discussed in the previous section.  

The analogy with two-fluid model for heavy fermions\cite{NPF,NPFcurro} can be better understood
from an alternative approach to three-band Hubbard model\cite{emery,GS,varma}, that is perturbative
in p-d hybridization, $t_{pd}$.
In case of exactly one hole per Cu and strong Hubbard Coulomb on-site repulsion $U_d$, the low-energy
Hamiltonian for the three-band Hubbard model reduces to the Heisenberg spin Hamiltonian with exchange
coupling
\begin{equation}
J_H = \frac{4 t_{pd}^4}{(E_{CT} + U_{pd})^2}\, \left( \frac{1}{U_d}\, + \frac{2}{2 E_{CT} + U_p}\, \right),
\end{equation}
since the charge excitations have a gap $E_{CT}^g \simeq 1.7 eV$. 
Numerical calculations for the three-band Hubbard 
model\cite{hybersten} extract the value $J_H \simeq 0.13 eV$, consistent with that observed in neutron scattering experiments.
The oxygen holes and the nearest $Cu$ spin interact via a strong Kondo exchange interaction, which can be
obtained in perturbation theory in $t_{pd}/E_{CT}$,
\begin{equation}
J_K = t_{pd}^2 \left(\frac{1}{E_{CT}}\, + \frac{1}{U_d - E_{CT}} \right).
\end{equation}
The perturbative analysis of the three-band Hubbard model reduces it to the spin-fermion model via the Schrieffer-Wolff\cite{SW}
transformation,
\begin{equation}
H_{SF} = \sum_{j,j', \sigma} t_{jj'} c^{\dagger}_{j \sigma} c_{j' \sigma}  +
\sum_{j, j', i, \sigma, \sigma'} J_K (j,j',i) c^{\dagger}_{j \sigma} \bm{\sigma}_{\sigma \sigma'} \bm{S}_i c_{j', \sigma'} +
J_H \sum_{i, i'} \bm{S}_i \bm{S}_{i'}.
\label{SFmodel}
\end{equation} 
Here the index $i$($j$) runs over Cu (O) sites, $c^{\dagger}_{j \sigma}$, $c_{j \sigma}$ are creation and annihilation operators,
respectively, for holes in oxygen p bands. The Kondo interaction is, in general, non-local\cite{GS}.
The Hamiltonian Eq.(\ref{SFmodel}) is the Kondo-Heisenberg model, known for 
the periodic Anderson model in many physical systems,  including
magnetic semiconductors\cite{dietl}, heavy fermions (the Kondo lattice model)\cite{SW}, and high-T$_c$
superconductors\cite{emery,GS}. The heavy fermion materials and high-$T_c$ materials correspond to
different limits of the model Eq.(\ref{SFmodel}). In heavy fermion superconductors the Kondo
interaction $J_K$ is initially comparatively small, and gets renormalized by the conduction band
as the temperature is lowered, leading to the onset of coherence at $T^*$, and the two-fluid
heavy fermion behavior\cite{NPF,NPFcurro,barzykin} with a temperature-dependent fraction of spin liquid 
(spin impurity contribution). 
The high-$T_c$ superconductors, on the other hand, are in the 
strong-coupling Zhang-Rice\cite{ZR} limit of the same microscopic model, Eq.(\ref{SFmodel}). 
The Kondo interaction is very large, $J_K \sim 0.7 eV$, which leads to a formation and propagation 
of the ``Zhang-Rice singlets'',  reducing the microscopic Hamiltonian to an  effective
one-band $t-J$ model\cite{RVB}. The projection operators of the Zhang-Rice singlet\cite{ZR} 
automatically imply that $\chi_{pd} = \chi_{pp} = 0$. Thus, as suggested by MMP\cite{MMP},
different nuclei then feel the presence of a single temperature-dependent spin component residing on copper, 
or $\chi_{dd}$. The Zhang-Rice singlet picture (and the t-J model), however, inevitably gets violated at higher doping, as
a single large Fermi surface develops, that corresponds to antibonding p-d hybridized  band. 
A linear decrease of effective $J(x)$ with increased doping $x$ is ``high-energy physics'',
as is evident from, for example, an adiabatic treatment of the t-J model\cite{PWA}.
Thus, the fraction of spin liquid\cite{BPprl} 
observed in ``low energy'' experiments on the cuprate superconductors will be temperature-independent.
In heavy fermion metals it is the mixed $c-f$ contribution, $\chi_{cf}(T) = (1 - f(T)) \chi_{FL}(T)$,
that gives rise to the heavy Fermi liquid contribution\cite{NPFcurro}. In high-temperature superconductors,
on the other hand, one normally uses the Zhang-Rice singlet band and the upper Hubbard band of the three-band Hubbard
model as lower and upper Hubbard bands of the effective single-band Hubbard Hamiltonian. 
Using the analogy the heavy fermion two-fluid model\cite{NPF,NPFcurro} and the two-fluid model for high-T$_c$ 
superconductors, the Fermi liquid contribution in the cuprates should arise from $\chi_{pd}$ and $\chi_{pp}$ parts of the bulk
spin susceptibility, which, strictly speaking, indicates a violation of the Zhang-Rice singlet picture at high doping,
given by Eqs. (\ref{ddpart}), (\ref{pdpart}).
Since the effective mass of carriers is not heavy in high-$T_c$ superconductors, 
one would expect both $\chi_{pp}$ and $\chi_{pd}$ contributions to be present. 
This result contrasts strongly with the pure Zhang-Rice\cite{ZR} limit in which all 
contributions arise from $\chi_{dd}$\cite{MMP}.
The key answer to this very important question 
is held by the new Knight shift experiments\cite{NPFcurro,HS}, since then Knight shifts on different nuclei should see a different
temperature dependence, at least below the superconducting temperature, $T_c$. 

\section{Conclusion}

Our review of scaling behavior in the cuprate superconductors makes evident the critical role that the localization of Cu spins by 
strong on-site repulsive Coulomb interactions plays in determining the properties of the magnetically underdoped superconductors, 
those whose planar hole concentrations are $<0.20$. While there is much more in the way of details that experiment can and will 
provide, from additional c-axis  NMR experiments on 1-2-3 materials that enable one to follow the emergence of a distinct Fermi 
liquid contribution,  to ARPES,  NMR , and STM experiments that fill in the details of the emergence of the gapped hot quasiparticle
state at $T^*$, we believe that the overall low-energy physical picture of a normal state described by two fluids whose 
interaction leads to its remarkable properties has now been firmly established. 

	It is equally clear that much much more remains to be done, beginning with the development of a better microscopic foundation 
for the physical picture that experiment has "forced" upon us, and an improved treatment of transport processes based upon it.  
We mention here but a few of the many remaining challenges and mysteries that surround these fascinating materials:

(i) There are striking similarities in the doping dependence of $T_c$ in the 1-2-3 and 2-1-4 materials despite the difference in their 
spin-fermion couplings and the magnitudes of their respective transition temperatures. Does this tell us that these differences are 
primarily in how the hot quasiparticles couple to spin fluctuations and that the spin-fermion couplings for the cold quasiparticles that 
bring on their superconductivity can somehow be scaled for these two classes of materials?

(ii) Although the spin liquid ceases to be a significant  "player" for doping levels greater than, say, $0.20$, the existence of a 
smooth cross-over in most physical properties of the normal state, including the superconducting transition, as one goes from 
magnetically  underdoped to magnetically overdoped materials, suggests that its disappearance must be accompanied by the 
emergence of sufficiently strong antiferromagnetic correlations in the Fermi liquid that enable its properties to merge smoothly 
with those induced by the spin liquid. Can these be detected experimentally and understood microscopically? Might the disappearance 
of the spin liquid be accompanied by the emergence of a single-band Hubbard model that is valid for the overdoped materials? 

(iii) Do the gapped hot quasiparticles found below $T^*$ in the 1-2-3 materials become effectively localized and so become part of the 
spin liquid?  Is there any way to make their gapping more productive, in terms of their becoming part of the superconducting condensate, 
thereby opening the way to a superconducting transition at the considerably higher transition  temperature, $T^*$?

(iv) Can the lessons learned from the cuprates be transferred to other materials, and so lead over time to the discovery of still higher 
temperature superconductors?

\section{Acknowledgements}

We would like to thank E. Abrahams, N. Curro, L.P. Gor'kov, J.
Haase, J. Schmalian, and C.P. Slichter for stimulating
discussions, and acknowledge support for VB from NHMFL through the 
NSF Cooperative agreement No. DMR-008473 and for DP from the Institute for Complex 
Adaptive Matter and the US Department of Energy.

\end{document}